\documentclass[ twocolumn]{aastex62}
\usepackage[T1]{fontenc}
\usepackage{ae,aecompl}
\usepackage{amssymb}% http://ctan.org/pkg/amssymb
\usepackage{pifont}% http://ctan.org/pkg/pifont
\received{xxxx}
\revised{xxxx}
\accepted{xxxx}
\submitjournal{ApJ}

\shorttitle{Thickness of HI layers in BLUEDISK}
\shortauthors{Randriamampandry et al.}
\begin{document}

\title{The Bluedisk survey: thickness of H{\sc i} layers in gas rich spiral galaxies}

\correspondingauthor{Toky H. Randriamampandry}
\email{rtoky@pku.edu.cn}

\correspondingauthor{Jing Wang}
\email{jwang@pku.edu.cn}
\author{Toky H. Randriamampandry}

\author{Jing Wang}
\affiliation{Kavli Institute for Astronomy and Astrophysics, Peking University
5 Yiheyuan Road, Haidian District, Beijing 100871,People's Republic of China}
%\nocollaboration{(AAS Journals Data Scientists collaboration)}

\author{K. Moses Mogotsi}
%\altaffiliation{Creator of AASTeX v6.2}
\affiliation{South African Astronomical Observatory, PO Box 9, Observatory, Cape Town, South Africa}
\affiliation{Southern African Large Telescope, P.O. Box 9, Observatory, 7935, Cape Town, South Africa}
\begin{abstract}
We use an empirical relation to measure the H{\sc i} scale height of relatively H{\sc i} rich galaxies using 21-cm observations. The galaxies were selected from the BLUEDISK, THINGS and VIVA surveys. We aim to compare the thickness of the H{\sc i} layer of unusually H{\sc i} rich with normal spiral galaxies and find any correlation between the H{\sc i} scale height with other galaxies properties. We found that on average the unusually H{\sc i} rich galaxies have similar H{\sc i} disk thickness to the control sample and the galaxies selected from the THINGS and VIVA surveys within their uncertainties. Our result also show that the average thickness of the neutral hydrogen inside the optical disk is correlated with the atomic gas fraction inside the optical disk with a scatter of $\sim$0.22 dex. A correlation is also found between the H{\sc i} scale height with the atomic-to-molecular gas ratio which indicates the link between star formation and the vertical distribution of H{\sc i} which is consistent with previous studies. This new scaling relation between the H{\sc i} scale height and atomic gas fraction will allow us to predict the H{\sc i} scale height of a large number of galaxies but a larger sample is needed to decrease the scatter. 

\end{abstract}

%

\keywords{galaxies: kinematics and dynamics --- 
dark matter --- disk scale height --- gas fraction}


\section{Introduction} 
\label{sec:intro}
Neutral hydrogen gas (H{\sc i} hereafter) is the most abundant element in the universe and the primary fuel for star formation. For nearby galaxies, H{\sc i} is observed at 21 cm in emission and its distribution is often used to study the distribution of dark matter in spiral and dwarf galaxies ( e.g. \citealt{1981AJ.....86.1825B,2008AJ....136.2648D}) assuming that it traces the gravitational potential. Another application is the relationship between the gas surface density and star formation surface density, the so-called Kennicutt-Schmidt law (see \citealt{1998ARA&A..36..189K} for a review). These studies are based on the assumption that the thickness of the H{\sc i} layer is negligible. 

Knowing the three-dimensional distribution of H{\sc i} allows us to study the shape of the dark matter halo (see \citealt{1995AJ....110..591O,1996AJ....112..457O}) and the relationship between the gas volume density and star formation, recently known as volumetric star formation law (\citealt{2019A&A...632A.127B,2019A&A...622A..64B}, see \citealt{1959ApJ...129..243S} for theoretical background). Recent studies have also demonstrated the role of the gas disk thickness on gravitational instabilities of galactic disk for high and low redshift galaxies (e.g., \citealt{2010MNRAS.407.1223R,2014MNRAS.442.1230R,2012MNRAS.425.1511H}) previously assumed to be negligible.

In observations, the thickness of the H{\sc i} layer is difficult to measure for less inclined galaxies and direct measurements are only possible for edge-on galaxies (eg NGC891, \citealt{1979A&A....74...73S,1981A&A....99..298V,1997ApJ...491..140S}). However, the measurements might still be affected by projection effects produced by a warp or flaring of the outer disk. Other methods include using the H{\sc i} power spectrum \citep{2009MNRAS.397L..60D} and the spectral correlation function \citep{2001ApJ...555L..33P} which require sub-kiloparsec spatial resolution observations. The size and distribution of H{\sc i} shells have also been used as an indirect method to measure the H{\sc i} thickness. The maximum diameter of the shell produced by supernovae explosion in OB association region is believed to be comparable to the thickness of the H{\sc i} layers (e.g \citealt{1998MNRAS.299..249S}).

In theory, detailed modeling is usually used assuming that H{\sc i} is in hydrostatic equilibrium (e.g. \citealt{1995AJ....110..591O,1996AJ....112..457O}). The H{\sc i} disk thickness is determined by the equilibrium between the confining pressure due to the gravitational potential and uplifting pressures produced by gas motions \citep{iorio2018}. In this framework, the vertical density distribution of the gas is described by the stationary Euler equation \citep{iorio2018}. The density profile is obtained by solving the Euler equation and assuming that the velocity dispersion is constant along the vertical direction and the gravitational potential of a rotating gas is symmetric. The gravitational potential is computed for each component (star, dark matter, gas) and the scaleheight is varied until the model reproduces the observations using an iterative method. \cite{2019A&A...622A..64B} applied this method on 12 galaxies selected from the THINGS survey (see also \citealt{2008ARep...52..257A,2011MNRAS.415..687B}).

Recently, \cite{2019ApJ...882....5W} estimated the vertical scale height of molecular hydrogen of five luminous and ultra-luminous galaxies assuming vertical equilibrium. {\bf \cite{2019ApJ...882....5W} used the original formulation by \cite{1942ApJ....95..329S} which stated that the confining pressure due to gravity is equal to the uplifting forces produced by the interstellar medium (ISM) and formulated an empirical relation based on the measured velocity dispersion and surface densities which are observable quantities to estimate the gas scale height.} They also included the effects of other factor such as magnetic field and cosmic rays on the gas disk thickness (e.g., \citealt{2005A&A...436..585D,2016ApJ...816L..19G,2018ApJ...862...55H}).

Here, we applied a modification of this method on a sample of 28 H{\sc i} rich and control galaxies from BLUEDISK \citep{2013MNRAS.433..270W}, 14 spiral galaxies from the VIVA \citep{2009AJ....138.1741C} and 12 from the THINGS \citep{2008AJ....136.2563W} surveys. 
The BLUEDISK survey \citep{2013MNRAS.433..270W} was primarily designed to investigate the origin and existence of gas accretion in nearby galaxies. The survey consists of 23 galaxies with unusually large HI mass fraction according to the fundamental plane by \cite{2012A&A...544A..65C} and control sample which match in term of their stellar mass, stellar surface density and NUV-r color indexes. Several work have been done using BLUEDISK data such as the morphological studies of the H{\sc i} gas using moment 0 maps (\citealt{2013MNRAS.433..270W,2014MNRAS.441.2159W}). Therefore, we will make use of these measurements in this work. In this paper, we focus on the thickness of the HI gas layer and its relation to the gas fraction and other galaxies global properties. Our goal is to compare the thickness of the H{\sc i} layer of unusually H{\sc i} rich with normal spiral galaxies and derive scaling relations which will be useful to study the thickness of H{\sc i} layers for a large sample of galaxies that will be available in the near future.

The paper is structured as follow: section 2 describes the sample selection and method, the result of the tilted ring analysis, the measurements of HI scale height and the correlation between the scale height with other galaxy properties are presented and discussed in section 3. A summary is given in section 4.

\begin{figure*}
\begin{center}
\includegraphics[width=19cm]{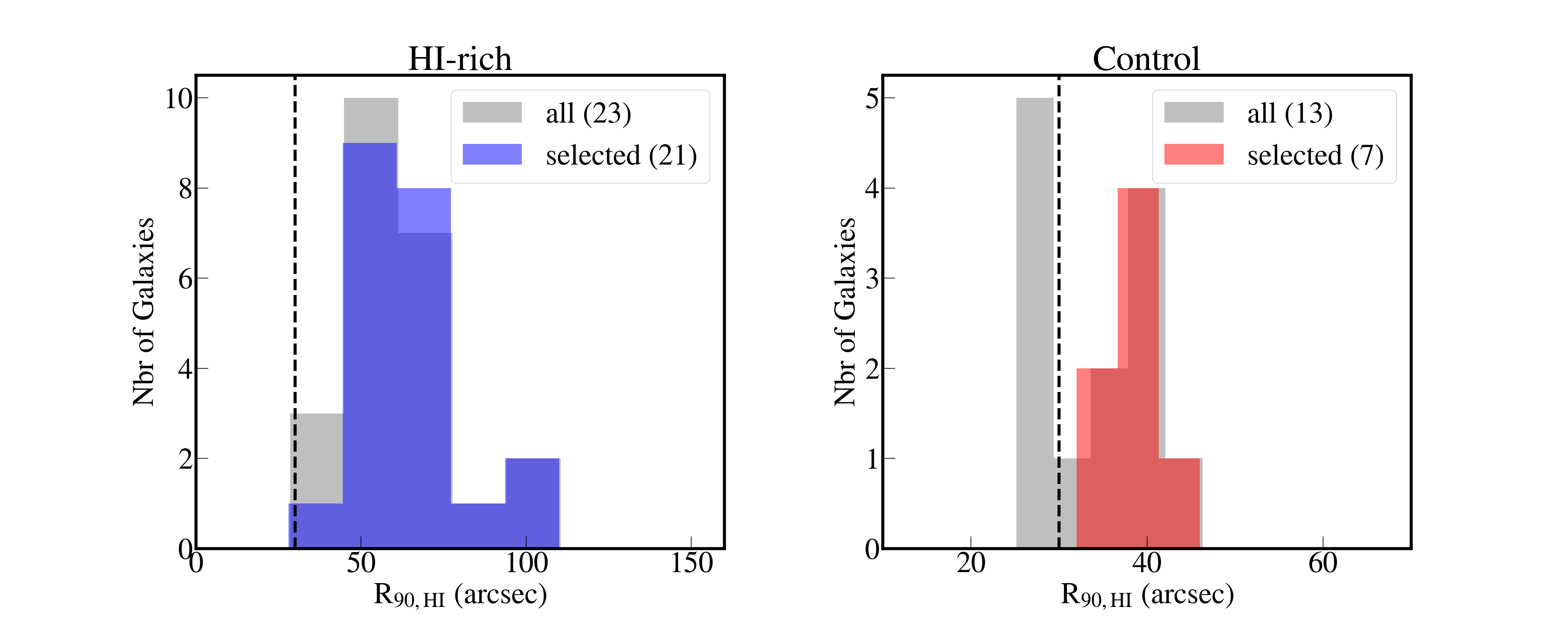}
\caption{Histogram of the H{\sc i} size distribution of the galaxies modeled using $^{3D}$BAROLO (grey) and the selected galaxies. The HI-rich sample galaxies are on the left panel and the control sample on the right panel. }
\label{fig1}
\end{center}
\end{figure*}

\begin{figure}
\begin{center}
\includegraphics[width=9cm]{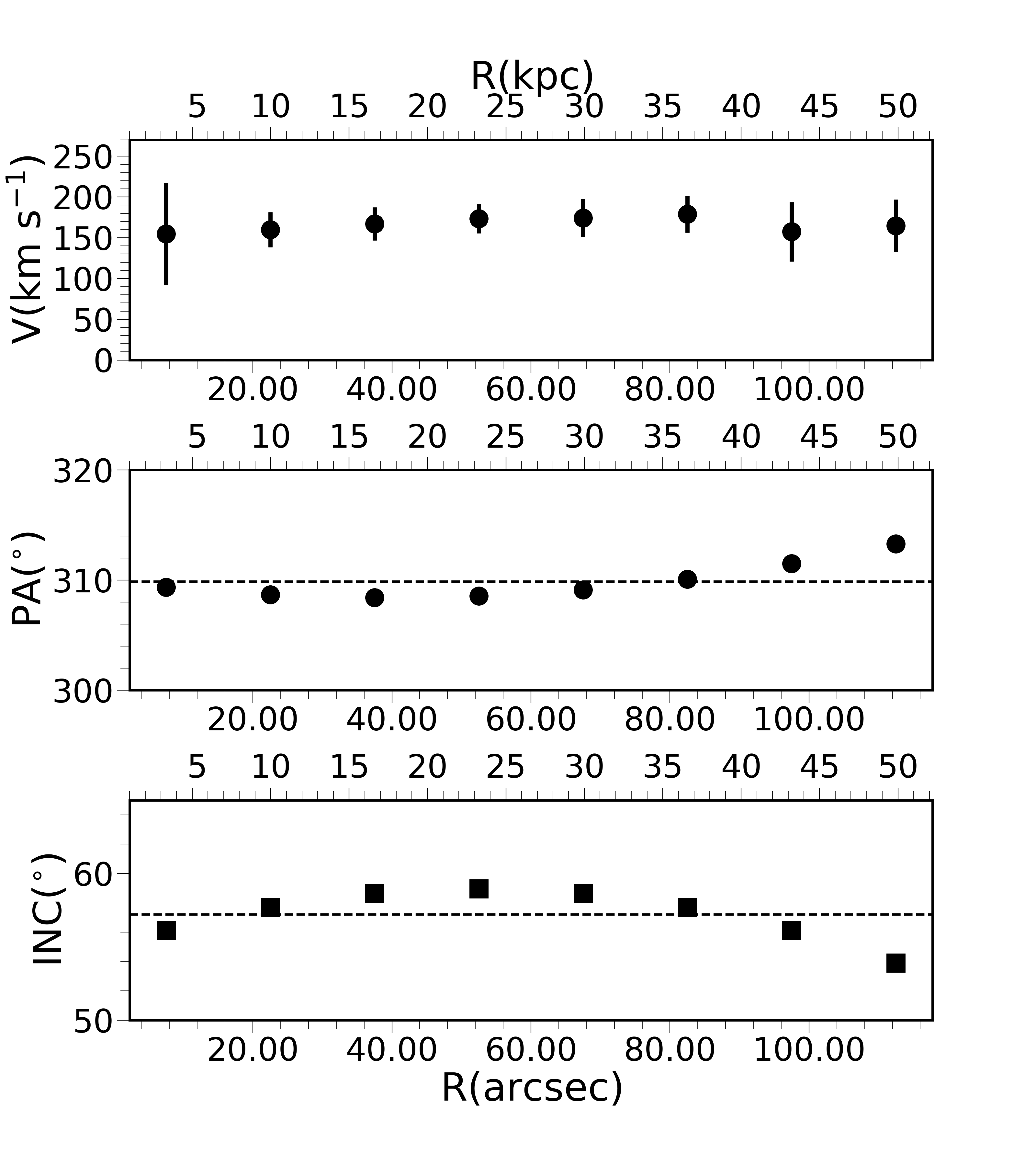}
\caption{Tilted ring analysis results for ID15; Top panel: the derived rotation curves are shown as black points with errorbars. The middle and bottom panels are the position angle and inclination respectively, the median values are shown as dashed lines. }
\label{fig2}
\end{center}
\end{figure}

\begin{figure*}
\begin{center}
\includegraphics[width=15cm]{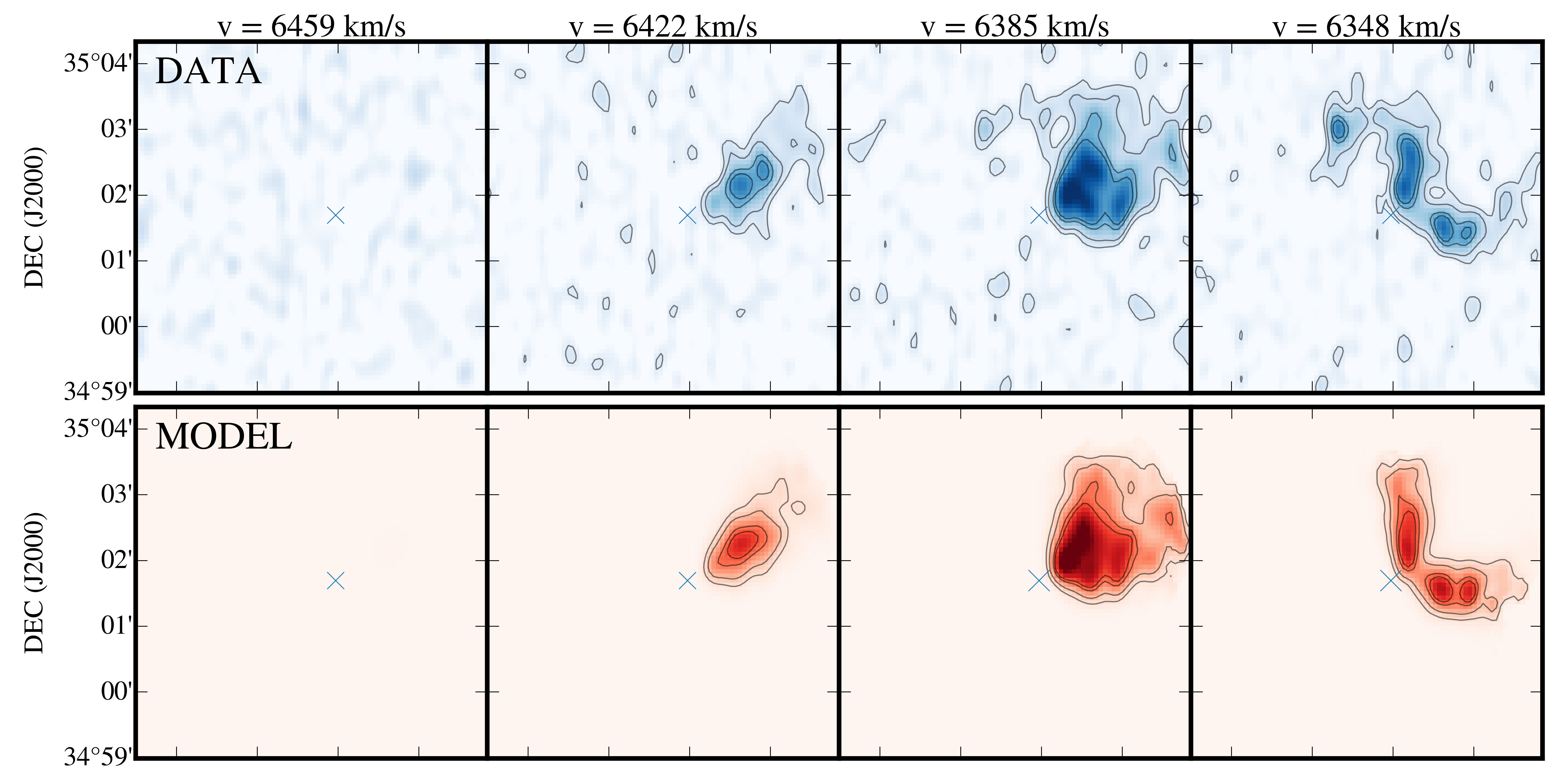}
\includegraphics[width=15cm]{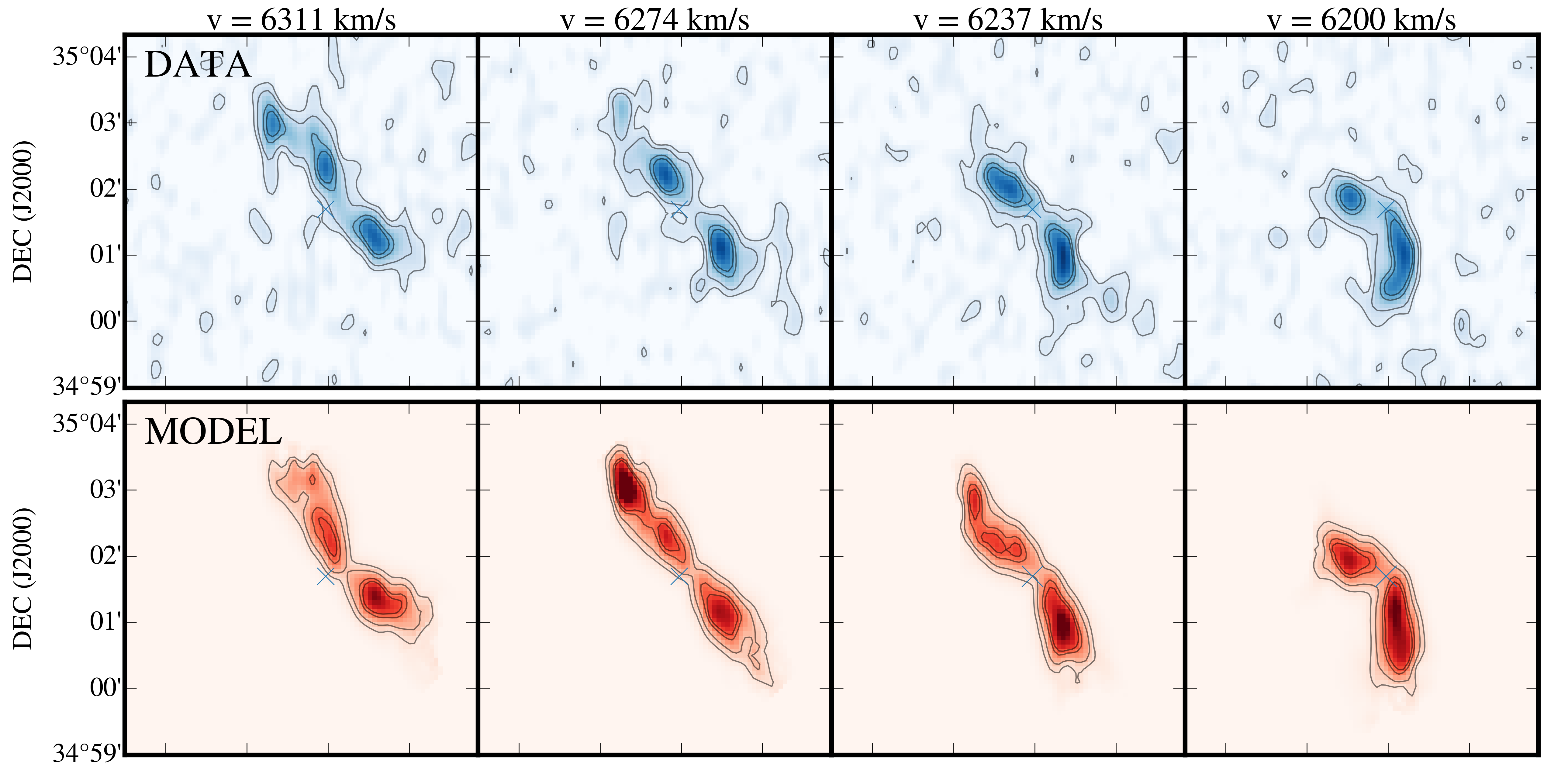}
\includegraphics[width=15cm]{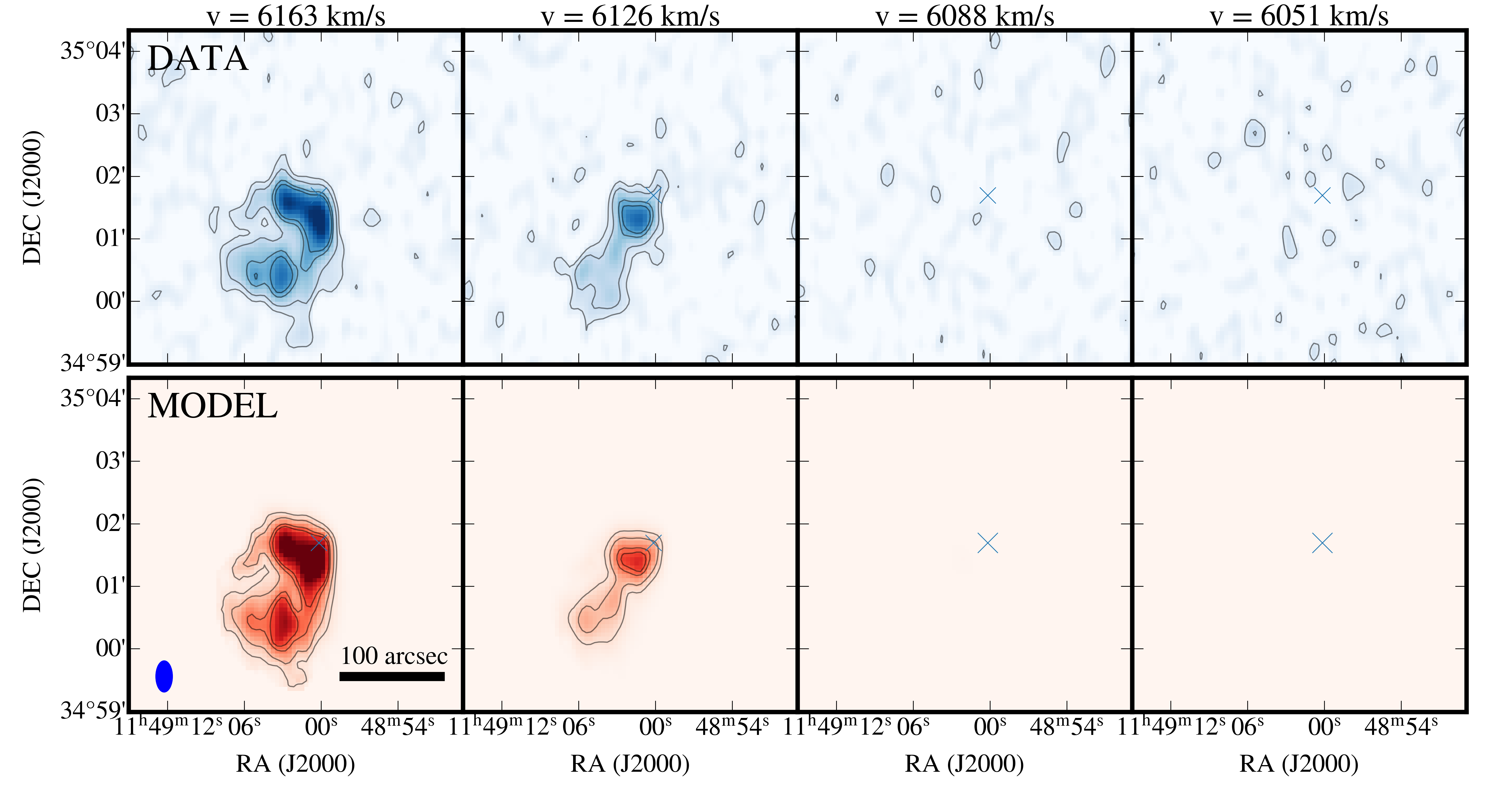}
\caption{Channel maps for ID15; Top rows: data, bottom rows: 3D-BAROLO model. Only every three channel is shown, the beam and scale bar are plotted on the bottom left panel. The contours are 2,4,8 and 10 $\sigma$ where 1 $\sigma$ $\sim$ 0.0023 Jy beam$^{-1}$.  }
\label{fig3}
\end{center}
\end{figure*}

\begin{figure}
\begin{center}
\includegraphics[width=8.5cm]{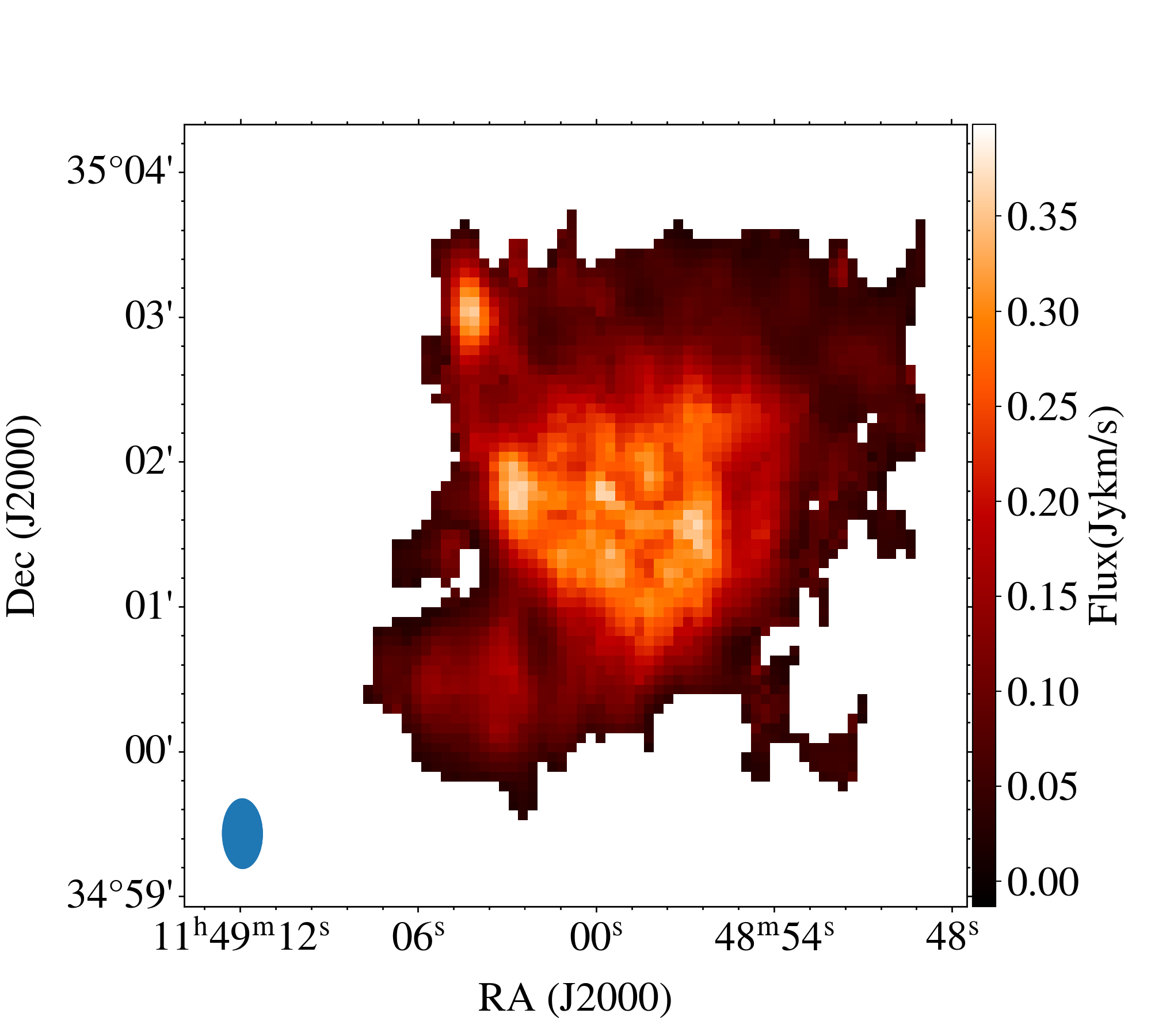}
\includegraphics[width=8.5cm]{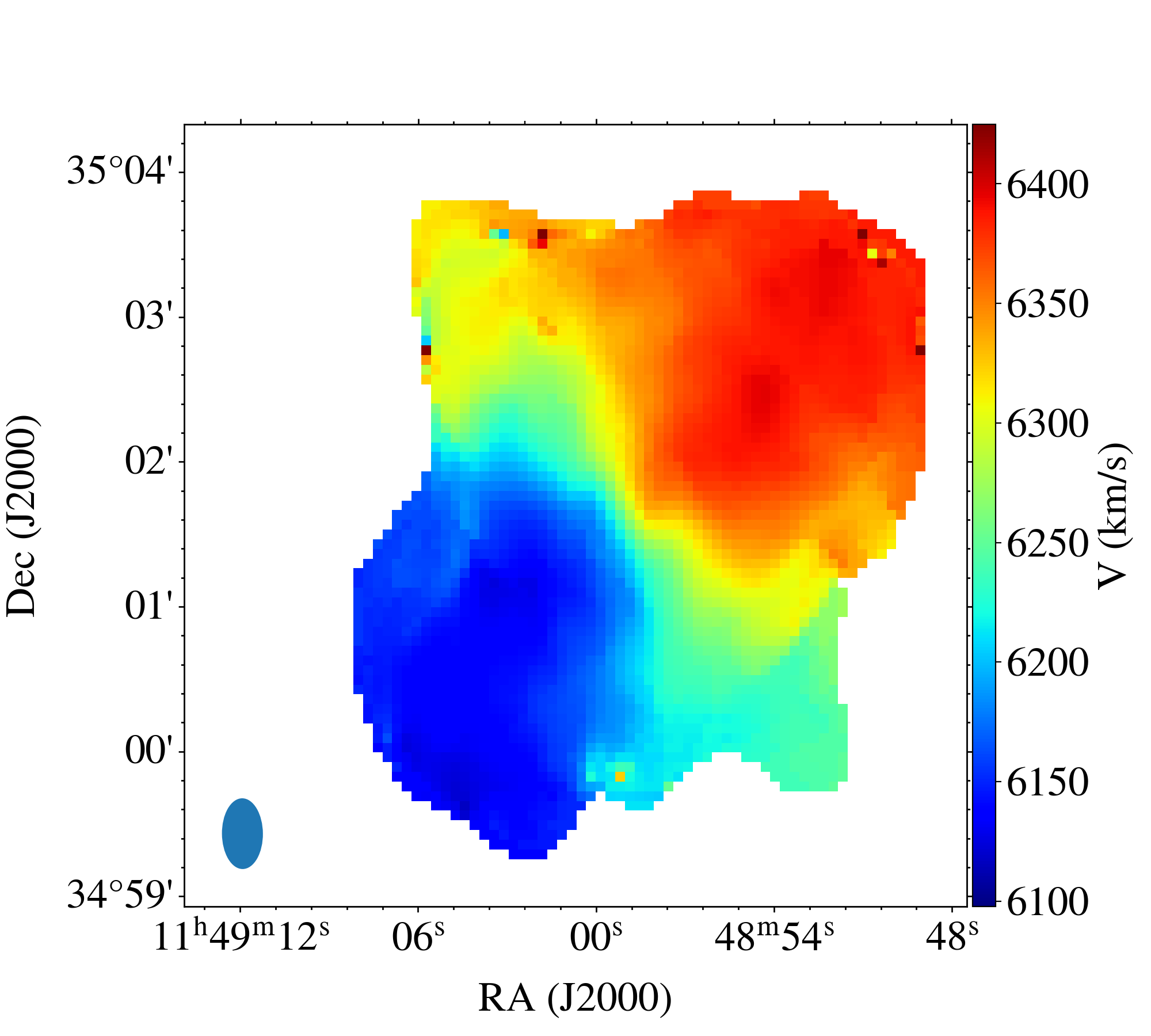}
\caption{Intensity map (top) and velocity field (bottom) for ID15. The beam is shown on the bottom left corner.}
\label{fig4}
\end{center}
\end{figure}

\begin{figure}
\begin{center}
\includegraphics[width=9cm]{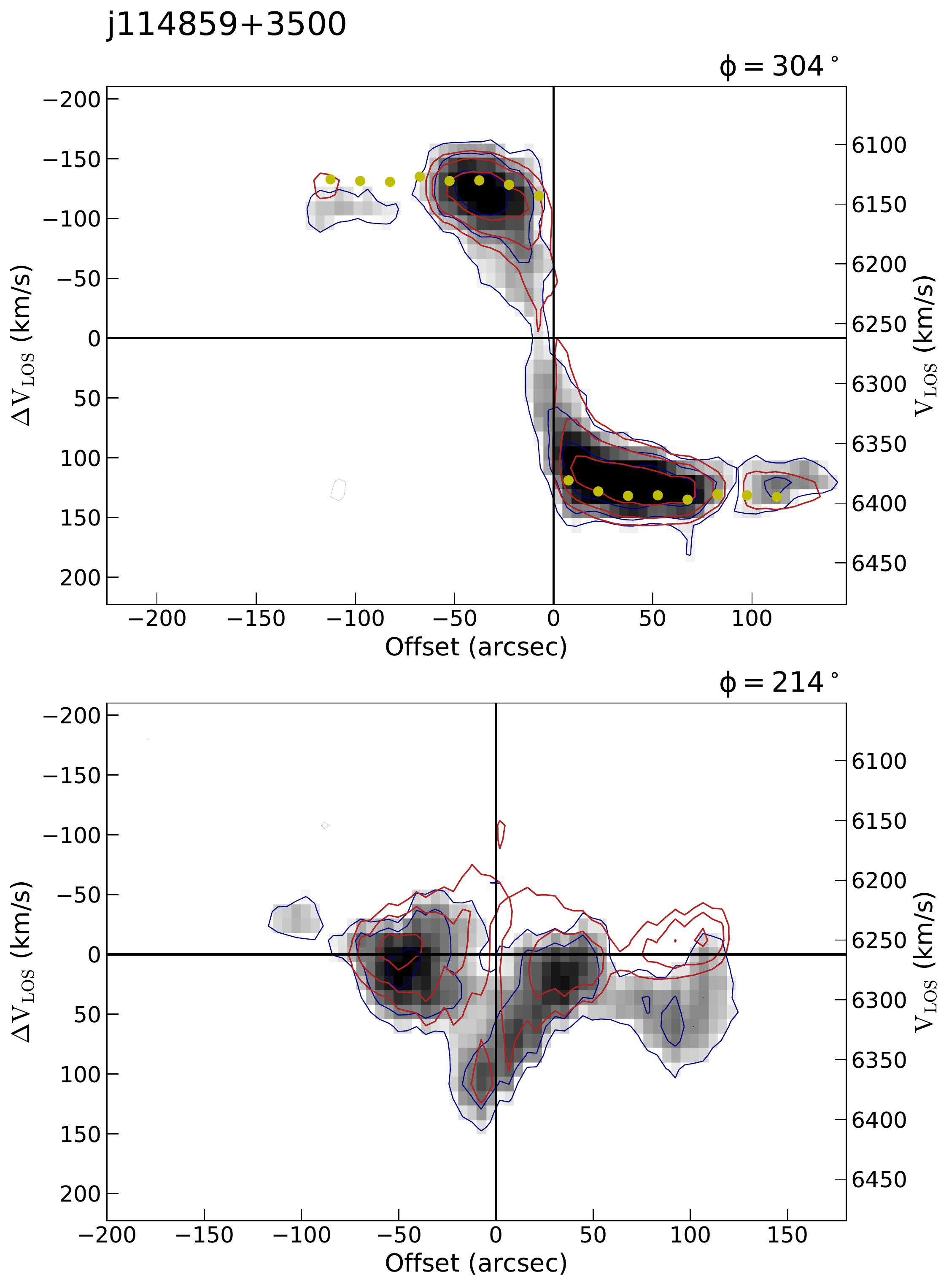}
\caption{Position-velocity diagram for ID15; Top panel: along the major axis, bottom panel: along the minor axis. The rotation curve are plotted on top. }
\label{fig5}
\end{center}
\end{figure}

\begin{figure}
\begin{center}
\includegraphics[width=9cm]{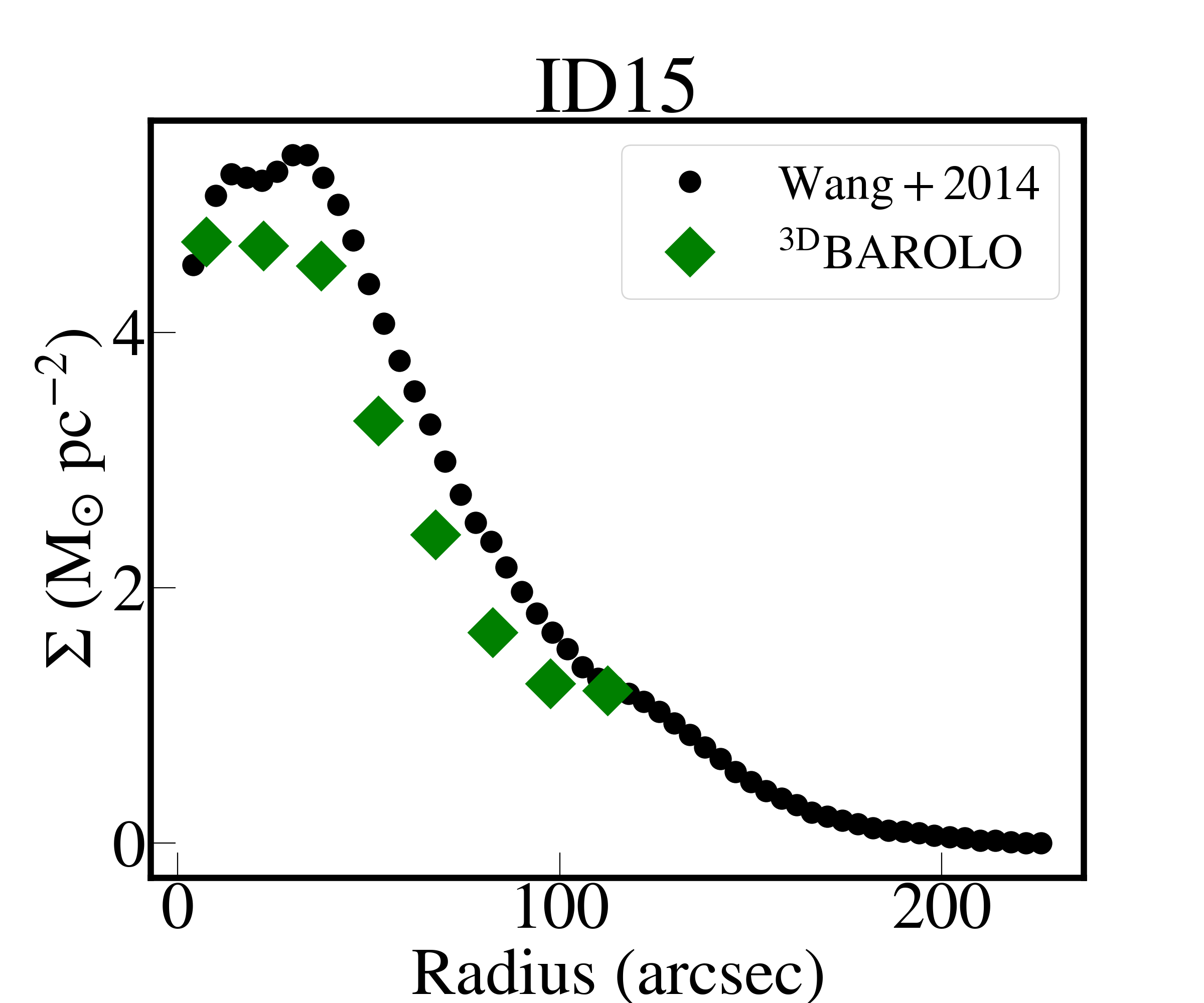}
\caption{ Comparison between the H{\sc i} surface density profiles from \cite{2014MNRAS.441.2159W}(balck cirles) and $^{3D}$BAROLO (green squares) for ID15. }
\label{fig6}
\end{center}
\end{figure}

\section{Data and analysis}
\subsection{Sample selection}
Our main sample are selected from the BLUEDISK H{\sc i} survey \citep{2013MNRAS.433..270W}. More details about the survey is given in \cite{2013MNRAS.433..270W}; here we give a brief summary. The BLUEDISK H{\sc i} survey constitutes of galaxies that are unusually H{\sc i} rich based on the fundamental plane of \cite{2012A&A...544A..65C} and control sample which match in term of their stellar mass (M$_{*}$), stellar surface density ($\Sigma_{*}$) and color index (NUV-r). The galaxies were observed using the WRST interferometer. We excluded galaxies that are interacting or in close pairs following \cite{2013MNRAS.433..270W,2014MNRAS.441.2159W}.  We further excluded galaxies that show complex H{\sc i} distributions such as ID10 and ID39 (see \citealt{2013MNRAS.433..270W}). The kinematic analysis tool $^{3D}$BAROLO (di Teodoro et al. 2015) also requires that the galaxy should be moderately inclined and have at least two resolution elements along the semi-major axis (e.g. \citealt{2016MNRAS.462.3628R,2017MNRAS.466.4159I}). We therefore select moderately inclined galaxies that have R$_{90,HI}$ (the radius that contains 90\% of the H{\sc i} flux) larger than 30 arcsec. The R$_{90,HI}$ distribution of the galaxies modeled using $^{3D}$BAROLO and the selected sample are shown in Figure \ref{fig1}.
Our selected sample consist of 28 H{\sc i} rich and control galaxies from BLUEDISK \citep{2013MNRAS.433..270W}. The SFR are taken from \citet{2016MNRAS.463.1724C} for the BLUEDISK sample. They were measured from a combination of archival WISE 22 $\micron$m (Wright et al. 2010) and GALEX FUV (Martin et al. 2005). The molecular gas masses are from CO follow-up observations of the BLUEDISK galaxies \citet{2016MNRAS.463.1724C}. The CO(2-1) and CO(0-1) lines observations were done with IRAM using the HEterodyne Receiver Array (HERA) instrument for 26 galaxies and the Eight Mixer Receiver (EMIR) instrument for 11 galaxies. The data were reduced using GILDAS standard packages. The reader are referred to \citet{2016MNRAS.463.1724C} for more details about the observations and data reduction.

We further added moderately inclined galaxies from the VIVA \citep{2009AJ....138.1741C} (14 galaxies) and THINGS \citep{2008AJ....136.2563W} (12 galaxies) surveys for comparison with BLUEDISK. We select galaxies that are not strongly warped and have similar stellar mass with the BLUEDISK sample.
\subsection{Tilted ring analysis}
A tilted ring model is usually used to extract kinematics and orientations information from emission line observations. The galaxy is divided into a concentric rings (e.g \citealt{1989A&A...223...47B}) and the rotation velocities and other orientation parameters are derived for each ring. 
The line-of-sight velocity is given as:
\begin{equation}
V(x,y)=V_{sys}+V_{c}sin(i)cos(\theta)+V_{r}sin(i)cos(\theta)
\end{equation}
where V$_{sys}$, V$_{c}$ and V$_{r}$ are the systemic, circular and radial velocities respectively, {\it i} is the inclination angle and $\theta$ is the azimuthal angle in the plane of the galaxy, related to the major axis position angle $\Phi$ by:

\begin{equation}
cos(\theta)=\frac{-(x-x_{0})sin(\Phi)+(y-y_{0})cos(\Phi)}{R}
\end{equation}
and
\begin{equation}
sin(\theta)=\frac{-(x-x_{0})cos(\Phi)+(y-y_{0})sin(\Phi)}{R sin(i)}
\end{equation}
The tilted ring method is usually applied to velocity maps for high angular resolution observations (e.g. \citealt{2008AJ....136.2648D}). However it is severely affected by beam smearing for low resolution observations. Moreover, the derived rotation curves and velocity dispersion also depends on the method used to derive the velocity maps \citep{2008AJ....136.2648D}.
Three-dimensional tilted ring mitigates this problem since it uses all the information in the data, it also derives the rotation velocities and velocity dispersion simultaneously. Several 3D tilted ring software are available in the literature (eg. TIRIFIC \citealt{2007A&A...468..731J},GBKFIT \citealt{2016MNRAS.455..754B}, $^{3D}$BAROLO \citealt{2015MNRAS.451.3021D}). Here we use the 3D-BAROLO sofware to derive the rotation curve, velocity dispersion and HI surface density profiles of the BLUEDISK, VIVA and THINGS galaxies.
\subsection{H{\sc i} disk scale height}
%Theoretically, the gas disk thickness is measured by assuming that the gas layer is in hydrostatic equilibrium \citep{1942ApJ....95..329S}. In this framework, the vertical distribution of the gas layers in a rotating disk is described by the stationary Euler equation \citep{1995AJ....110..591O,1996AJ....112..457O,2019A&A...632A.127B,2019A&A...622A..64B}:
%\begin{equation}
%\frac{\partial \Phi (R,z)}{\partial z} = - \frac{1}{\rho(R,z)} \frac{\partial P (R,z)}{\partial z}
%\end{equation}
%where $P(R,z)$ is the gas pressure due to combination of thermal and turbulence motions.
%The gas is assumed to be isotropic (ie $\sigma_{z}=\sigma_{x}=\sigma_{y}$=$\sigma$), therefore, the gas pressure is given by:
%
%\begin{equation}
%P (R,z)=\sigma^{2}(R)\rho(R,z)
%\end{equation}
%
%The gas volume density is obtained by solving equation 4 
%
%\begin{equation}
%\rho(R,z)=\rho(R,0) exp[ \frac{\partial \Phi (R,z)-\partial \Phi (R,0)}{\sigma^{2}(R)}]
%\end{equation}

The H{\sc i} disk scale height is usually measured assuming that the gas is in hydrostatic equilibrium \citep{1995AJ....110..591O,1996AJ....112..457O,2019A&A...632A.127B,2019A&A...622A..64B}. This method requires prior knowledge of the dark matter distribution obtained through rotation curve decomposition in addition to the gas velocity dispersion and surface densities \citep{2019A&A...632A.127B,2019A&A...622A..64B}. However, mass modeling by decomposing the observed rotation curve is limited to few galaxies with high resolution emission lines observations (see \citealt{1981AJ.....86.1825B,2008AJ....136.2648D}).

\cite{2019ApJ...882....5W} used a semi-empirical formula to measure the molecular scale height of five luminous and ultra-luminous infrared galaxies. By solving the equation of equilibrium for an isothermal gas \citep{1942ApJ....95..329S} and equating the confining pressure due to gravity P$_{grav}$ with the uplifting pressure produced by the gas motions, magnetic fields and cormic rays P$_{ISM}$ they showed that the scale height could be expressed entirely with observationally-derived  variables \citep{2019ApJ...882....5W}. 

The gas scale height is therefore expressed as (equation 6 in \citealt{2019ApJ...882....5W}).
\begin{equation}
h(R)=\frac{\sigma^{2}(R)}{\pi G \Sigma_{gas}} \times (\frac{1+\alpha+\beta}{1+g_{galaxy}/g}) \times (\frac{\Sigma_{gas}}{\Sigma_{total}})
\end{equation}

where $\sigma(R)$, $\Sigma_{gas}$ and $\Sigma_{total}$ are the velocity dispersion, gas surface density and the total surface density within the gas layer; $g_{galaxy}/g$ is the ratio between the total gravitational acceleration of the galaxy and the gravitational acceleration due to the disk mass; $\alpha$ is the ratio between magnetic field and turbulence and thermal support and $\beta$ is cosmic rays to turbulence and thermal support. Using magneto-hydrodynamic simulation, \cite{2015ApJ...802...99K} estimated the ratio between the vertical magnetic pressure to the turbulence plus thermal support $\alpha$ $\sim$ 0.3. However, magnetic fields are more associated with dense molecular gas and not with H{\sc i}. Using a sample of 20 nearby spiral galaxies, \cite{2015ApJ...799...35V} found that the magnetic fields strength is correlated with molecular gas surface density and star formation rate but not with atomic gas. For dwarf irregular, \cite{2017A&A...603A.121C} noted that the magnetic fields are very weak (< 4.2 $\mu$G). For individual galaxies, it have been found also that the magnetic field are more associated with dense molecular gas than with atomic hydrogen gas (e.g. \citealt{2013A&A...557A.129T,2013A&A...552A..19T}). Our sample consist of H{\sc i}-rich and normal spiral galaxies. Therefore, we will assume that the effect of magnetic field on atomic gas disk thickness is negligible compare to turbulence and thermal support. 
The coefficient $\beta$ is also negligible ($\beta$ $\sim$ 0.0, see \citealt{1966ApJ...145..811P}). We also assume that the average ratio between the gas and stellar surface densities ($\Sigma_{gas}/\Sigma_{total}$) could be approximated by the ratio between the H{\sc i} mass within the optical disk to the stellar mass and that the molecular gas have small contribution to the total mass. The H{\sc i} mass within the optical disk is measured from H{\sc i} intensity map by converting the total flux within the optical disk into mass using the following equation:
\begin{eqnarray}
M_{HI}/M_{\odot}=2.356 \times 10^{5}S_{HI}D_{Mpc}^{2},
\end{eqnarray}
where $S_{HI}$ is the total flux inside the optical disk in Jy Km s$^{-1}$ and D is the distance in Mpc.
The H{\sc i} within the optical disk can also be predicted using the method developed by \cite{2020ApJ...890...63W} for low spatial resolution and single dish HI observations. \cite{2014MNRAS.441.2159W} investigated the shape of the HI surface density profiles using a large sample of nearby spiral and dwarf galaxies. They found that the profiles are well fitted using an exponential function and the outer part of the average profile of all the galaxies in their sample is almost universal. \cite{2020ApJ...890...63W} used the average profile from \cite{2014MNRAS.441.2159W} and estimated the fraction of HI mass outside the optical radius for galaxies selected from the xGASS sample. The HI mass within the optical disk is then given by the difference between the total HI mass and the HI outside the optical disk.  The readers are referred to \cite{2020ApJ...890...63W} for more details about the method and the assumption used to estimate the HI mass within the optical disk.

Using the above assumptions and following \cite{2019ApJ...882....5W},  $g_{galaxy}/g$ could be written as:
 
\begin{equation}
g_{galaxy}/g = \frac{\frac{(1+\alpha+\beta)*\sigma_{HI,0}^{2}}{V_{max}^{2}}}{2\times(\frac{M_{dyn}}{M_{HI,in}+M_{star}})^{2}}
\end{equation}
and 
\begin{equation}
\frac{\Sigma_{HI,in}}{\Sigma_{total}} \sim f_{HI,in}=\frac{M_{HI,in}}{M_{HI,in}+M_{star}}
\end{equation}
where M$_{dyn}$ is the dynamical mass,$f_{HI,in}$ is the H{\sc i} fraction within the optical disk, $\sigma_{HI,0}$ is the central HI velocity dispersion obtain by fitting the radial profile of the velocity dispersion with an exponential function ($\sigma(R)=\sigma_{HI,0} \times exp(-R_{\sigma}/R)$).

Finally, the H{\sc i} scale height is given by

\begin{equation}
h_{HI}(R) \sim K\times \frac{\sigma^{2}(R)}{\Sigma_{HI}}
\end{equation}

where K is given by 

\begin{equation}
K=\frac{1.0}{\pi*G*(1+g_{galaxy}/g)}\times f_{HI,in}
\end{equation}

\begin{table*}
 \caption{General properties of the BLEUDISK sample}
 \label{tab1}
 %\scriptsize
 \begin{center}
 \begin{tabular}{lllllllllll}
  \hline \hline
ID &RA&DEC&Distance&D$_{25}$&PA&INC&V$_{max}$&$\sigma_{HI,0}$& <h$_{HI,in}$>&\\
 &($^{\circ}$)&($^{\circ}$)&Mpc&arcsec&($^{\circ}$)&($^{\circ}$)& (km s$^{-1}$)&(km s$^{-1}$)&(kpc)&\\
1 &2&3&4&5&6&7&8&9&10&11\\\\
  \hline  \hline

ID1&123.591766 &  39.251354&114.59&75.96&125.02	&36.1&	116.2$\pm$5.3&13.1$\pm$1.4&0.66 $\pm$ 0.20&HI-rich \\[2pt]
ID2&127.194809& 40.665886&101.70&81.30&	198.6&	66.5&	208.2$\pm$	4.8&	13.7$\pm$	1.9&	0.47 $\pm$	0.27&HI-rich\\[2pt]
ID4	&129.641663&30.798681&106.48&88.61&132.4	&35.1	&114.3$\pm$	1.5&	10.7$\pm$	1.0&	0.30 $\pm$	0.13&HI-rich\\[2pt]
ID5	&132.318298 &36.119797&104.73&65.27&120.5&	34.6&	126.6$\pm$	8.2&	18.2$\pm$	5.0&	0.53 $\pm$	0.22&HI-rich\\[2pt]
ID6	&132.344025&36.710327&104.12&103.79&336.8	&58.8&	228.6$\pm$	3.0&	11.2$\pm$	1.4&	0.21 $\pm$	0.03&HI-rich\\[2pt]
ID8	&137.177567&44.810658&110.86&66.18&325.7&	50.1&	161.1$\pm$	9.8	&14.7$\pm$	3.2&	0.51 $\pm$	0.05&HI-rich\\[2pt]
ID12	&154.042709&58.427002&106.04   &  82.89& 93.9&	62.5&	178.9$\pm$	20.7&	12.3$\pm$	2.4&	0.28 $\pm$	0.08&HI-rich\\[2pt]
ID14	&176.739746&50.702133&98.98 &    87.33&70.09	&65.05&	201.6$\pm$	10.2&	12.24$\pm$	2.36&	0.15 $\pm$	0.03&HI-rich\\[2pt]
ID15	&177.247757&35.016048&88.93  &  111.76&309.8&	57.2&	170.5$\pm$	3.5&	11.6$\pm$	0.9&	0.31 $\pm$	0.06&HI-rich\\[2pt]
ID16	&193.014786&51.680046&112.75  &   54.06&91.02&	48.11&	196.28$\pm$	23.17&	20.04$\pm$	5.6&	0.73 $\pm$	0.28&HI-rich\\[2pt]
ID17	&196.806625&58.135014&114.09  &   61.00&290.0	&47.0&	159.7$\pm$	10.2&	11.7$\pm$	1.9&	0.10 $\pm$	0.01&HI-rich\\[2pt]
ID18	&199.015060&35.043518&96.49  &   71.17&82.67&	65.73&	202.48$\pm$	12.23&	13.25$\pm$	1.29&	0.24 $\pm$	0.01&HI-rich\\[2pt]
ID19	&212.631851&38.893559&106.16  &   57.90&55.45	&68.60	&163.38$\pm$	5.84&	13.91$\pm$	4.30&0.35 $\pm$	0.07&HI-rich\\[2pt]
ID20	&219.499756&40.106197&108.42  &   77.39&90.6	&46.5&	185.8 $\pm$	7.16 &	20.70$\pm$	7.16 &	1.14 $\pm$	0.26&HI-rich\\[2pt]
ID21	&241.892578&36.484032&123.20   &  46.28&332.50	&74.3&	75.15$\pm$	1.70 &	12.62 $\pm$	1.87 &	0.26 $\pm$	0.03&HI-rich\\[2pt]
ID22	&250.793503&42.192783&117.32  &   86.07&10.87	&78.85&	263.14 $\pm$ 0.74 &	17.33 $\pm$	2.75 &	0.39 $\pm$	0.09&HI-rich\\[2pt]
ID24	&259.156036&58.411900&122.22  &   80.72&357.6&	59.6&	193.05 $\pm$	3.84 &	12.42 $\pm$	1.21 &	0.40 $\pm$	0.07&HI-rich\\[2pt]
ID26	&111.938042&42.180717&96.01  &   52.96&235.7	&52.5&	206.9 $\pm$	0.34 &	13.18$\pm$	6.03 &	0.22 $\pm$	0.10&HI-rich\\[2pt]
ID30	&138.603149&40.777924& 115.96   &  70.32&166.03& 	78.2&	172.3$\pm$	1.7&	15.6$\pm$	7.5 &	0.55 $\pm$	0.35&HI-rich\\[2pt]
ID35	&149.420227&45.258678&100.71  &   76.99&13.25&	77.95	&158.88$\pm$	2.45 &	12.80 $\pm$	1.52 &	0.10 $\pm$	0.01&HI-rich\\[2pt]
ID47	&244.382523&31.194477& 99.88  &   65.82 &263.9	&58.28	&221.18 $\pm$	18.09 &	13.98 $\pm$	1.35 &	0.18 $\pm$	0.01&HI-rich\\[2pt]
ID9	&138.742996&51.361061&113.88   &  73.23&13.01&	64.21&	216.38 $\pm$	2.06 &	11.03 $\pm$	2.35 &	0.60 $\pm$	0.01&control\\[2pt]
ID11	&152.726593&45.950371&100.03    & 73.74&271.48&	62.95&	291.17 $\pm$	49.53 &	14.81 $\pm$	1.77 &	0.10$\pm$	0.01&control\\[2pt]
ID23	&251.811615&40.245079&122.02   &  79.17&142.05&	49.63&	373.97 $\pm$	27.53 &	16.73 $\pm$	0.35 &	0.34 $\pm$	0.09&control\\[2pt]
ID25	&262.156342&57.145065& 113.99  &   58.66&103.12&	62.6	&144.3 $\pm$	23.4 &	12.14 $\pm$	0.16 &	0.12 $\pm$ 	0.02&control\\[2pt]
ID37	&153.797638&56.672085&107.92  &   62.86 &185.75	&41.92&	235.17 $\pm$	13.47 &	16.73 $\pm$	4.24 &	0.27 $\pm$	0.01&control\\[2pt]
ID40	&168.563553&34.154381&112.65  &   44.02&189.64	&44.23	&169.50 $\pm$	2.43 &	12.43 $\pm$	1.87 &	0.05 $\pm$	0.01&control\\[2pt]
ID43	&203.374603&40.529671&111.69  &   45.31&249.34	&55.56	&222.33 $\pm$	2.35 &	17.67 $\pm$	3.81 &	0.41 $\pm$	0.05&control\\[2pt]
  
 \hline

 \end{tabular}
   \end{center}
\end{table*}

%\begin{figure}
%\begin{center}
%\includegraphics[width=9cm]{ID15_thickness.png}
%\caption{ Surface density, velocity dispersion and scaleheight of the HI disk of ID15; Blue: HI, black: psuedo-total gas from Wang et al. (2014.}
%\label{fig2}
%\end{center}
%\end{figure} 
%
%\begin{figure*}
%\begin{center}
%\includegraphics[width=7cm]{ID15_density.png}
%\includegraphics[width=7cm]{ID17_density.png}\\
%\includegraphics[width=7cm]{ID35_density.png}
%\includegraphics[width=7cm]{ID47_density.png}
%\caption{Velocity dispersion radial profiles of four HI-rich galaxies. The points with errorbars are the measurement the red line is the exponential fit to the data. The central velocity dispersion is shown on bottom left corner. The remaining galaxies are in appendix.}
%\label{fig2}
%\end{center}
%\end{figure*}
%
%\begin{figure*}
%\begin{center}
%\includegraphics[width=8cm]{ID15_dispersion.png}
%\includegraphics[width=8cm]{ID17_dispersion.png}\\
%\includegraphics[width=8cm]{ID35_dispersion.png}
%\includegraphics[width=8cm]{ID47_dispersion.png}
%\caption{Velocity dispersion radial profiles of four HI-rich galaxies. The points with errorbars are the measurement the red line is the exponential fit to the data. The central velocity dispersion is shown on bottom left corner. The remaining galaxies are in appendix.}
%\label{fig2}
%\end{center}
%\end{figure*}

\begin{figure*}
\begin{center}
\includegraphics[width=18.5cm]{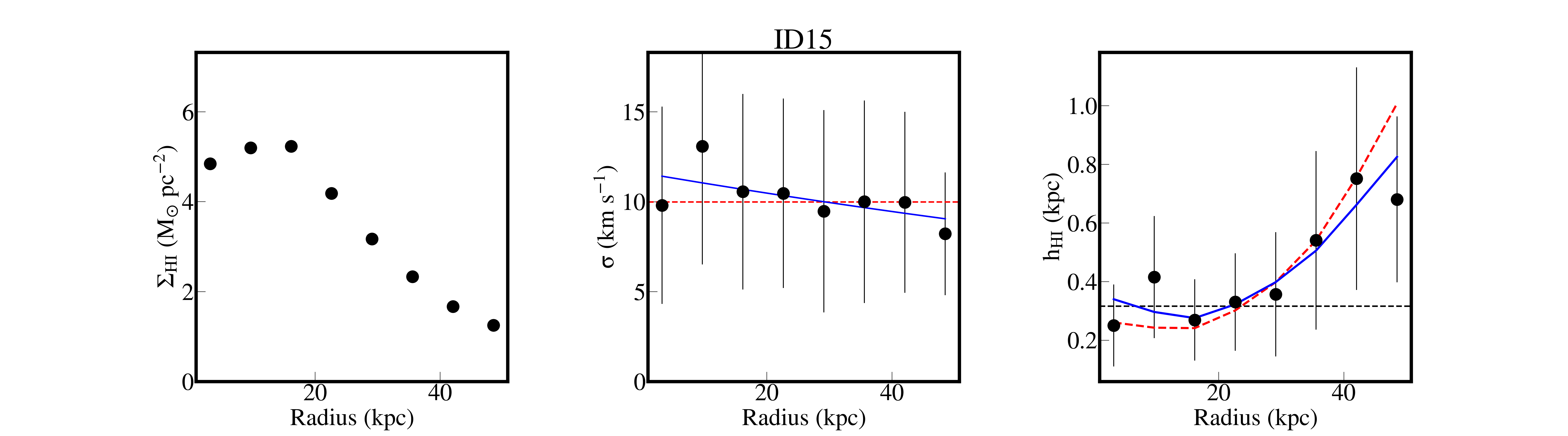}
\includegraphics[width=18.5cm]{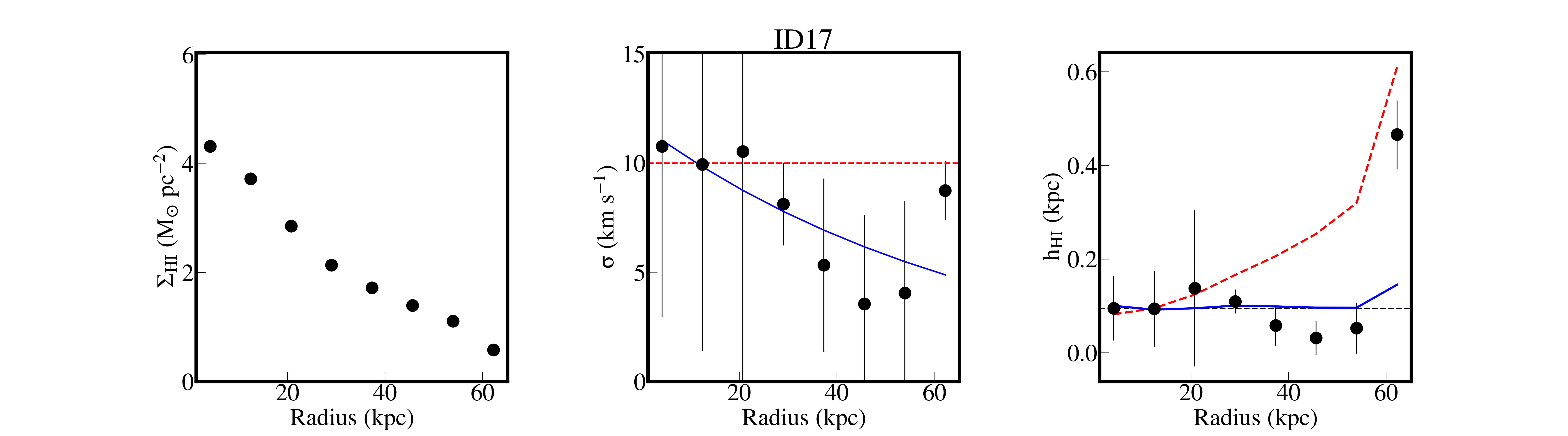}
\includegraphics[width=18.5cm]{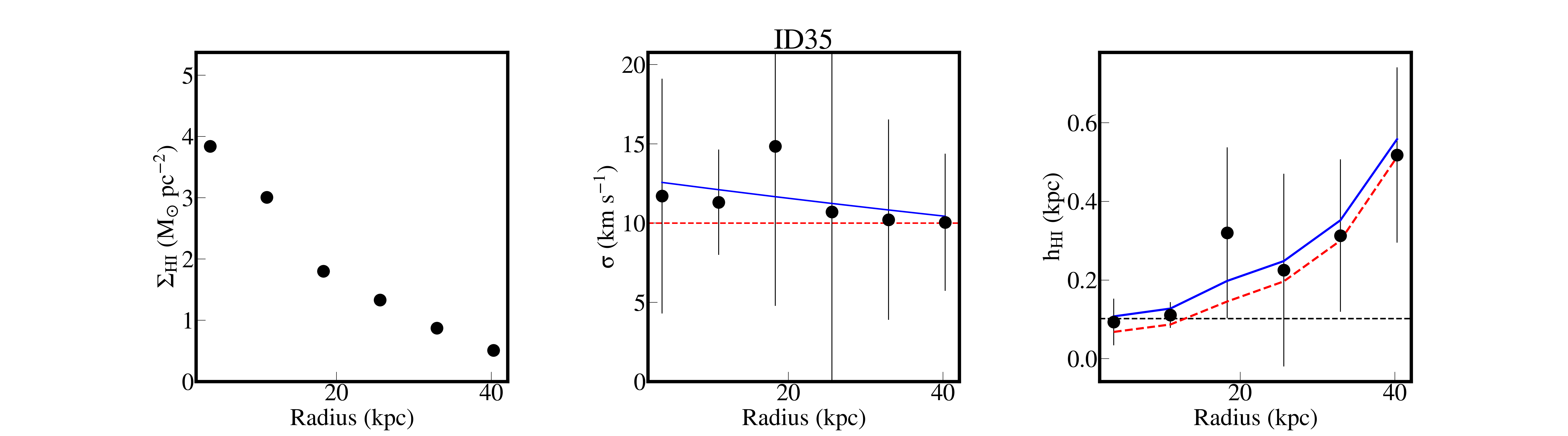}
\includegraphics[width=18.5cm]{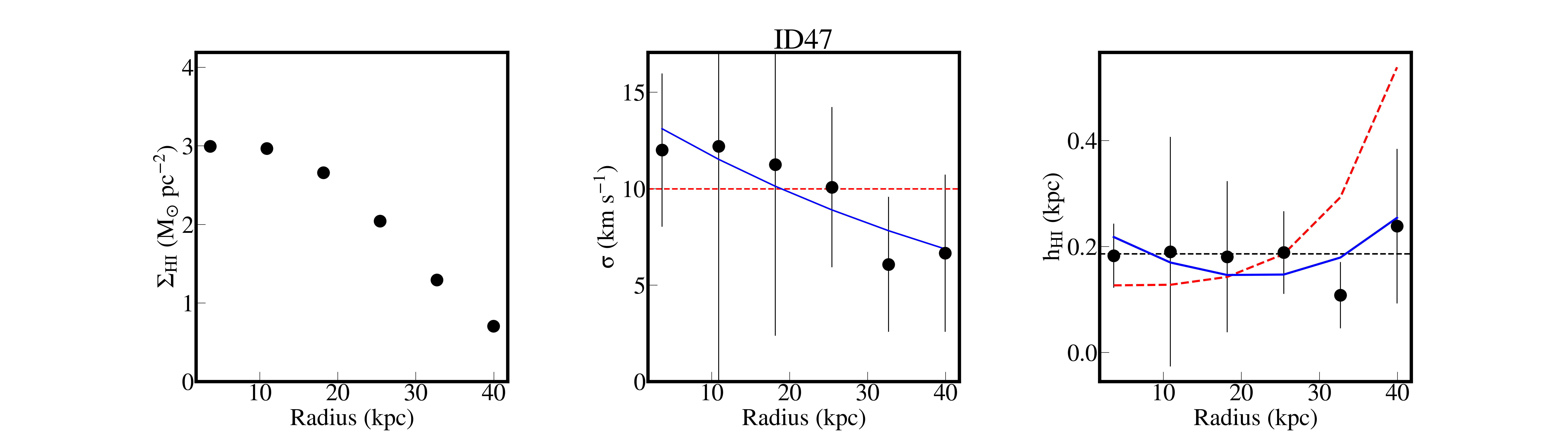}
 \caption{HI surface density, velocity dispersion and HI  Scale-height of four HI-rich galaxies. The density profiles obtain from $^{3d}$BAROLO are on the first column; the velocity dispersion profiles derived using $^{3d}$BAROLO are on the middle panels, the blue curves are an exponential fit to the observation, the dashed lines indicates $\sigma_{HI}=10 km/s$. The HI scale height calculated using equation 10 are on the third column, the red dashed line is the scale height assuming a constant velocity dispersion, the blue lines is the scale height when the exponential fit is used for the velocity dispersion and the black circles with errorbar are the measured scale heights. The dashed horizontal lines are the average scale height inside the optical radius.}
\label{fig7}
\end{center}
\end{figure*}

\begin{figure*}
\begin{center}
\includegraphics[width=18.5cm]{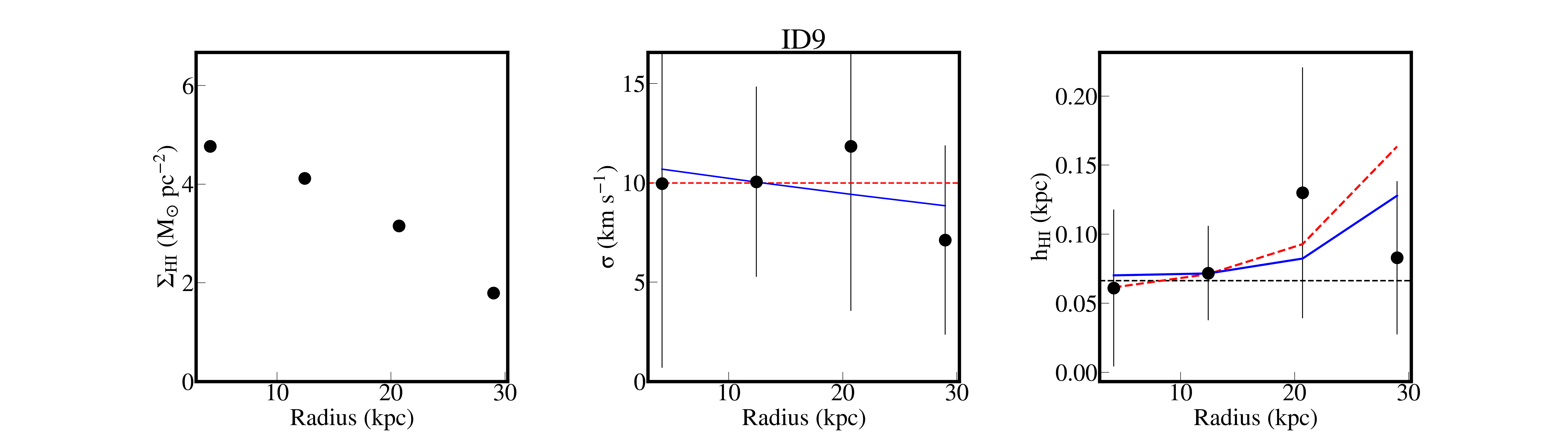}
\includegraphics[width=18.5cm]{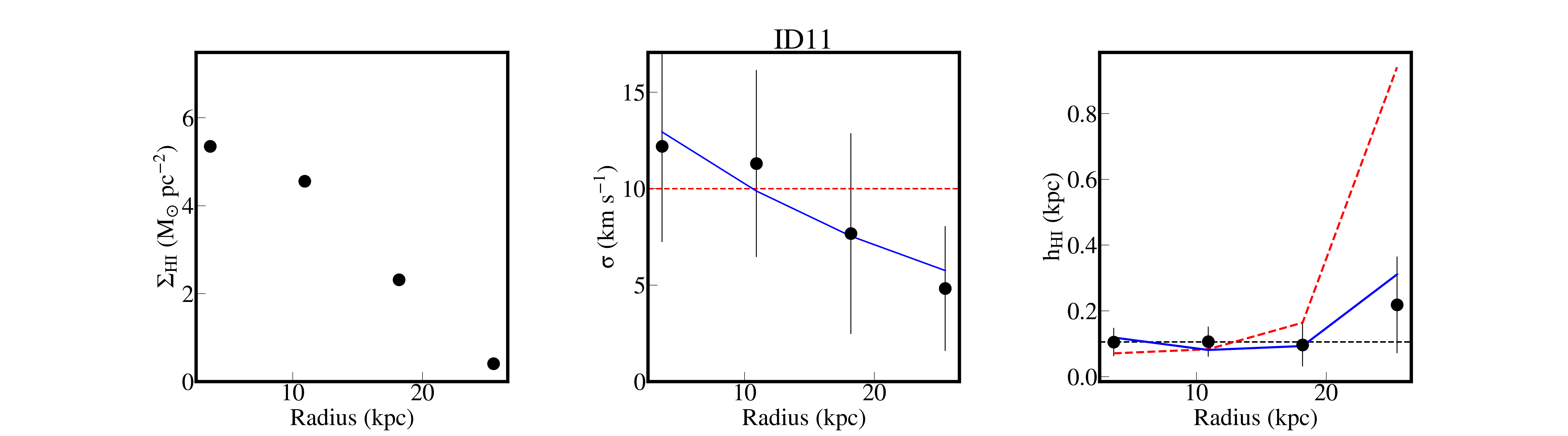}
\includegraphics[width=18.5cm]{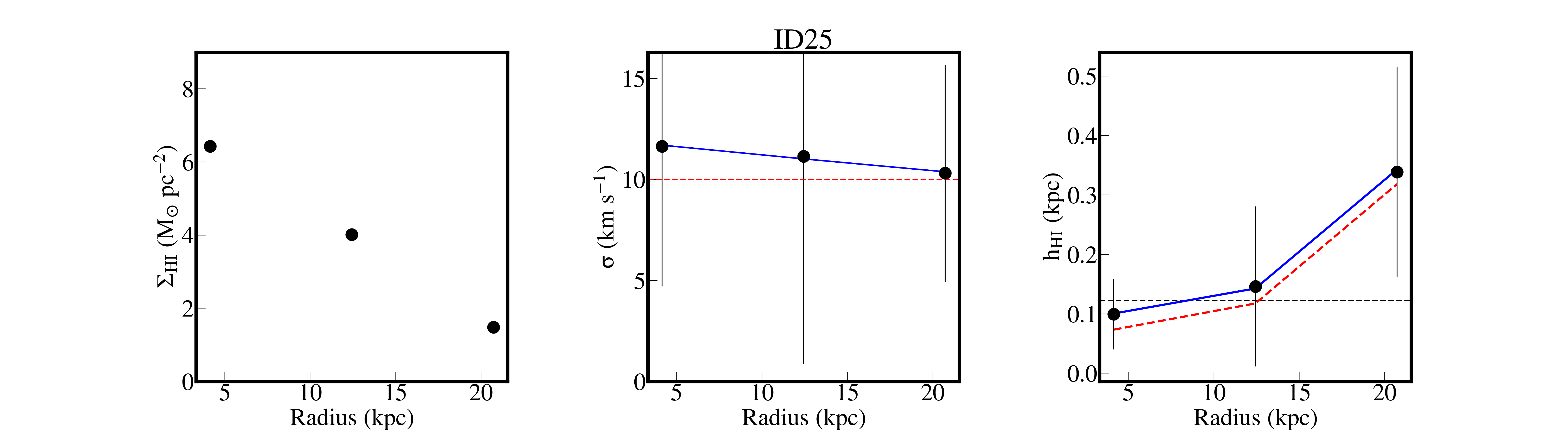}
\includegraphics[width=18.5cm]{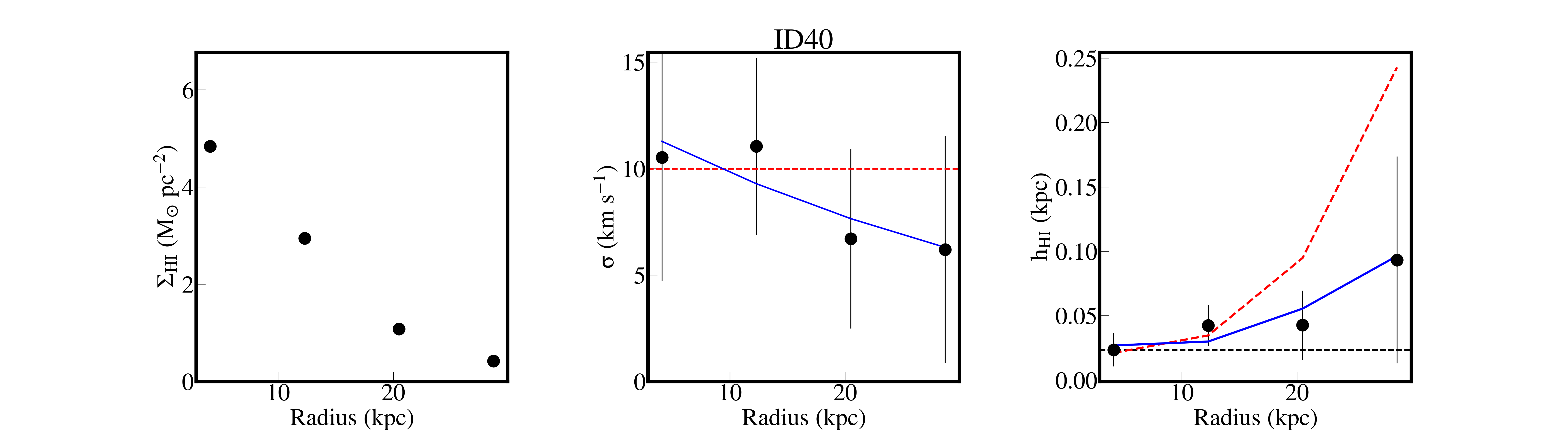}

\caption{Same as figure 4 but for control galaxies.}
\label{fig8}
\end{center}
\end{figure*}

\begin{figure}
\begin{center}
\includegraphics[width=8.cm]{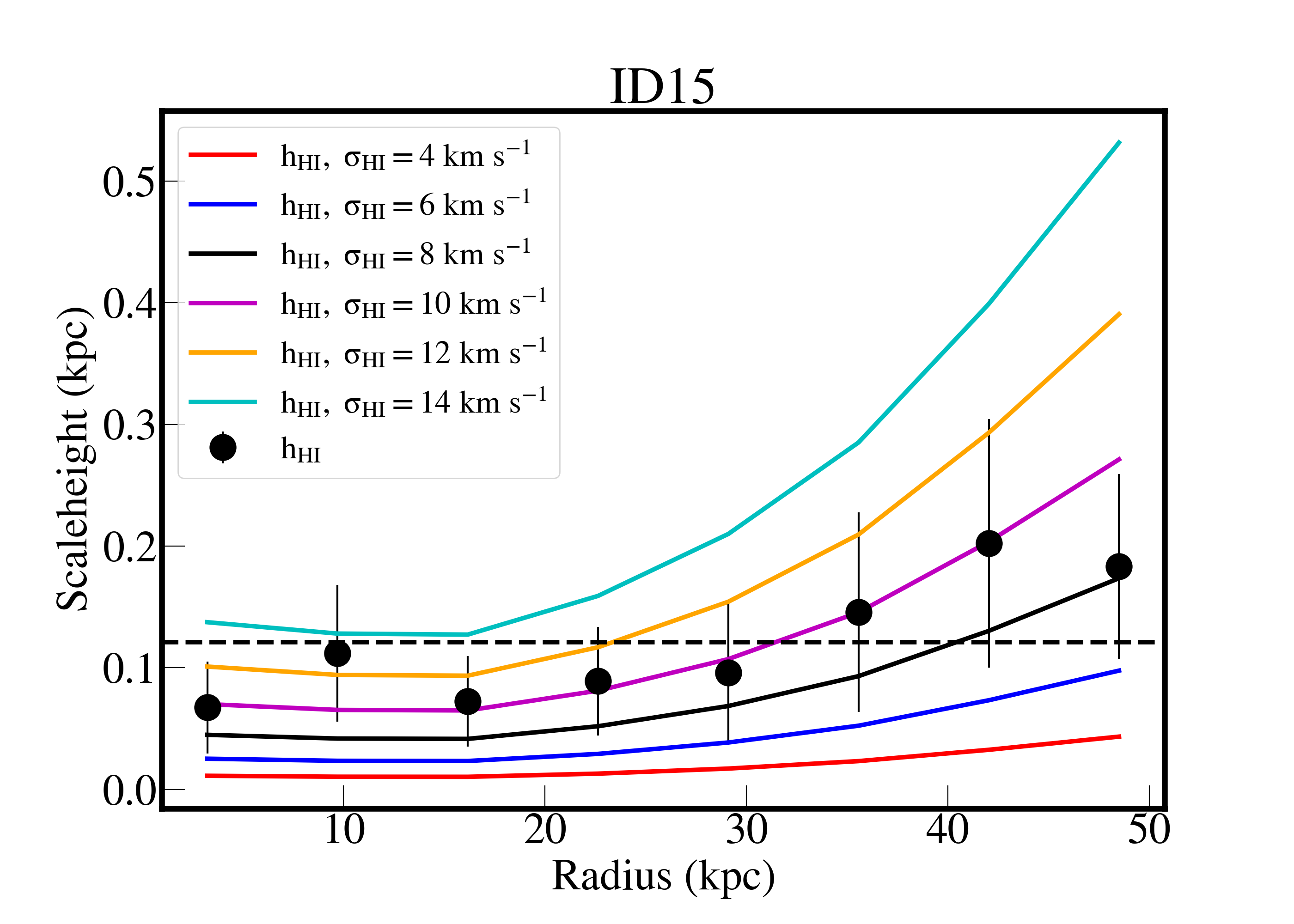}\\
\includegraphics[width=8.cm]{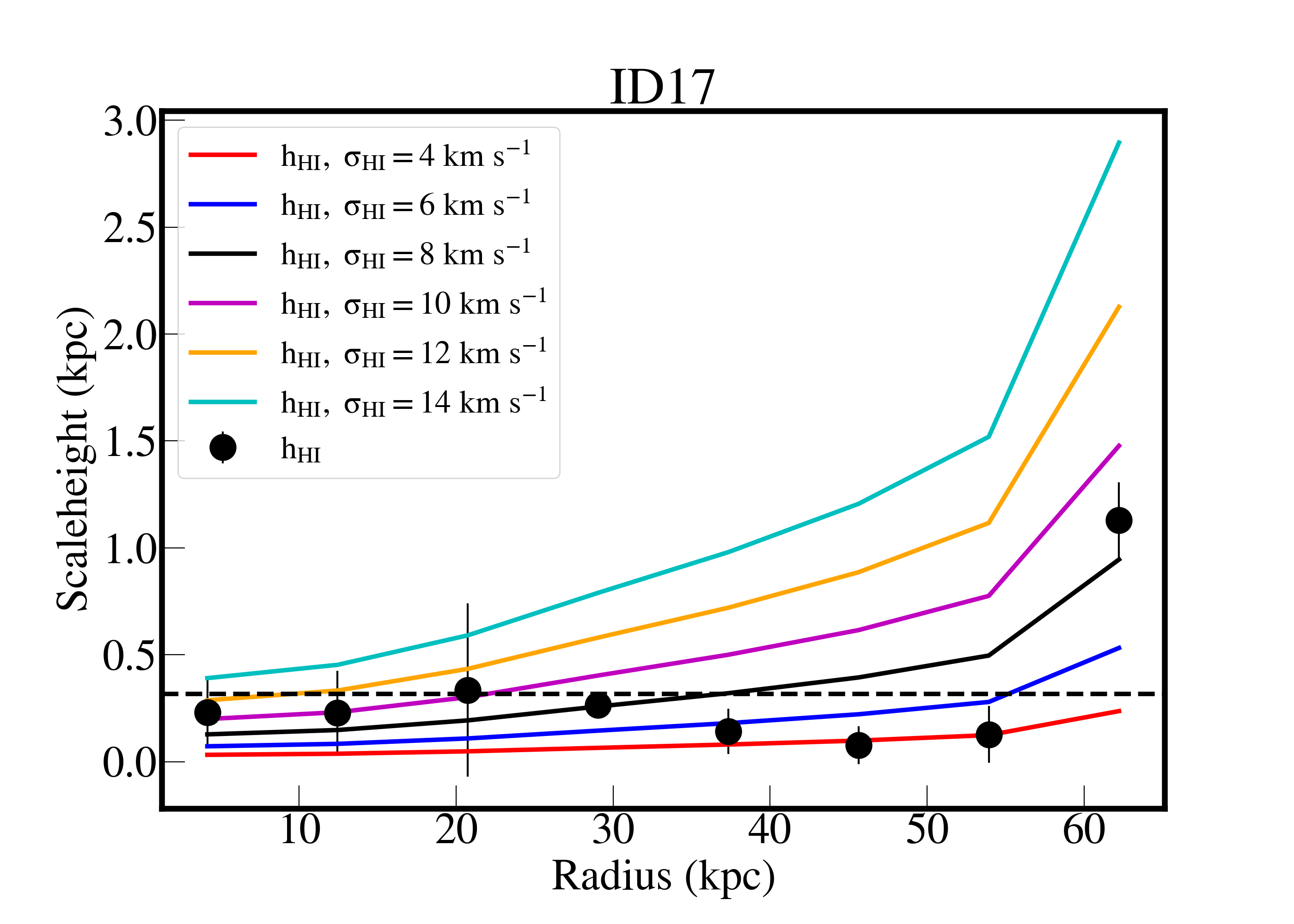}
\caption{Radial variation of the H{\sc i} scale height using different values for the velocity dispersion for ID15 (top) and ID17 (bottom). The dashed lines are scale height for fixed velocity dispersion and the filled circles when radial profile of the velocity dispersion is used.}
\label{fig9}
\end{center}
\end{figure} 

\begin{figure}
\begin{center}
\includegraphics[width=8.5cm]{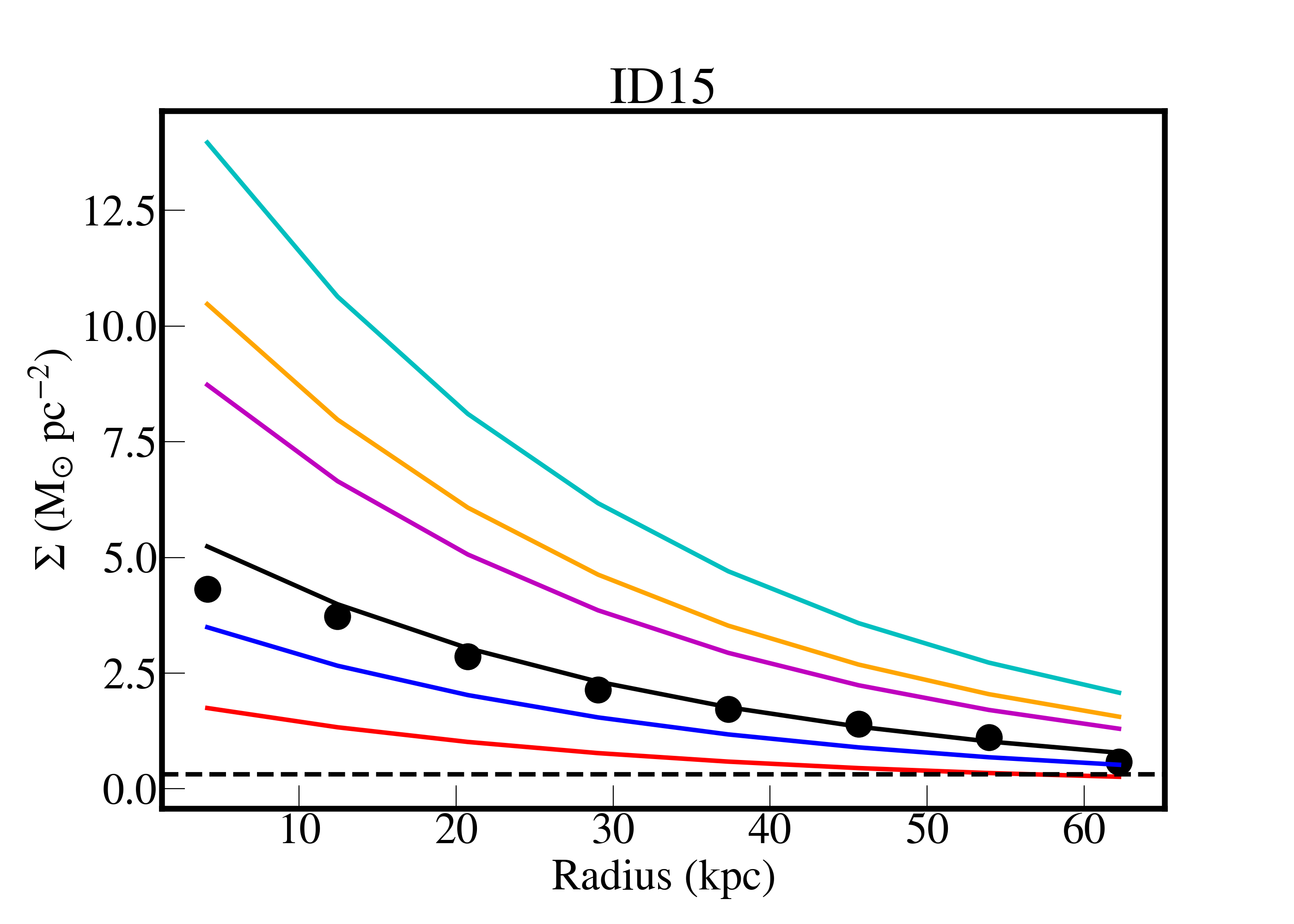}\\
\includegraphics[width=8.5cm]{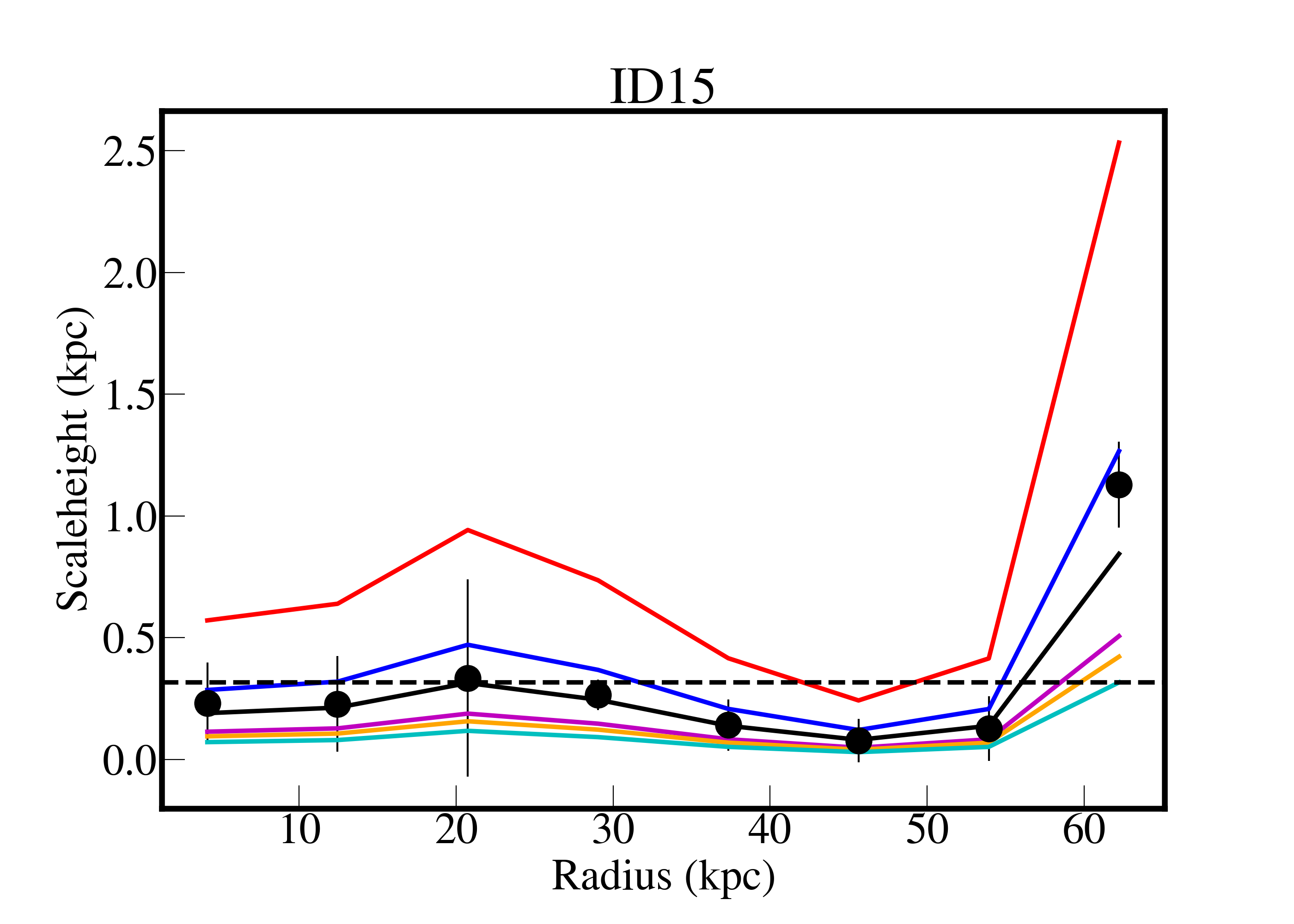}
\caption{Top panel: different models of the H{\sc i} density profile of ID15 (see text for details). The corresponding scale heights for each profile are on the bottom panel. The filled circles are the observations and the dashed horizontal line is the average scale height }
\label{fig10}
\end{center}
\end{figure} 

\begin{figure*}
\begin{center}
\includegraphics[width=8cm]{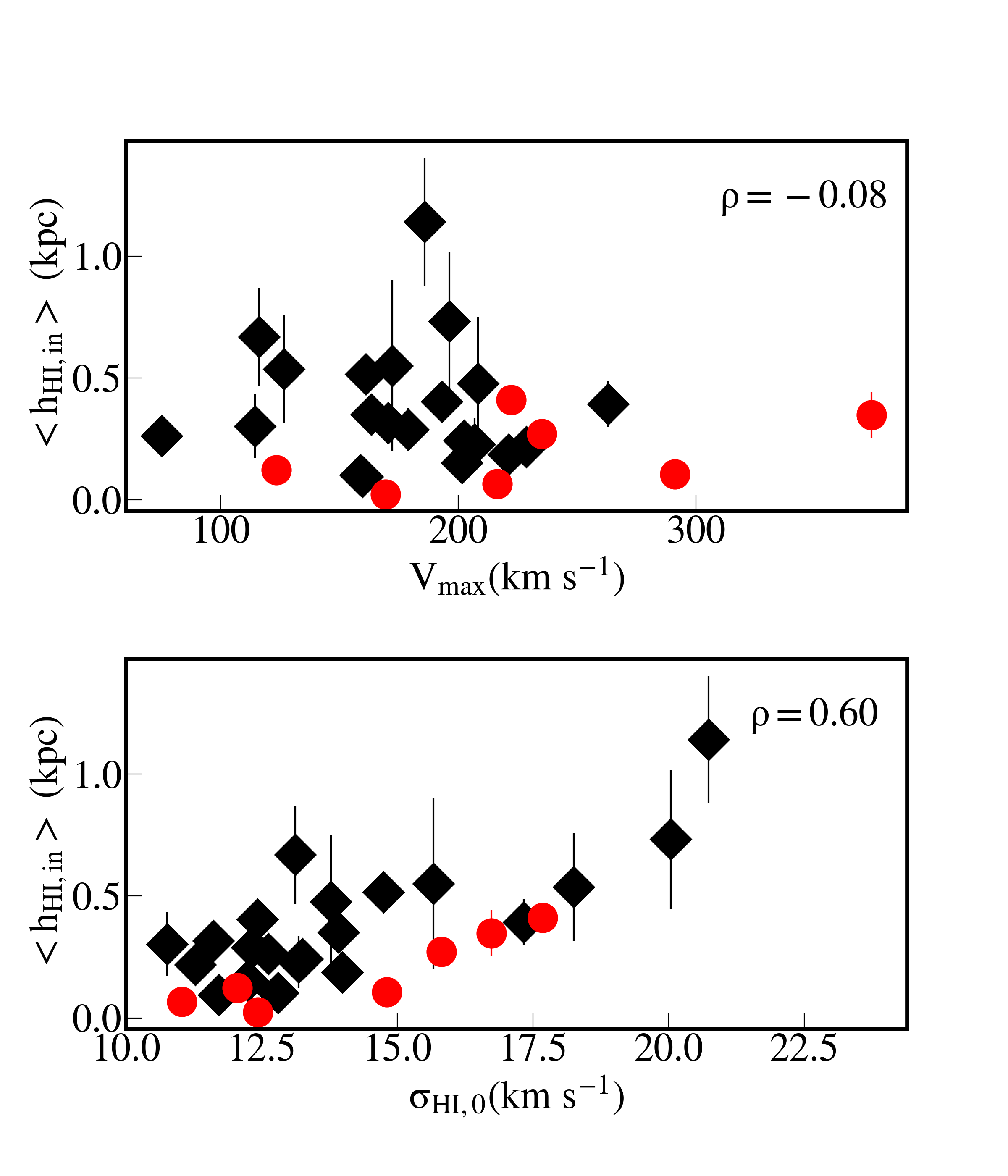}
\includegraphics[width=8cm]{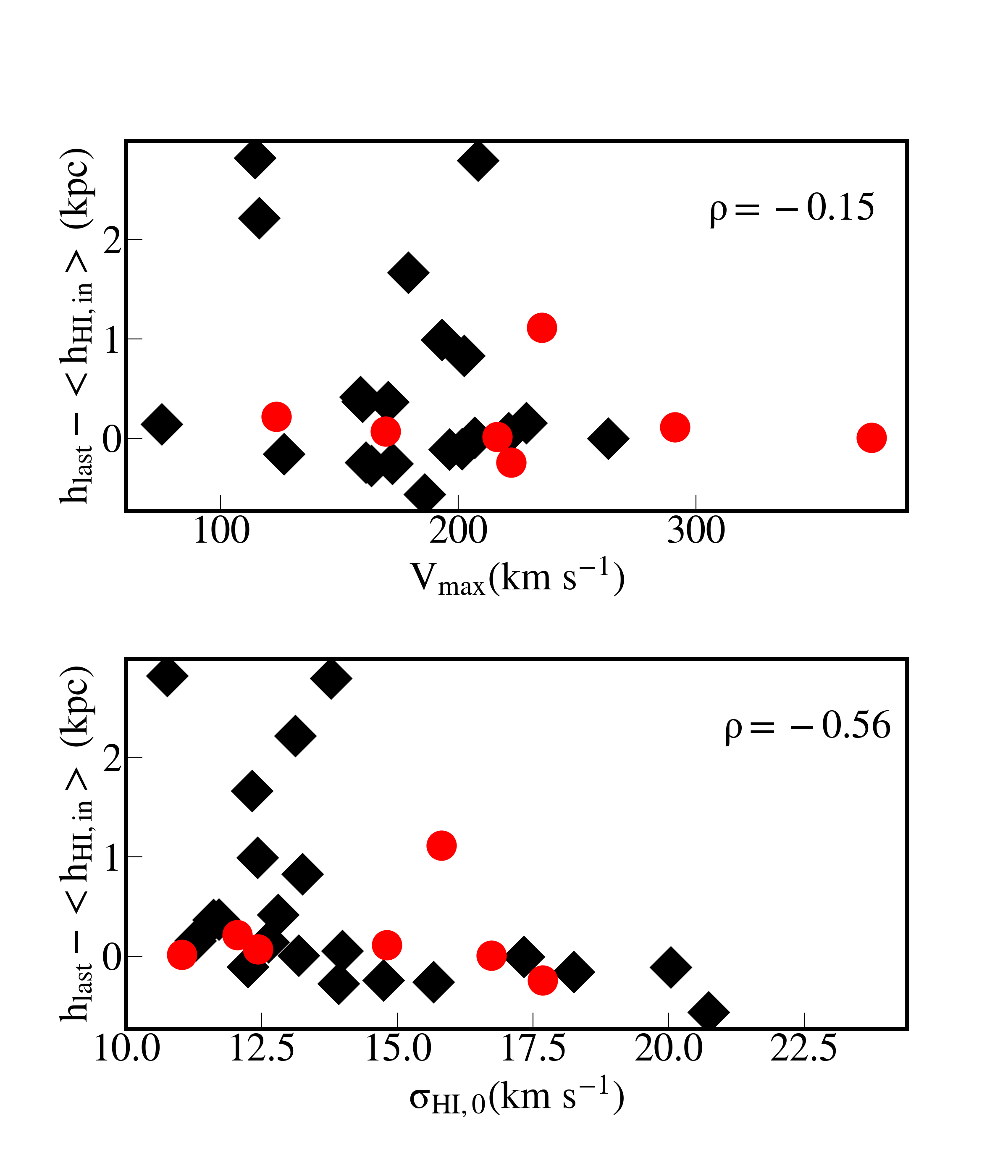}
\caption{Left panels: The average H{\sc i} scale height inside the optical disk is plotted against the maximum circular velocities (top) and central H{\sc i} velocity dispersion (Bottom). Right panels: the difference between the average H{\sc i} scale height inside the optical radius and the outer disk scale height (h$_{last}$) is plotted against the maximum circular velocities (top) and central H{\sc i} velocity dispersion (Bottom). The black diamonds are the H{\sc i}-rich galaxies and the red circles are the control galaxies. The Pearson correlation coefficient are shown on each panel.}
\label{fig11}
\end{center}
\end{figure*}

\begin{figure*}
\begin{center}
\includegraphics[width=20cm]{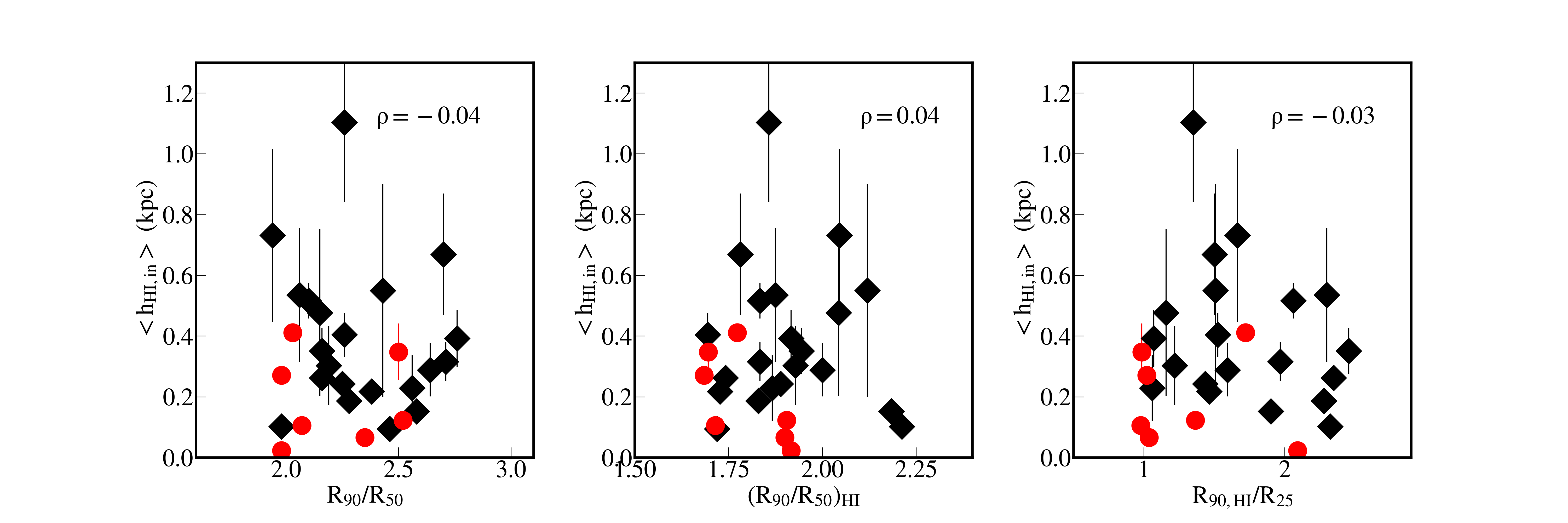}
\caption{Correlations between the average H{\sc i} scale height inside the optical radius and the optical concentration, the H{\sc i} concentration and the ratio between the H{\sc i} and optical extend of the disk. The black diamonds are the H{\sc i}-rich galaxies and the red circles are the control galaxies. The Pearson correlation coefficient are shown on each panel.}
\label{fig12}
\end{center}
\end{figure*} 

\begin{figure}
\begin{center}
\includegraphics[width=8cm]{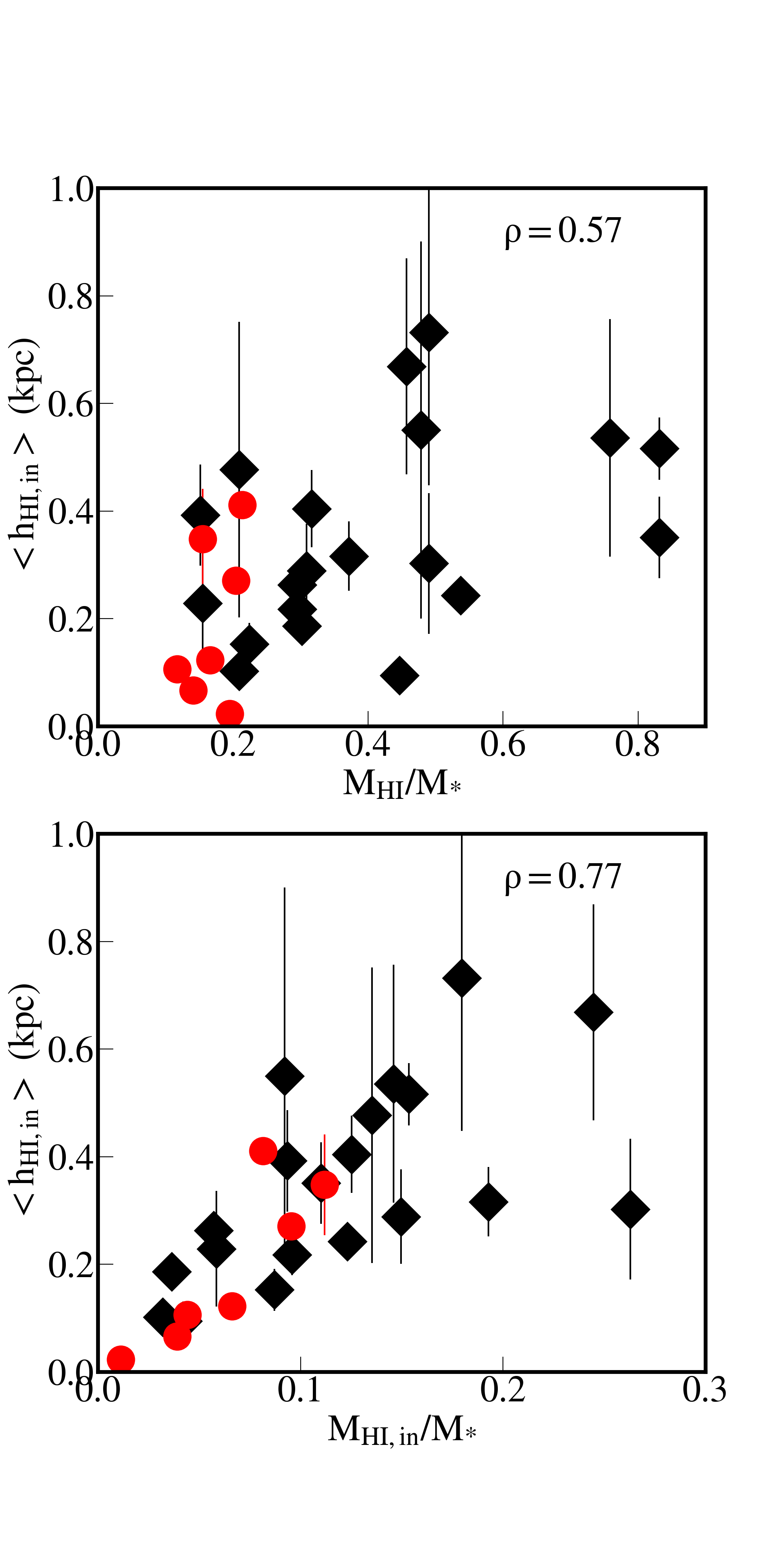}
\caption{Top panel: correlation between the average H{\sc i} scale height inside the optical radius with total H{\sc i} to stellar mass ratios. Bottom panel: correlation between the average H{\sc i} scale height inside the optical radius with H{\sc i} inside the optical radius to stellar mass ratios. The black diamonds are the H{\sc i}-rich galaxies and the red circles are the control galaxies. The Pearson correlation coefficient are shown on each panel.}
\label{fig13}
\end{center}
\end{figure} 

\begin{figure}
\begin{center}
\includegraphics[width=9.5cm]{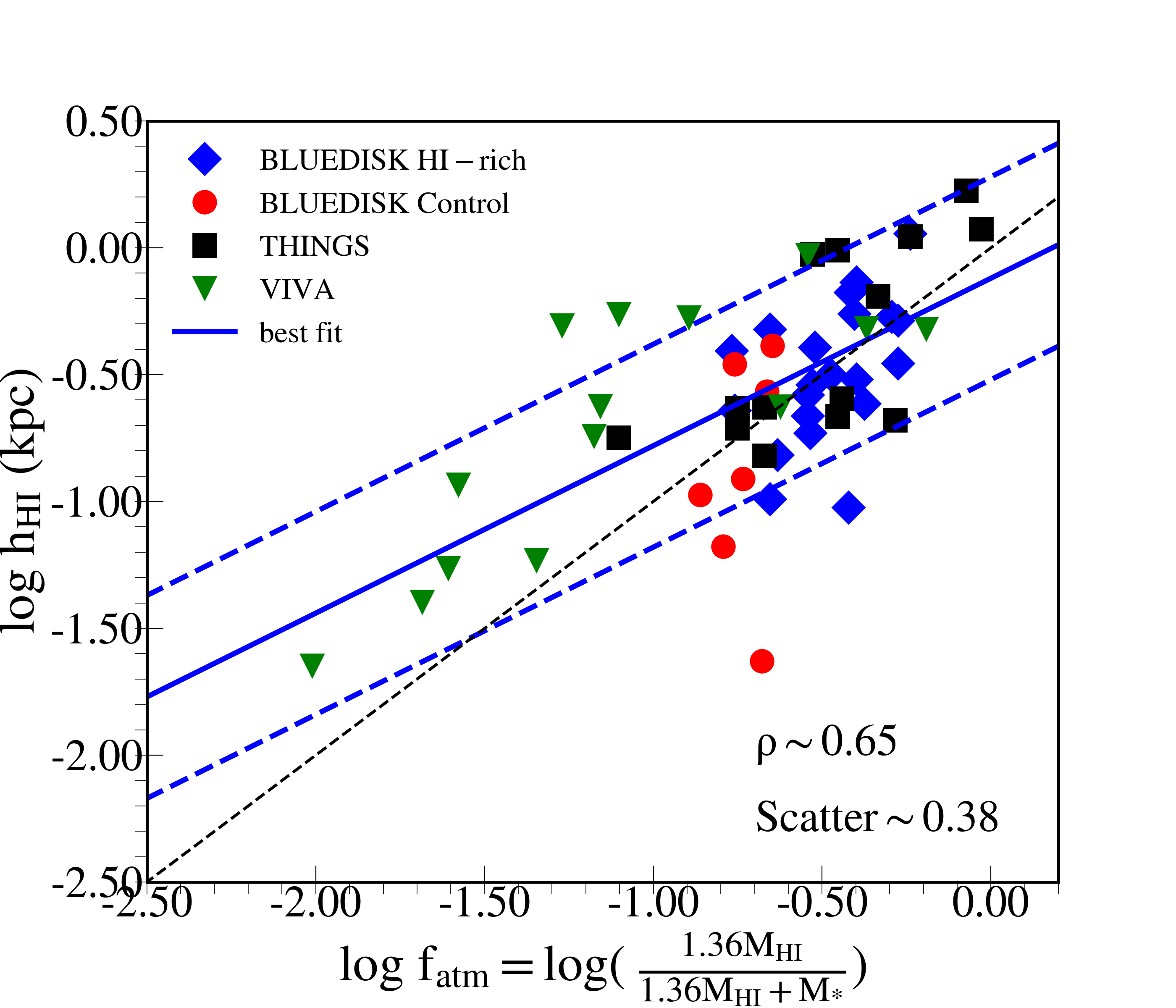}
\includegraphics[width=9.5cm]{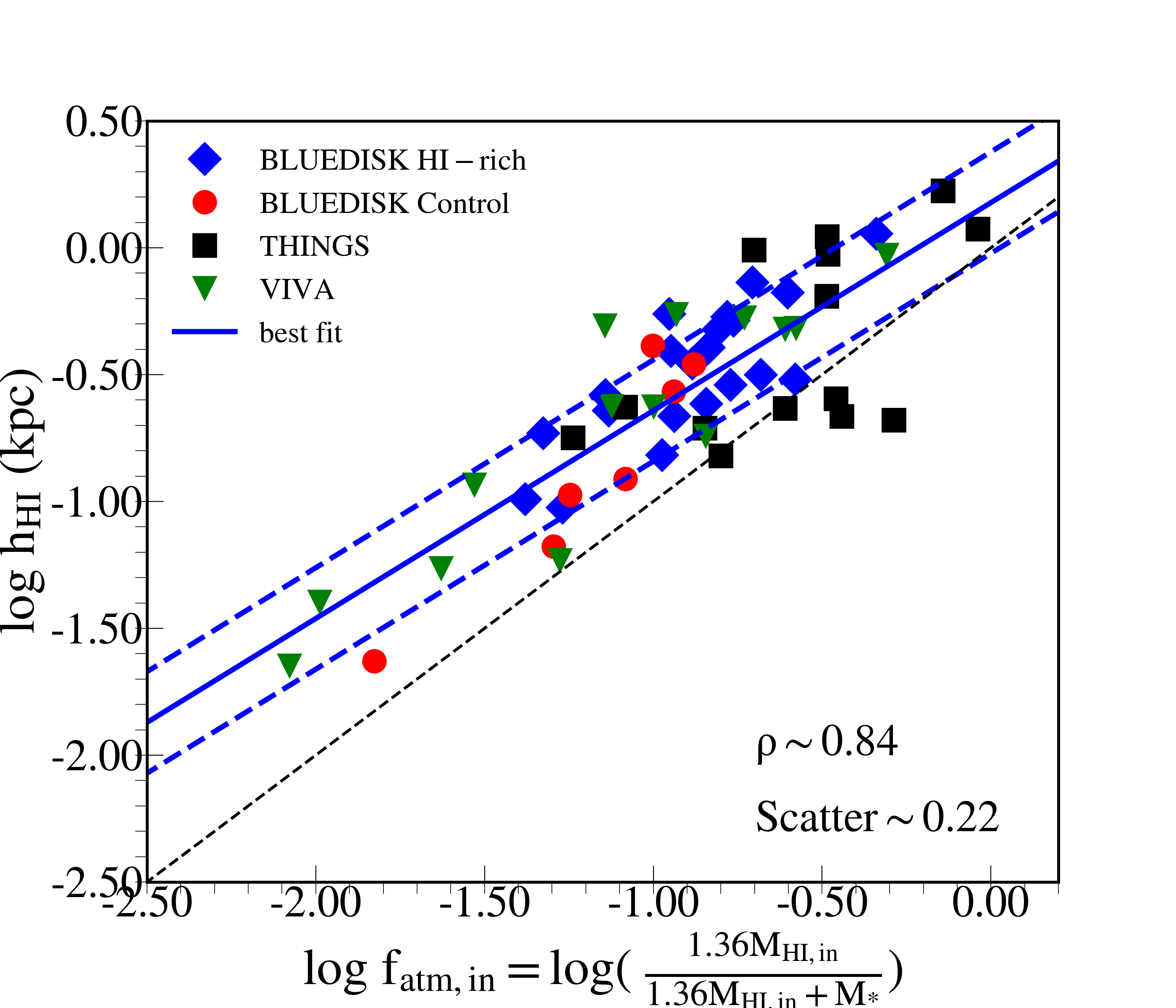}\\
\caption{Relations between H{\sc i} scale height and the atomic gas fraction (top panel)  and with the atomic gas fraction within the optical disk (bottom panel). Lines and symbols are described on the top left corner for each panel and the Pearson correlation coefficients and 1-$\sigma$ scatter are on the bottom right corner. The dotted black lines are  the one-to-one relation.}
\label{fig14}
\end{center}
\end{figure} 

\begin{figure*}
\begin{center}
\includegraphics[width=8.5cm]{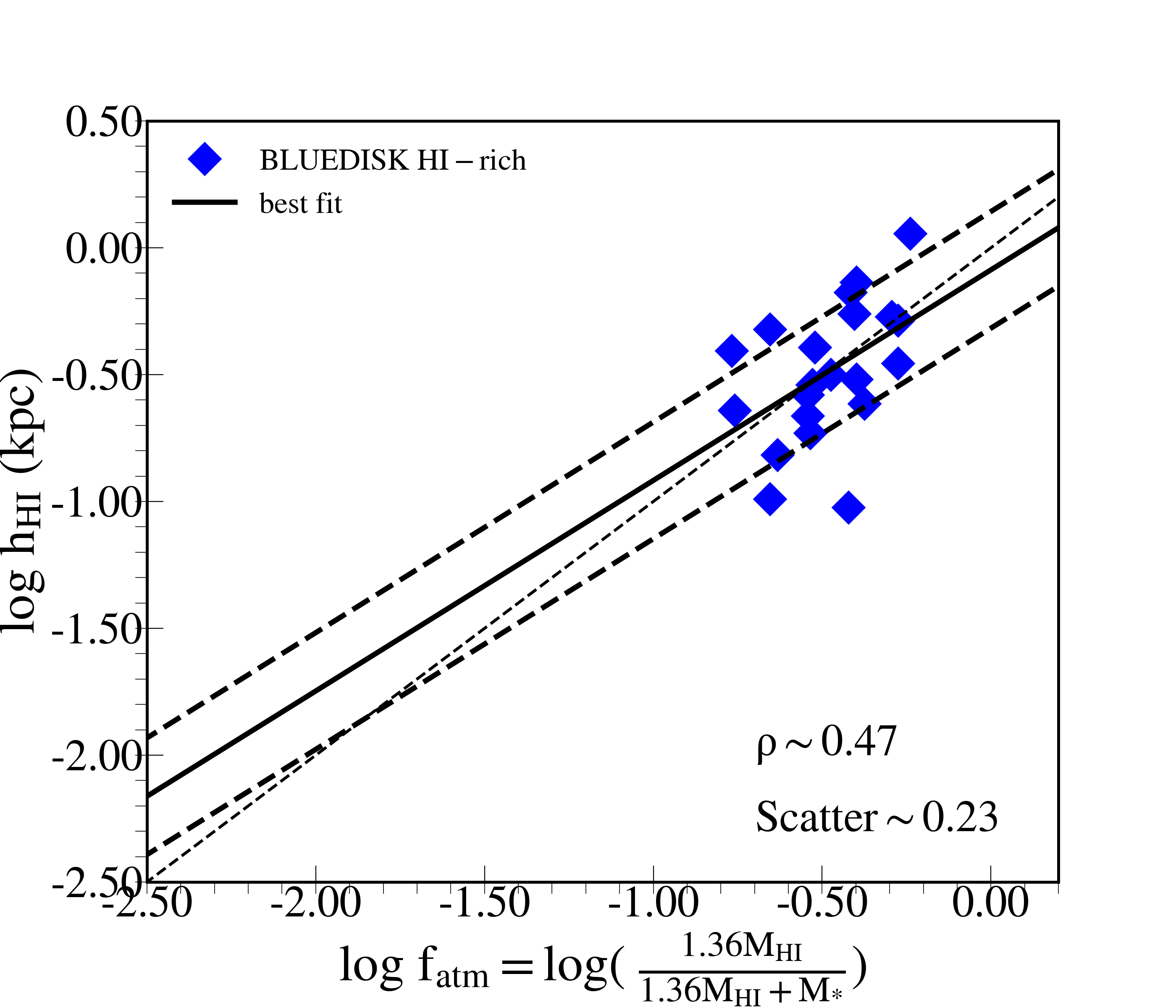}
\includegraphics[width=8.5cm]{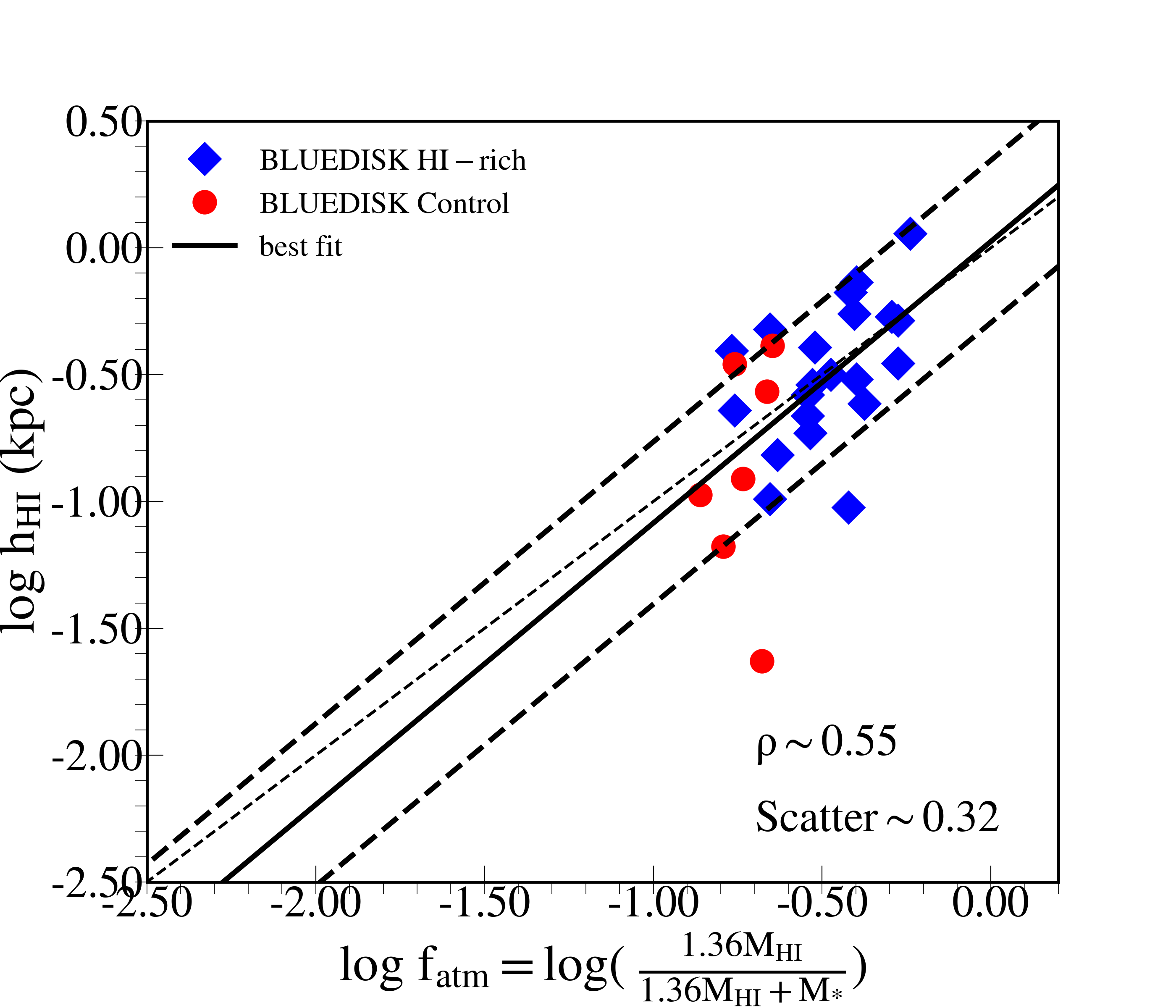}\\
\includegraphics[width=8.5cm]{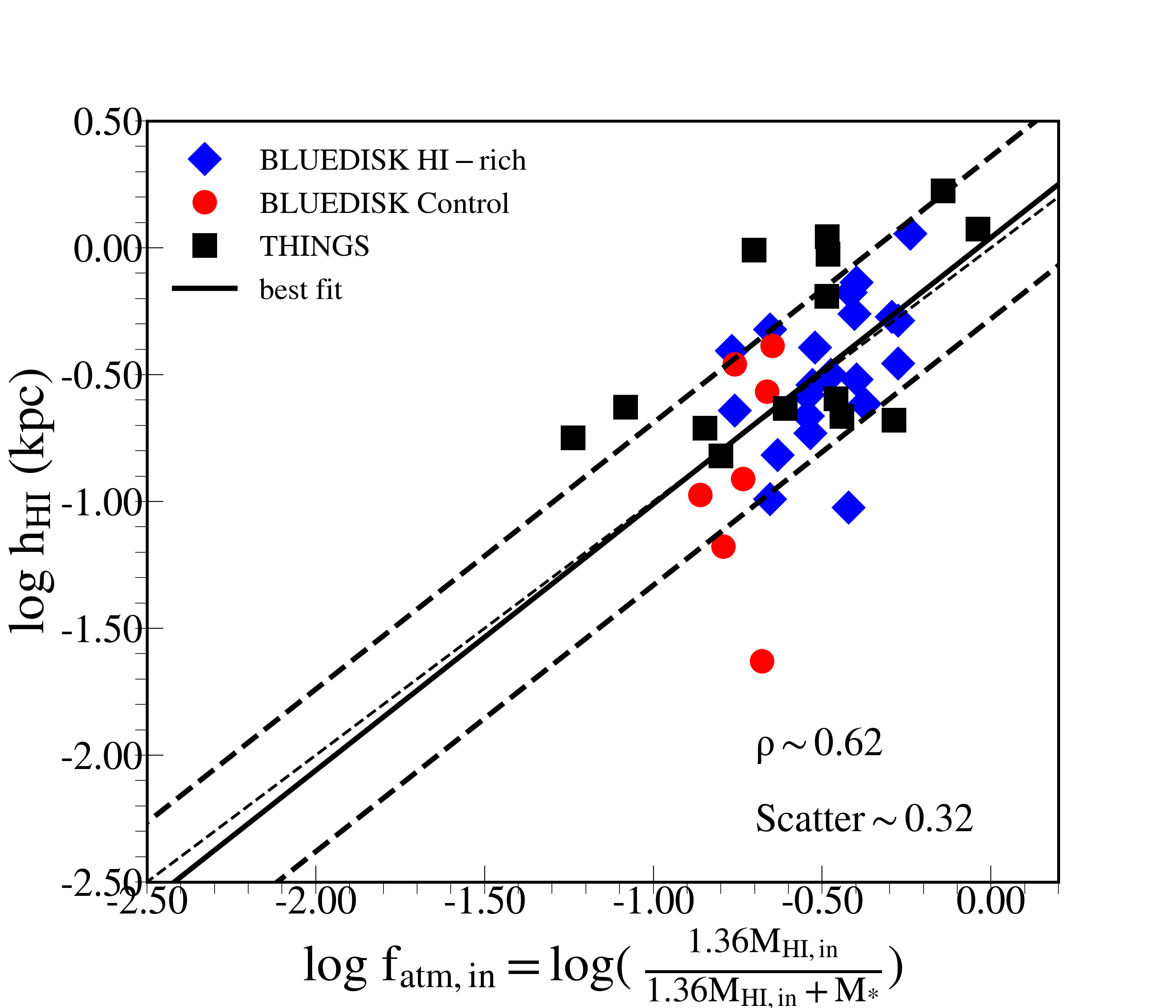}
\includegraphics[width=8.5cm]{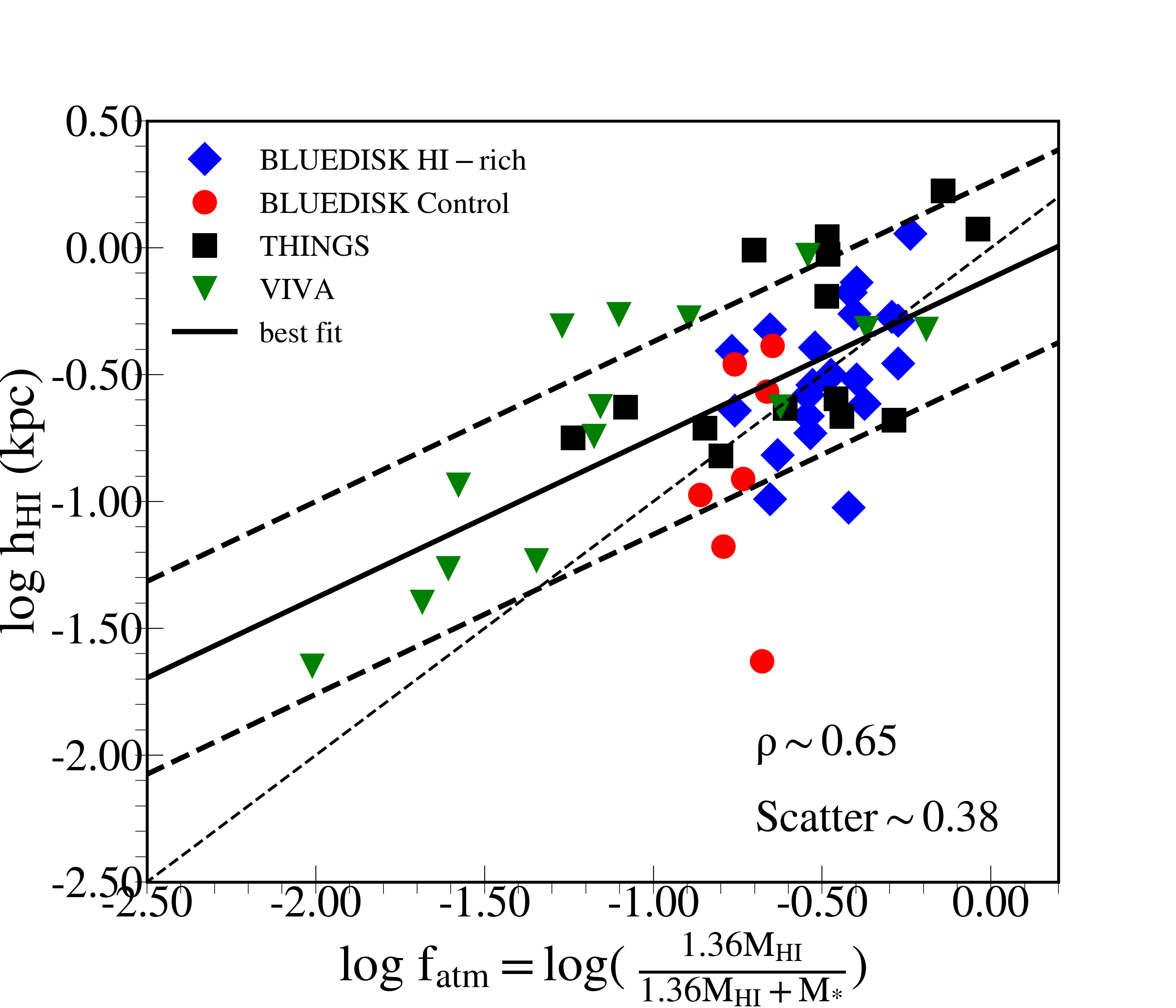}\\
\caption{Relations between H{\sc i} scale height and the atomic gas fraction derived using different combination of the sample. Lines and symbols are the same as in Figure \ref{fig14}.}
\label{fig15}
\end{center}
\end{figure*} 

\begin{figure*}
\begin{center}
\includegraphics[width=8.5cm]{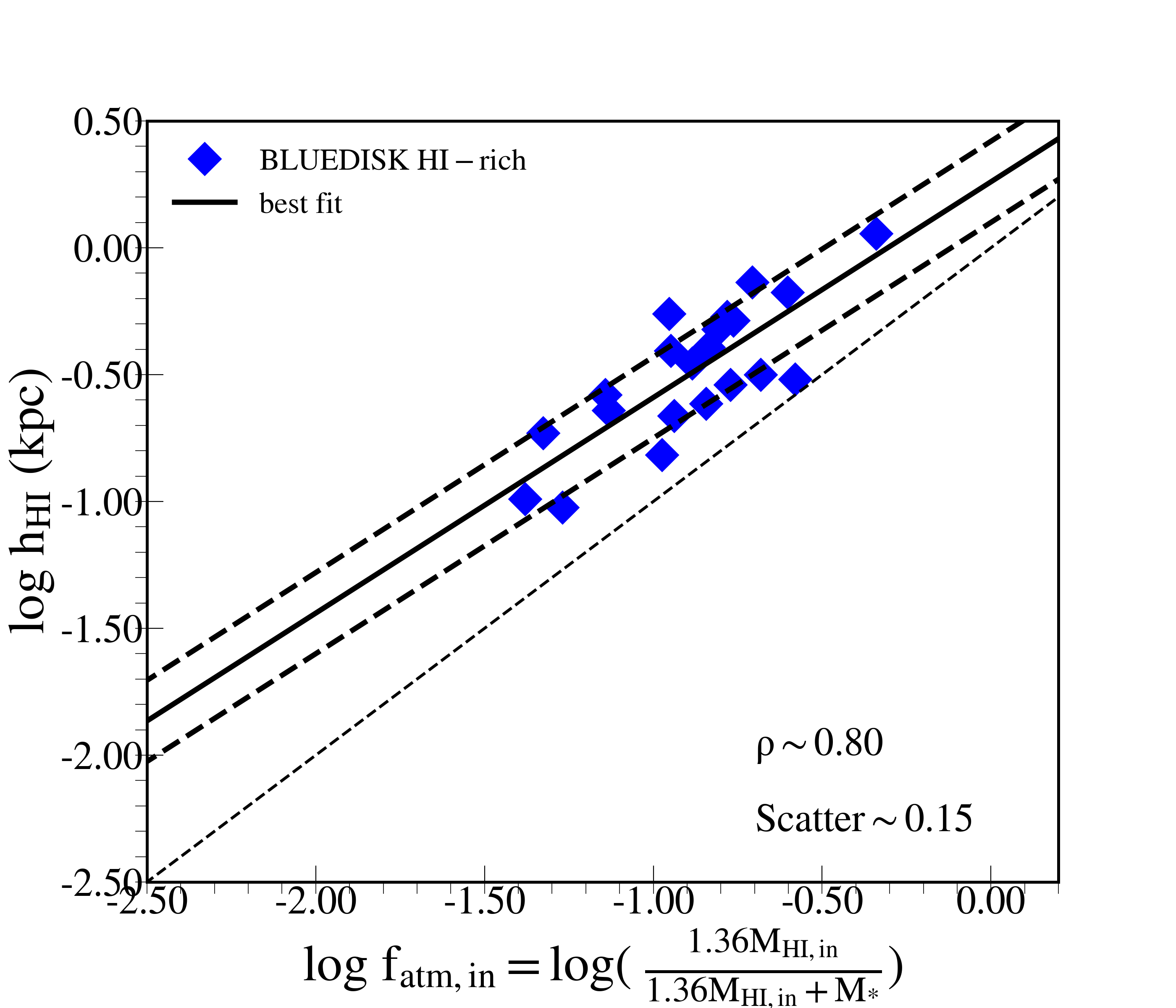}
\includegraphics[width=8.5cm]{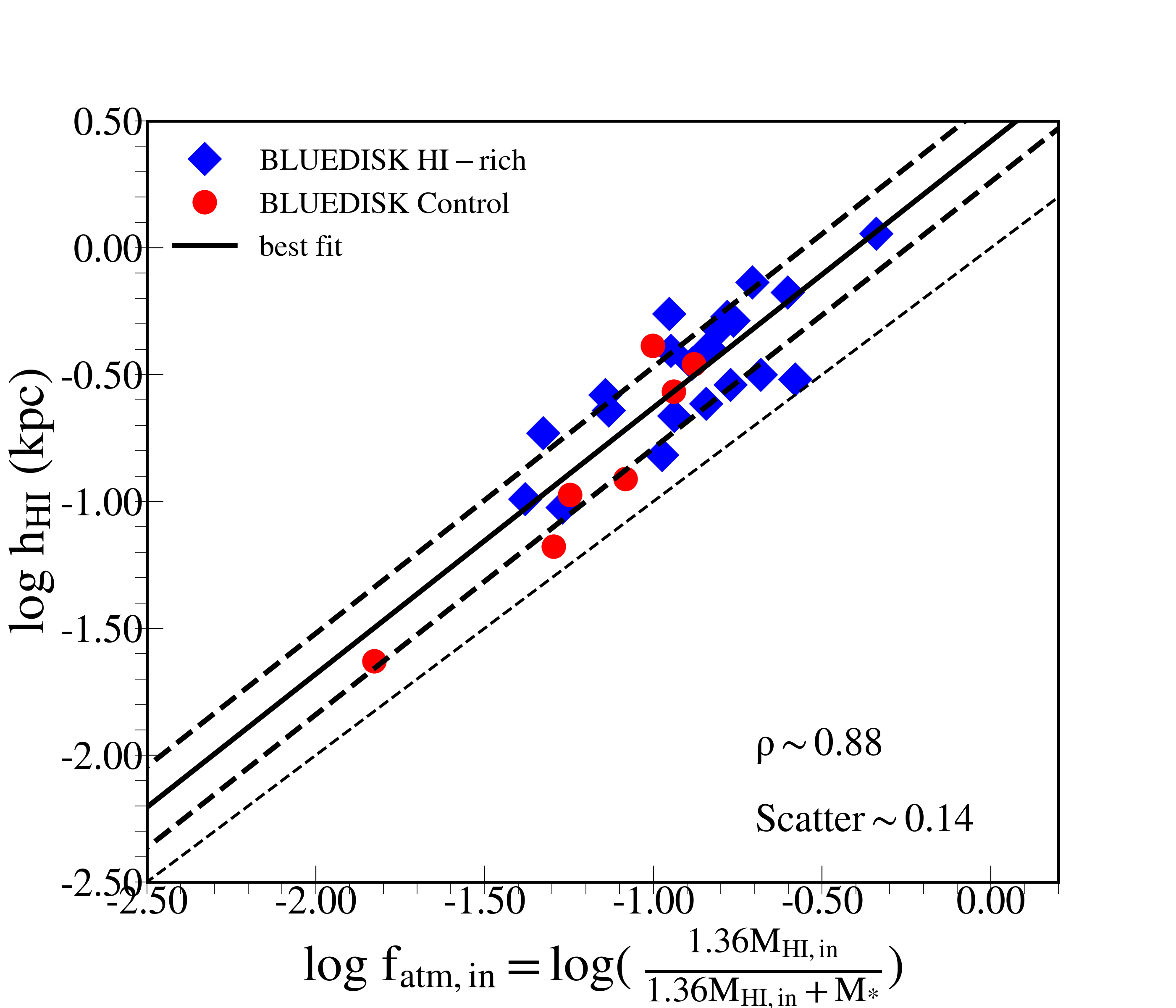}\\
\includegraphics[width=8.5cm]{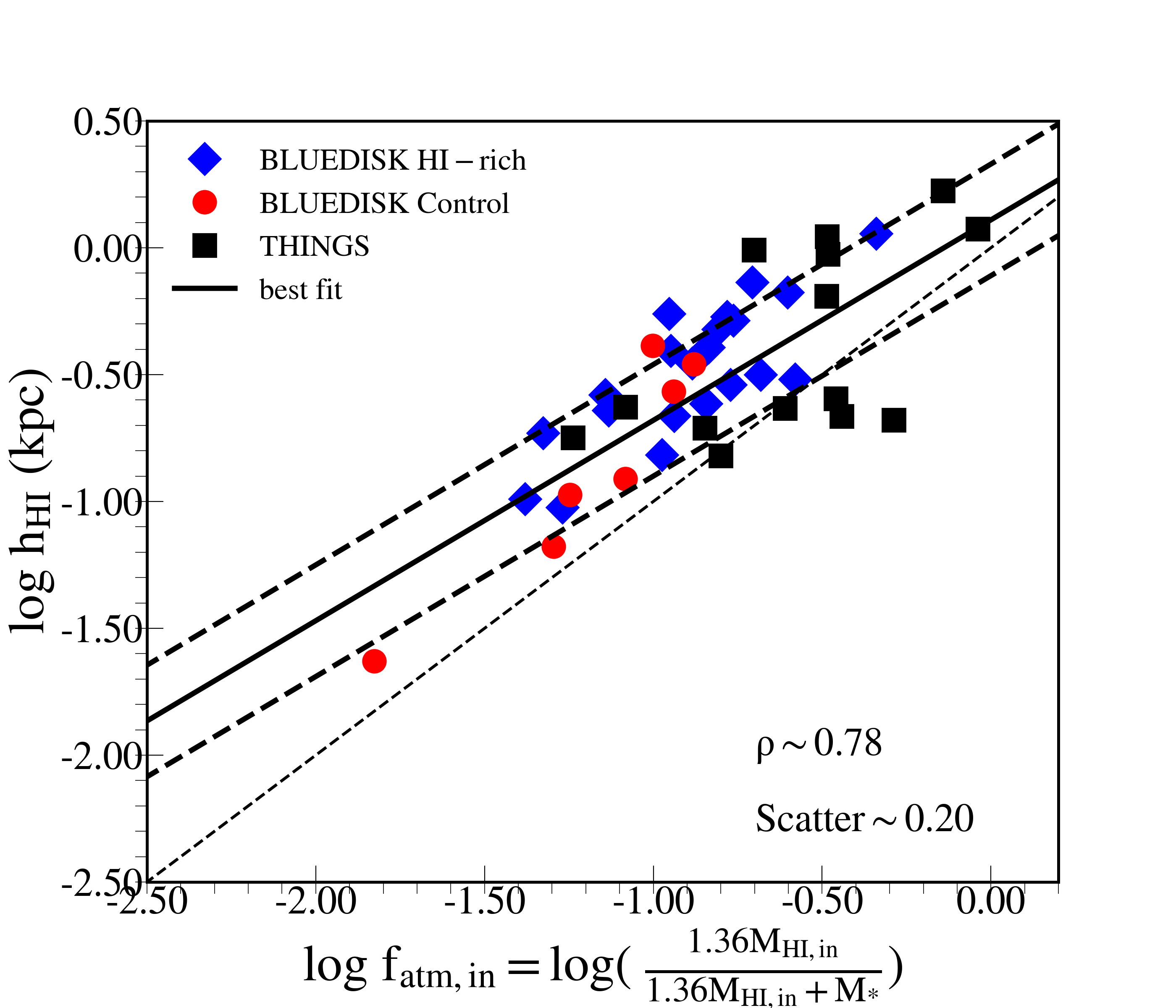}
\includegraphics[width=8.5cm]{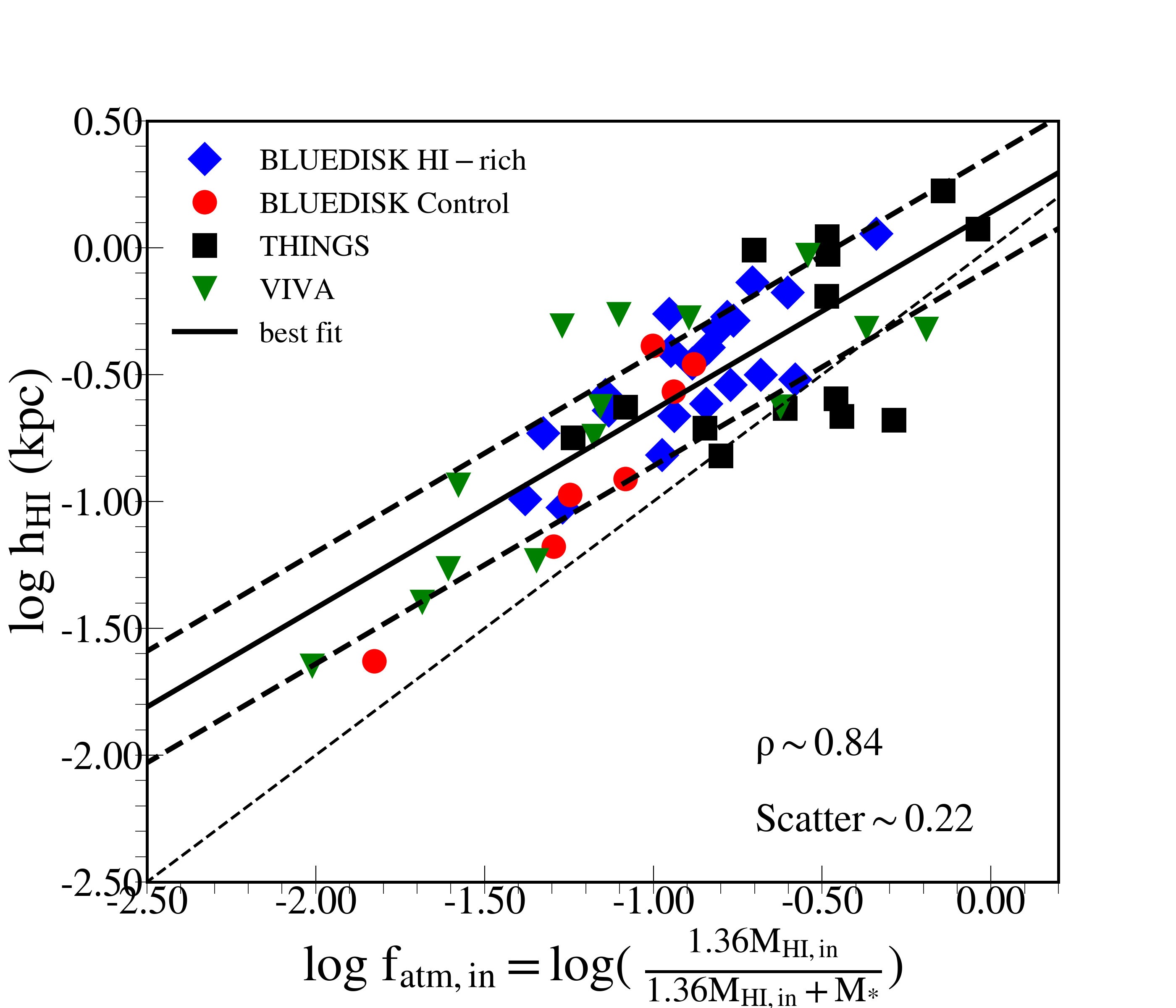}\\
\caption{Relations between H{\sc i} scale height and the atomic gas fraction inside the optical disk derived using different combination of the sample.  Lines and symbols are the same as in Figure \ref{fig14}.}
\label{fig16}
\end{center}
\end{figure*} 

\begin{table*}
 \caption{The H{\sc i} scale height-H{\sc i} fraction relation coefficients using the total H{\sc i} mass. log h$_{HI}$ = $\alpha$ log f$_{atm}$  + $\beta$}
 \label{tab2}
 %\scriptsize
 \begin{center}
 \begin{tabular}{llllll}
  \hline \hline
  Sample & Slope ($\alpha$)&Intercept ($\beta$)&rms Scatter & correlation coef. & p-value: \\\\
\hline \hline

BLUEDISK (HI-rich)&0.83$\pm$0.35	&-0.08$\pm$0.18 & 0.23	&0.47	&0.031\\[2pt]
BLUEDISK (HI-rich+control)&1.11$\pm$0.33& 0.025$\pm$0.18& 0.32& 0.55 &0.002\\[2pt]
BLUEDISK + THINGS&1.05$\pm$0.20& 0.034$\pm$0.12& 0.32& 0.62& 1.15e-5\\[2pt]
BLUEDISK +THINGS + VIVA&0.63$\pm$0.10& -0.12$\pm$0.085& 0.38& 0.65& 2.36e-7\\[2pt]
 \hline
 \end{tabular}
   \end{center}
\end{table*}

\begin{table*}
 \caption{The H{\sc i} scale height-H{\sc i} fraction relation coefficients using the H{\sc i} mass inside the optical disk. log h$_{HI}$ = $\alpha$ log f$_{atm, in}$  + $\beta$}
 \label{tab3}
 %\scriptsize
 \begin{center}
 \begin{tabular}{llllll}
  \hline \hline
    Sample & Slope ($\alpha$)&Intercept ($\beta$)&rms Scatter & correlation coef. & p-value\\\\
\hline \hline
BLUEDISK (HI-rich)&0.85$\pm$0.14& 0.26$\pm$0.13& 0.15& 0.80& 1.12e-5\\[2pt]
BLUEDISK (HI-rich+control)&1.05$\pm$0.11& 0.42$\pm$0.11 &0.14& 0.88& 6.28e-10\\[2pt]
BLUEDISK + THINGS&0.79$\pm$0.09& 0.11$\pm$0.08& 0.20& 0.78& 5.47e-10\\[2pt]
BLUEDISK +THINGS + VIVA&0.78$\pm$0.08& 0.14$\pm$0.08& 0.22& 0.84& 1.37e-13\\[2pt]
 \hline

 \end{tabular}
   \end{center}
\end{table*}

\begin{figure*}
\begin{center}
\includegraphics[width=8cm]{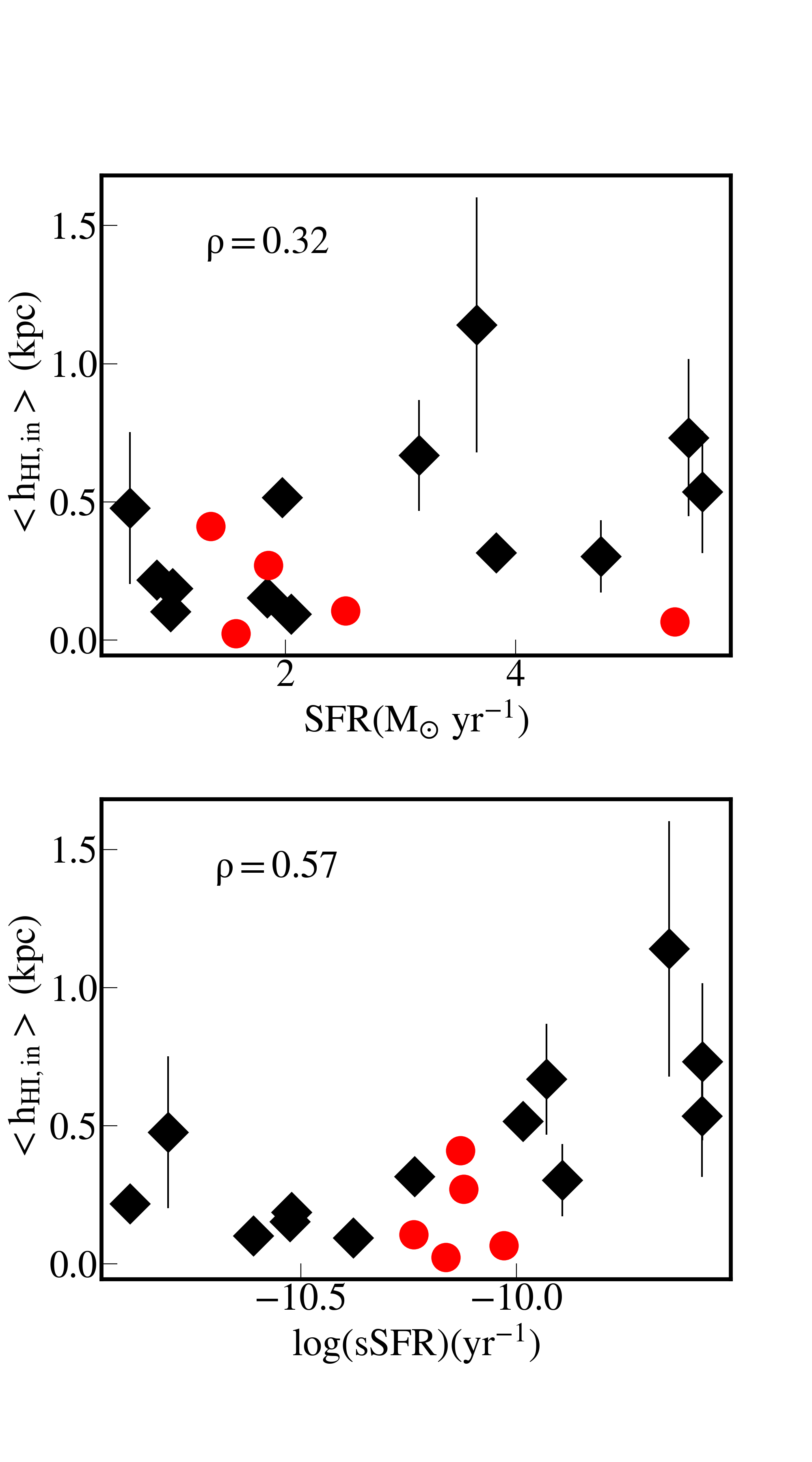}
\includegraphics[width=8cm]{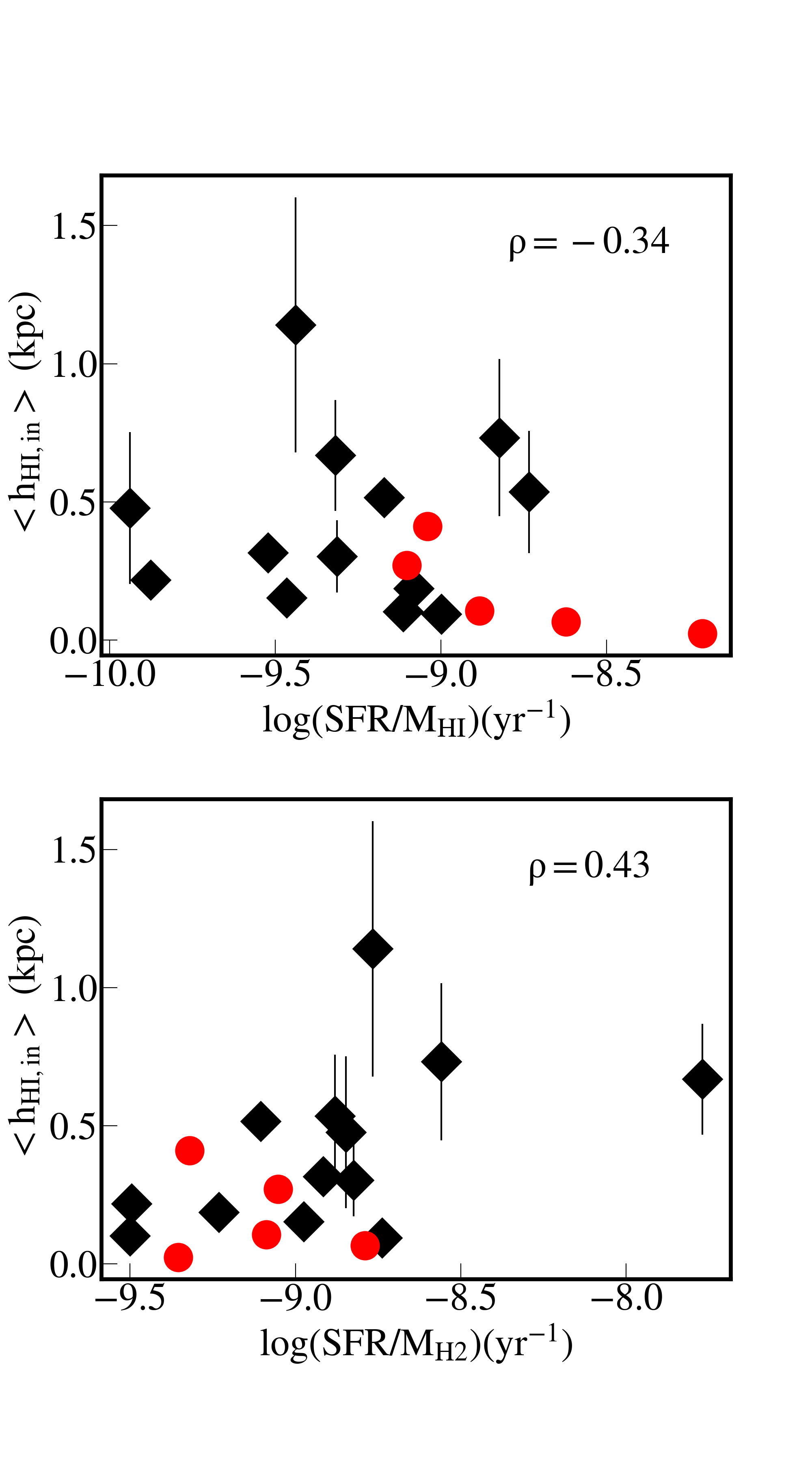}
\caption{Star formation rates, specific star formation rates and star formation per unit gas (H{\sc i} and molecular gas) as function of the average H{\sc i} scale height inside the optical disk. The black diamonds are the H{\sc i}-rich galaxies and the red circles are the control galaxies. The Pearson correlation coefficient are shown on each panel.}
\label{fig17}
\end{center}
\end{figure*} 

\begin{figure}
\begin{center}
\includegraphics[width=8cm]{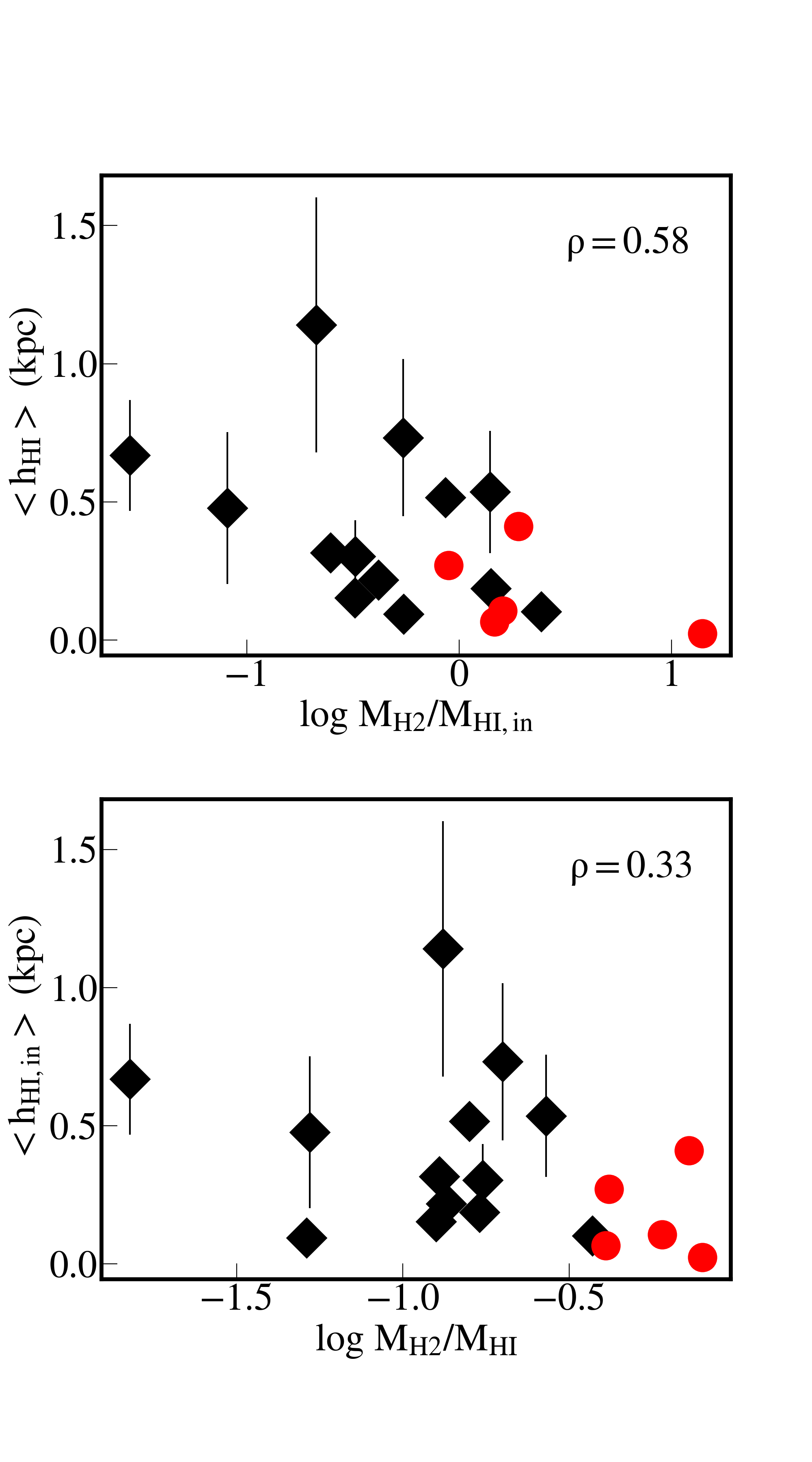}
\caption{Ratio between molecular hydrogen gas mass and H{\sc i} inside the optical radius are plotted against the average H{\sc i} scale height inside the optical disk (top) and total H{\sc i} to molecular gas ratio as function of the average H{\sc i} scale height inside the optical disk (bottom). The black diamonds are the H{\sc i}-rich galaxies and the red circles are the control galaxies. The Pearson correlation coefficient are shown on each panel.}
\label{fig18}
\end{center}
\end{figure}

\section{Results and Discussions}
\subsection{Tilted ring analysis}
 The tilted ring results for the H{\sc i}-rich galaxy ID15 is shown in Figure \ref{fig2}. The rotation curve is presented on the top panel. The errorbars are the quadratic sum of the statistical error and the difference between the approaching and receding side of the galaxy. The inclination and position angle are on the middle and bottom panels, their average values are shown as dashed lines and summarized in Table \ref{tab1}. An arctan function \citep{1997AJ....114.2402C} was fitted to the observed rotation curves in order to obtain the maximum circular velocity V$_{max}$ listed in Table \ref{tab1}. The channel maps are displayed in Figure \ref{fig3} for ID15 where the data is shown in blue and the 3D-BAROLO model in red. The contour levels correspond to 2,4,8 and 10 $\sigma$ where 1 $\sigma$ $\sim$ 0.0023 Jy Beam$^{-1}$. The moment 0 and moment 1 maps are presented in Figure \ref{fig4} and position velocity diagram in Figure \ref{fig5}. These figures show clear signature of non-circular motions along the minor axis of the galaxy. 
\subsection{H{\sc i} disk scale height}
Figure \ref{fig6} compares the \cite{2014MNRAS.441.2159W} density profile with the $^{3D}$BAROLO density profile. \citep{2014MNRAS.441.2159W} used one pixel radial bins ($\sim$ 4 arcsec) using the moment0 maps combined with photometric inclination and position angle, on the other hand, $^3D$BAROLO uses larger radial bins based on the angular resolution of the data ($\sim$ 15 arcsec) and uses the kinematic position and inclination angle listed in Table \ref{tab1}. The two density profiles are in good agreement expect in the inner region, this might be because the $^{3D}$BAROLO density profiles are averaged over larger radius compared to the \citep{2014MNRAS.441.2159W} profiles and$^{3D}$BAROLO fitted inclination and position angle as a function of radius, while \citep{2014MNRAS.441.2159W} used a constant inclination and position angle determined from the optical photometry. Therefore, we will use the $^{3D}$BAROLO density profiles throughout this paper. 

The velocity dispersion profiles are shown on the middle panels of Figure \ref{fig7} for the HI-rich sample and Figure \ref{fig8} for the control sample. They are fitted with an exponential function to obtain the central velocity dispersion. The fitting results are summarized in Table 1. The dashed line indicates a velocity dispersion of 10 km/s which is a typical value for the velocity dispersion of neutral hydrogen gas. The radial variation of the H{\sc i} scale height calculated using equation 10 are on the right panels of figure \ref{fig7} and \ref{fig8} for four HI-rich and control galaxies respectively. The scale height of the remaining galaxies are given in appendix (Figure \ref{figb1} to \ref{figb4} for H{\sc i}-rich and Figure \ref{figb5} for control galaxies.) 

 On average,  the H{\sc i}-rich galaxies have two times H{\sc i} gas fraction ($\sim$ 0.43) compared to the control galaxies ($\sim$ 0.17). The ratio between the H{\sc i} and optical disk size (R$_{HI}$/r$_{25}$) and M$_{HI,in}$/M$_{HI}$ where M$_{HI,in}$ is the H{\sc i} mass inside the optical disk also differ for the HI rich and control samples. The HI-rich galaxies have an average R$_{HI}$/r$_{25}$ $\sim$ 1.78 $\pm$ 0.54 and M$_{HI,in}$/M$_{HI}$ $\sim$ 0.24 $\pm$ 0.17. On the other hand, the control galaxies have an average R$_{HI}$/r$_{25}$ $\sim$ 1.31 $\pm$ 0.40 and M$_{HI,in}$/M$_{HI}$ $\sim$ 0.38 $\pm$ 0.18. The H{\sc i} disk scale height inside the optical disk for the HI-rich galaxies ($\sim$0.41 $\pm$ 0.29 kpc) is however comparable with the control galaxies ($\sim$0.29 $\pm$ 0.14 kpc) and galaxies from THINGS ($\sim$0.68 $\pm$ 0.58 kpc) and VIVA ($\sim$0.31 $\pm$ 0.25 kpc). The H{\sc i}-rich and control galaxies also have similar star formation rates (SFR $\sim$ 2.5 M$_{\odot}$ yr$^{-1}$) and molecular gas mass (M$_{H2}$=2.5 $\times$ 10$^{9}$ M$_{\odot}$) \citep{2016MNRAS.463.1724C}. Previous studies have also shown that H{\sc i} rich galaxies are inefficient at forming stars (e.g Cormier et al. 2016, Lemonias et al. 2014). One possible explanation is that the excess of H{\sc i} is mostly located outside the optical disk (see e.g \citealt{2020ApJ...890...63W}).

One caveats of our method is the angular resolution of the BLUEDISK data which lead to large uncertainties on the measured velocity dispersion. However, \cite{2016MNRAS.462.3628R,2017MNRAS.466.4159I,2019A&A...632A.127B} have shown that the velocity dispersion profiles derived using $^{3D}$BAROLO are reliable since it is corrected for beam smearing. $^{3D}$BAROLO also uses the full data cube instead of moment or Gaussian maps.

\subsection{HI disk flaring}

 It is clear from equation 10 that the scale height will systematically increase toward the outer part of the galaxy if the velocity dispersion is constant or have a small variation  with radius. This is because the surface density exponentially decrease with radius \citep{2014MNRAS.441.2159W}. This phenomena is commonly 
known as disk flaring and have been used to study the shape of dark matter halo \citep{1995AJ....110..591O,1996AJ....112..457O}. Previous studies have shown that the velocity dispersion profiles of nearby galaxies varies with radius \citep{2009AJ....137.4424T,2016AJ....151...15M}. The amplitude of the outer flaring thus depends on the velocity dispersion and H{\sc i} surface density profiles. All galaxies in the BLUEDISK sample show disk flaring when the velocity dispersion is fixed to 10 km s$^{-1}$. (e.g. \citealt{1995AJ....110..591O,1996AJ....112..457O}).

The radial profiles of the H{\sc i} scale height of the BLUEDISK galaxies can be divided into three categories;

\begin{itemize}
\item  galaxies with scale height that increase with radius are in the first category, $\sim$10 galaxies are in this category. Their velocity dispersion profiles are almost flat or have small variation with radius.
\item  galaxies with constant scale height are in the second category, $\sim$ 7 galaxies are in this category. Their velocity dispersion profiles decrease with radius and well fitted with an exponential function (examples: ID11 and ID23).
\item galaxies with complex scale height profiles are in the third category.
\end{itemize}

The spiral HI-rich galaxy ID17 is an example where the disk scale height is constant with radius. For this galaxy, the velocity dispersion decreases by 5 km/s between 40 to 100 arcsecs. ID47 is another galaxy that exhibit similar behavior. 

Figure \ref{fig9} shows the effect of velocity dispersion on the thickness of the HI{\sc i} layer. This figure demonstrates that Higher values for the velocity dispersion will lead to a thicker gas layer and more prominent outer flaring. Theoretically, the velocity dispersion is driven by stellar feedback and gas inflow \citep{2018MNRAS.477.2716K}. The flaring of H{\sc i} disks therefore should carry information about the radial distribution of most recent star formation and radial motion of the gas. This topic will be investigated more thoroughly in a future study. Thus, care should be taken when assuming a constant velocity dispersion. For example, Figure \ref{fig9} shows that a constant velocity dispersion of 8 km/s which is  commonly adopted in the literature (see \citealt{1995AJ....110..591O,1996AJ....112..457O}) is able to reproduce the HI scale height for ID17 but will lead to a thinner HI layer for ID15.

H{\sc i} surface density is another parameter that determine the thickness of the HI layer. Particularly in the outer disk where HI dominates the gravitational potential of the baryons (for H{\sc i} densities drop much slower than the stars or molecular gas, in relatively HI-rich galaxies). Figure \ref{fig10} shows different density profiles for ID15. Each profiles are modeled using an exponential function with the same scale length but different central surface density. We then compute the corresponding scale height using the same velocity dispersion profile but different surface density profiles. The results are shown on the bottom panel of Figure \ref{fig10}. It shows that the disk is thicker when the surface density is low and vice-versa. Environmental effect is also an important factor since it could modify the H{\sc i} surface density and velocity dispersion profiles. It also produces non-circular motions that could not be classified into thermal or turbulent motions. A detailed analysis of the effect of the environment on the H{\sc i} thickness will be deferred to the upcoming paper. 

The average H{\sc i} scale height within the optical disk are summarized in Table \ref{tab1}. 
The central velocity dispersion and the circular velocity V$_{max}$ are plotted against the average H{\sc i} scale height inside the optical disk on the left panels of Figure \ref{fig11} and the flaring amplitude h$_{last}$ - <h$_{HI,in}$> on the right panels. The bottom panels of this figure show clear trend with velocity dispersion. Galaxies with larger central H{\sc i} velocity dispersion are thicker on average and do not exhibit disk flaring.

\subsection{Correlations with galaxy global properties}
This section discusses relationships between the estimated H{\sc i} scale height inside the optical disk with global optical and H{\sc i} properties. We found no correlation between the R-band concentration index (R$_{90}$/R$_{50}$) and the H{\sc i} scale height as shown on the left panel of Figure \ref{fig12} with a Pearson correlation coefficient of $\rho$ $\sim$ -0.04 and a p-value of 0.95. We also did not find any correlation between the H{\sc i} scale height and the H{\sc i} concentration index (R$_{90}$/R$_{50}$)$_{HI}$ which is the ratio between the radius that contain 90\% and 50\% of the H{\sc i} fluxes ($\rho$ $\sim$ 0.04 and p-value 0.94) and the ratio between the radius of the H{\sc i} and optical disk ($\rho$ $\sim$ 0.03 and p-value 0.97). These results imply that for the Bluedisk galaxies, there are no relation between the vertical extend of the H{\sc i} and the central concentration of the stars and gas.

We also explore the relation between the H{\sc i} disk thickness with the atomic gas fraction. Figure \ref{fig13} plots the ratio between the H{\sc i} and stellar mass as a function of the H{\sc i} scale height. The bottom panel of this figure shows that the H{\sc i} scale height is strongly correlated with the H{\sc i} fraction estimated within the optical disk with a Pearson correlation of $\rho$ $\sim$ 0.77 and a p-value $\sim$ 10$^{-6}$. We further added galaxies form  the THINGS and VIVA H{\sc i} surveys to the relation  and used the atomic gas fraction instead of H{\sc i} fraction. The bottom panel of Figure \ref{fig14} shows a strong correlation between the atomic gas fraction inside the optical disk with the H{\sc i} disk thickness. 
We derive the following scaling relation between the H{\sc i} disk scale height and the atomic gas fraction measured inside the optical disk:
\begin{equation}
log <h_{HI,in}>=0.8196 \times log \ f_{atm,in}  + 0.178 \ (\pm 0.22)
\end{equation}
where the atomic gas fraction inside the optical disk is given by
\begin{equation}
log \ f_{atm,in} = log ( \frac{1.36 M_{HI,in}}{1.36M_{HI,in}+M_{*}})
\end{equation}

The 1-$\sigma$ scatter is smaller $\sim$ 0.22 dex if $f_{atm,in}$ is used compared to 0.38 dex if $log f_{atm}$ is used instead.
The relation between the HI scale height and atomic gas fraction can be expressed as 

\begin{equation}
<h_{HI}>=C_{1} \times f_{atm}^{C_{2}} 
\end{equation}

where $C_{1}$ and $C_{2}$ are constants.

\subsection{The H{\sc i} scale height-atomic gas fraction relation}
We checked if the relation between the H{\sc i} scale height and the atomic gas fraction relation measured inside the optical disk is produced to a specific sample. We derive the relation using different combination of the sample using the total atomic gas fraction in Figure \ref{fig15}. The fitting results are summarized in Table 2. We found no correlation between scale height and the total atomic gas fraction if only the BLUEDISK H{\sc i} rich galaxies are used in the fit with a p-value of 0.031. The fit improves when control galaxies are also added to the fit but the p-value of 0.002 still suggest no correlation. The scatter also increases by 0.1 dex. A correlation is found when both the THINGS and VIVA are also added to the fit. However, the relation becomes more flat with a slope of $\sim$ 0.6 compared to $\sim$ 1.1 when BLUEDISK galaxies alone are used. This change of slope is mainly driven by the VIVA galaxies that have lower H{\sc i} fraction but similar scale heights with the THINGS and BLUEDISK galaxies.
On the other hand, a clear correlation is found when the atomic gas fraction inside the optical disk is used. The fitting results are shown in Figure \ref{fig16} and summarized in Table 3 for the fit using the atomic gas fraction inside the optical disk. The slope of the relation also does not change significantly when adding more galaxies and the scatter is smaller compared to slope obtained when the total atomic gas fraction is used. Upcoming survey such as WALLABY will provide us with a large enough sample that span a wide range of H{\sc i} fraction and will give us a clear picture on the H{\sc i} scale height-atomic gas  fraction relation.

\subsection{Relation with star formation rates and molecular gas content}
In this section we explore the correlation between H{\sc i} scale height with the molecular gas content and SFR.  
In Figure \ref{fig17}, the star formation rates, star formation rates per unit gas (atomic and molecular) and specific star formation rates are plotted as a function of the HI scale height inside the optical disk. This figure shows that the scale height is correlated with the specific star formation rate and the star formation rate per molecular gas mass with a p-value smaller than 0.001. No correlation was found between the H{\sc i} thickness and the star formation rate nor the star formation rate per H{\sc i} mass. 
In Figure \ref{fig18}, the molecular to H{\sc i} ratio inside the optical disk (top) and the total molecular to H{\sc i} ratio (bottom) are plotted as a function of the H{\sc i} scale height. This figure shows that the H{\sc i} scale height is more correlated with the  H{\sc i} to the molecular inside the optical disk with correlation coefficient of 0.58 and a p-value of < 0.001 and not with the total molecular to H{\sc i} ratio. Note that theoretically, mid-plane pressure is a key factor in both setting the H{\sc i} thickness, and the conversion of H{\sc i} to the molecular (\citealt{2010ApJ...721..975O,2006ApJ...650..933B}). The role of mid-plane pressure in setting H$_{2}$ to H{\sc i} conversion has been confirmed by \citet{2008AJ....136.2782L}.

\section{Summary}

\label{sec:highlight}
 We used an empirical relation from \cite{2019ApJ...882....5W} to estimated the average H{\sc i} scale-height of spiral galaxies selected from BLUEDISK \citep{2013MNRAS.433..270W}, THINGS \citep{2008AJ....136.2563W} and VIVA \citep{2009AJ....138.1741C} H{\sc i} surveys. We investigated correlations between the HI scale height with other optical and HI global properties. 

Our main results are as follow:
\begin{itemize}
\item The HI-rich galaxies from BLUEDISK have similar H{\sc i} disk thickness inside the optical disk on average ($\sim$0.41 $\pm$ 0.29 kpc) to the control galaxies ($\sim$0.29 $\pm$ 0.14 kpc), THINGS ($\sim$0.68 $\pm$ 0.58 kpc) and VIVA ($\sim$0.31 $\pm$ 0.25 kpc) galaxies.
\item The H{\sc i} scale height radial profiles can be divided into three categories. The first category consist of galaxies where the H{\sc i} disk thickness increases with radius, the second category have a H{\sc i} scale height that is constant with radius and the third category have complex scale height radial profiles. The complexity may reflect a mixture of effects from gas inflow, the resulted radial distribution of gas surface density , stellar feedback and environmental effects.
\item The average H{\sc i} scale height within the optical disk is correlated  with the atomic gas fraction. Such correlations would be useful when estimating the H{\sc i} disk thickness of large sample of galaxies.
\item  The average scale height within the optical disk is also correlated with the central H{\sc i} velocity dispersion. However, a larger sample is needed before further interpretation.

\item The relation between $M_{H2}/M_{HI}$ and the H{\sc i} scale height confirm the role of gas disk thickness on star formation efficiency which have been previously reported ( e.g. \citealt{2008AJ....136.2782L}). 
\end{itemize}
Measuring the gas disk thickness is a challenging task and direct measurements are only available for handful of edge-on galaxies. A scaling relation is therefore crucial and will allow us to estimate the vertical extend of H{\sc i} in large sample of galaxies. The WALLABY survey will observe $\sim$ 500000 galaxies but most of these will be unresolved. Single dish radio telescope such as FAST will detect a large number of galaxies in HI. These upcoming survey will provide HI masses for a significant number of galaxies. We can therefore explore statistically the effect of the vertical extend of HI on star formation efficiency in nearby disk galaxies.

Dedicated survey of edge-on galaxies will increase the number of galaxies with measured scale height and will help us to better understand the vertical distribution of H{\sc i}. This will also allow us to test different theories of vertical equilibrium.
\acknowledgments

We thank the anonymous referee for the comments and suggestions. 

%\newpage

%\newpage
\appendix
\renewcommand\thefigure{\thesection.\arabic{figure}} 
\setcounter{figure}{0}
%\section{Effect of velocity dispersion and surface density profile on the H{\sc i} scale heights.}

\section{H{\sc i} scale heights of the remaining BLUEDISK galaxies}

\begin{figure*}
\begin{center}
\includegraphics[width=18.5cm]{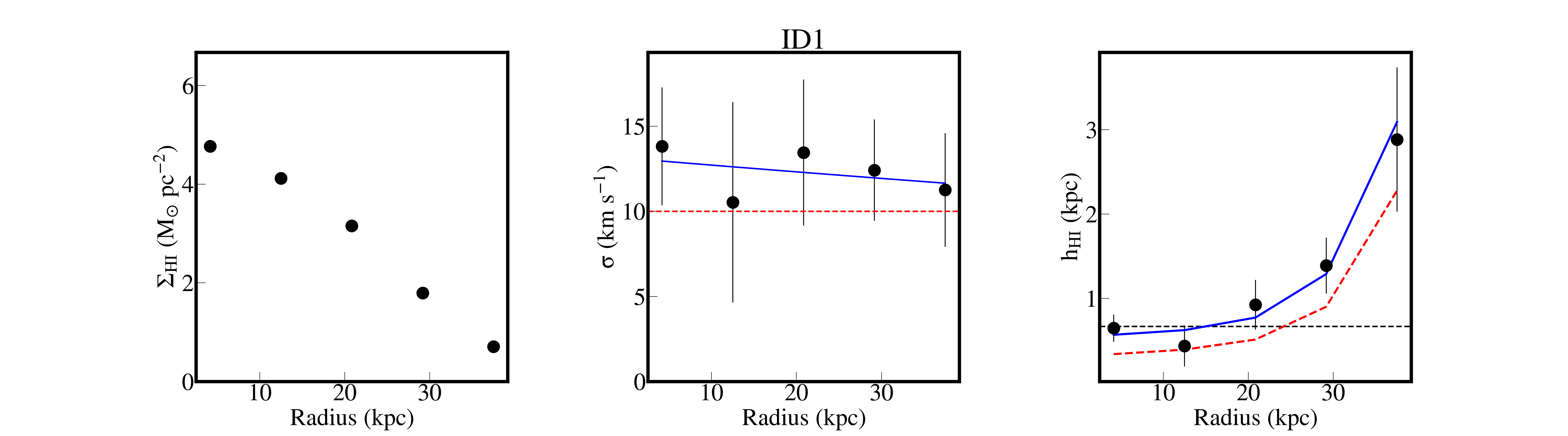}
\includegraphics[width=18.5cm]{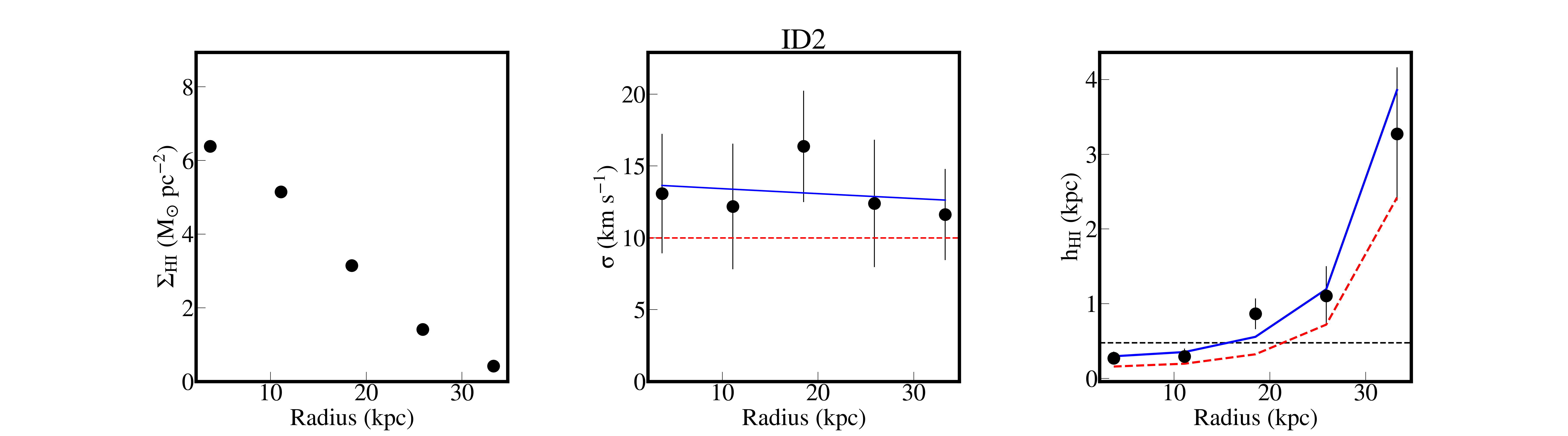}
\includegraphics[width=18.5cm]{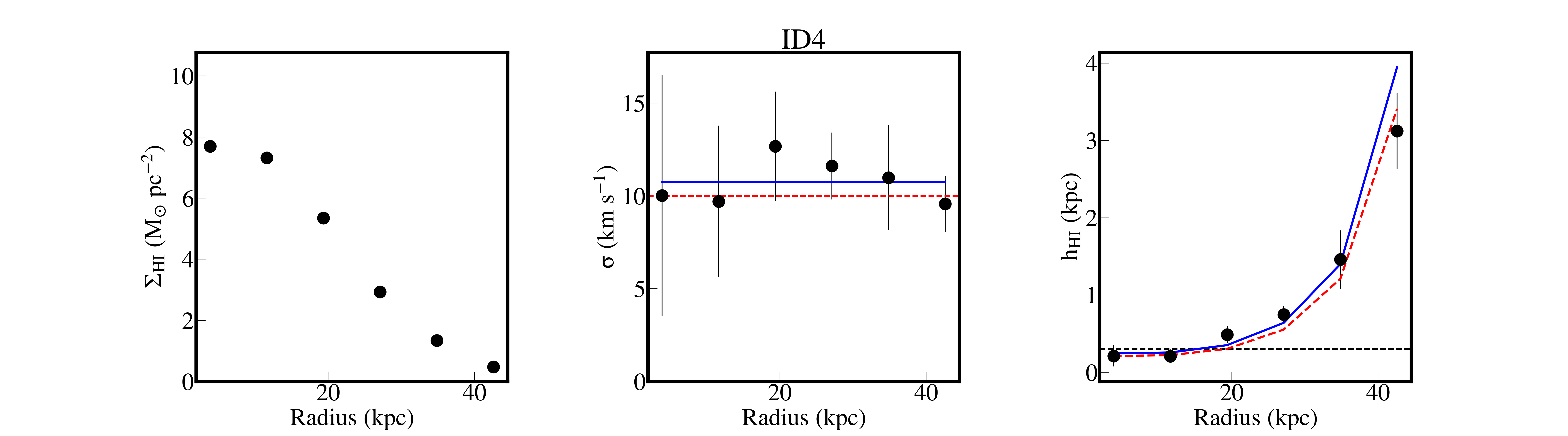}
\includegraphics[width=18.5cm]{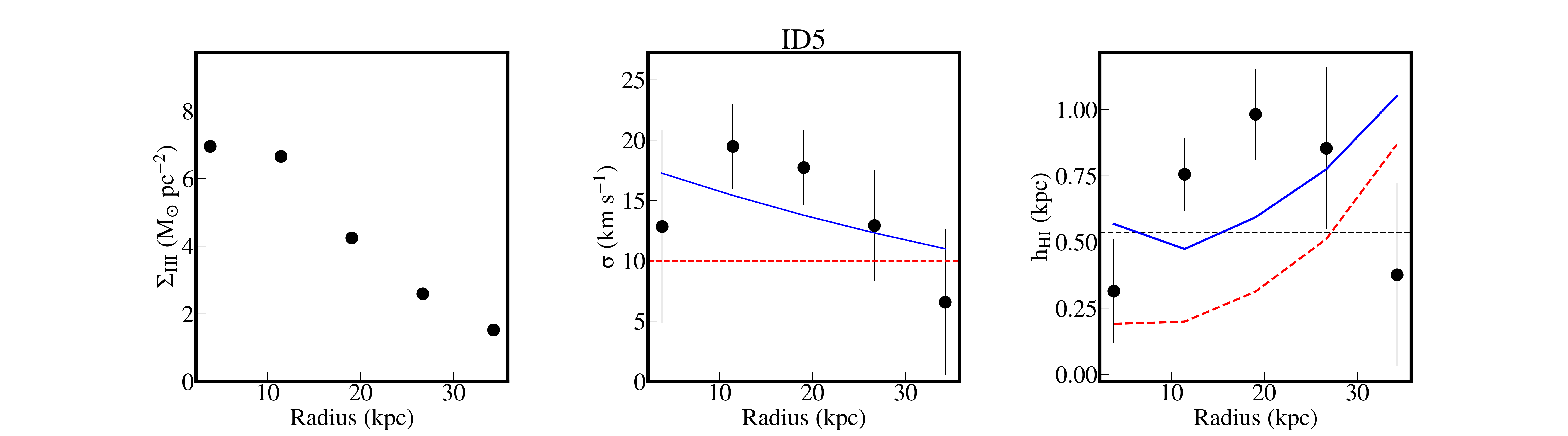}
\caption{HI scaleheight of the remaining HI-rich galaxies. See Figure 4 for details.}
\label{figb1}
\end{center}
\end{figure*}

\begin{figure*}
\begin{center}
\includegraphics[width=18.5cm]{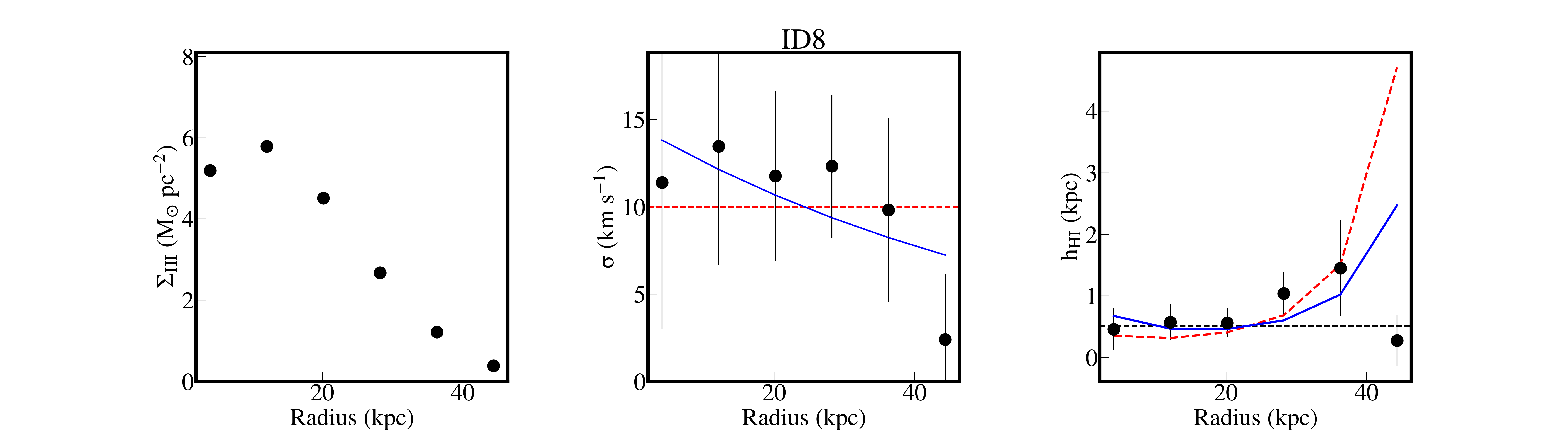}
\includegraphics[width=18.5cm]{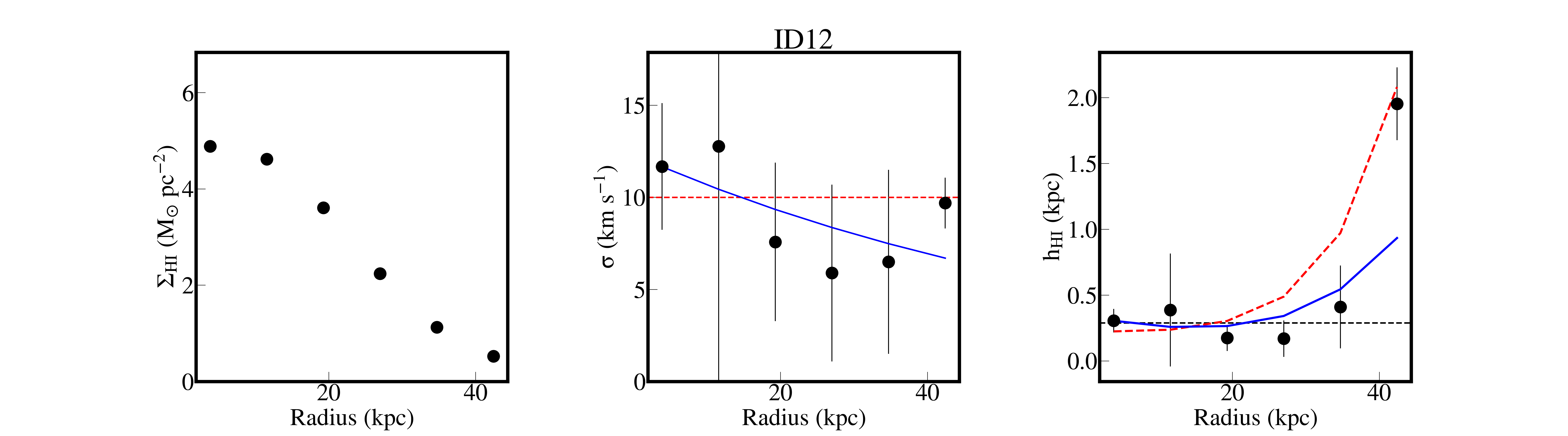}
\includegraphics[width=18.5cm]{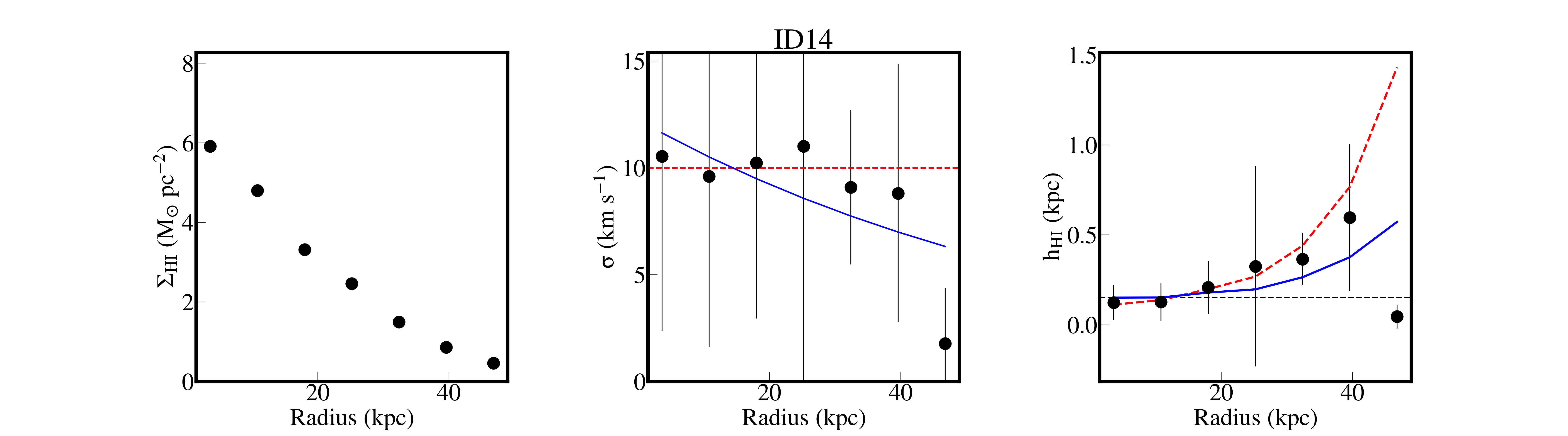}
\includegraphics[width=18.5cm]{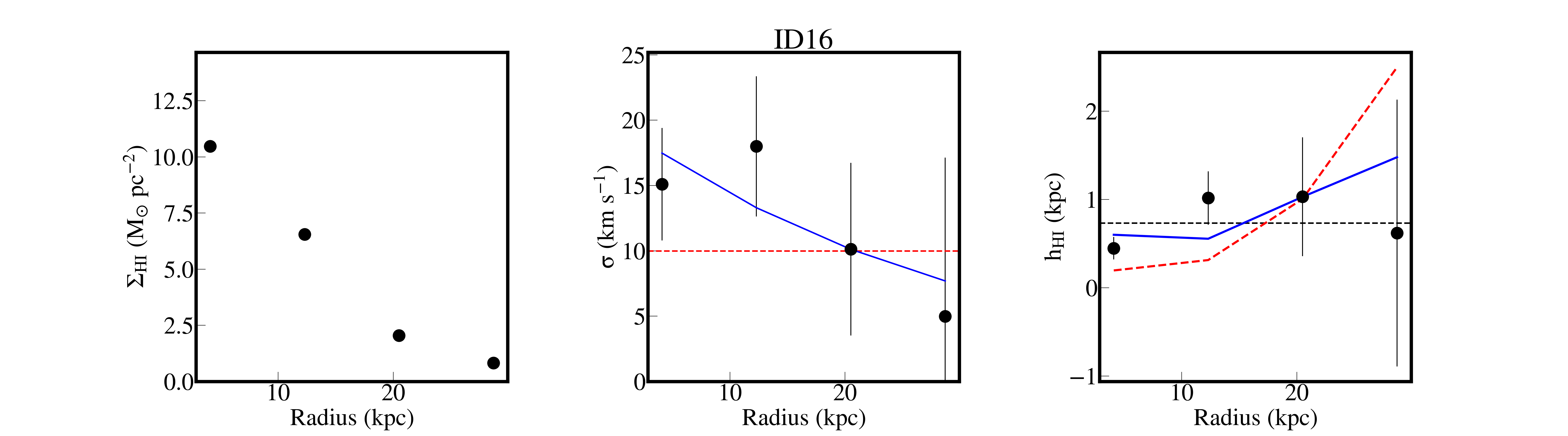}

\caption{HI scaleheight of the remaining HI-rich galaxies (cont.)}
\label{figb2}
\end{center}
\end{figure*}

\begin{figure*}
\begin{center}
\includegraphics[width=18.5cm]{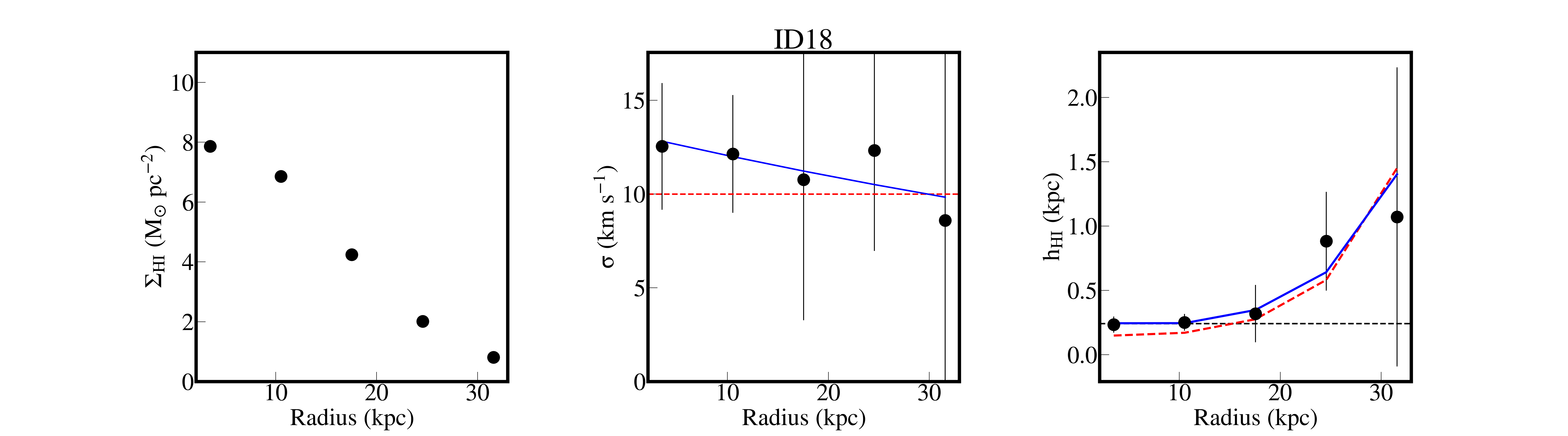}
\includegraphics[width=18.5cm]{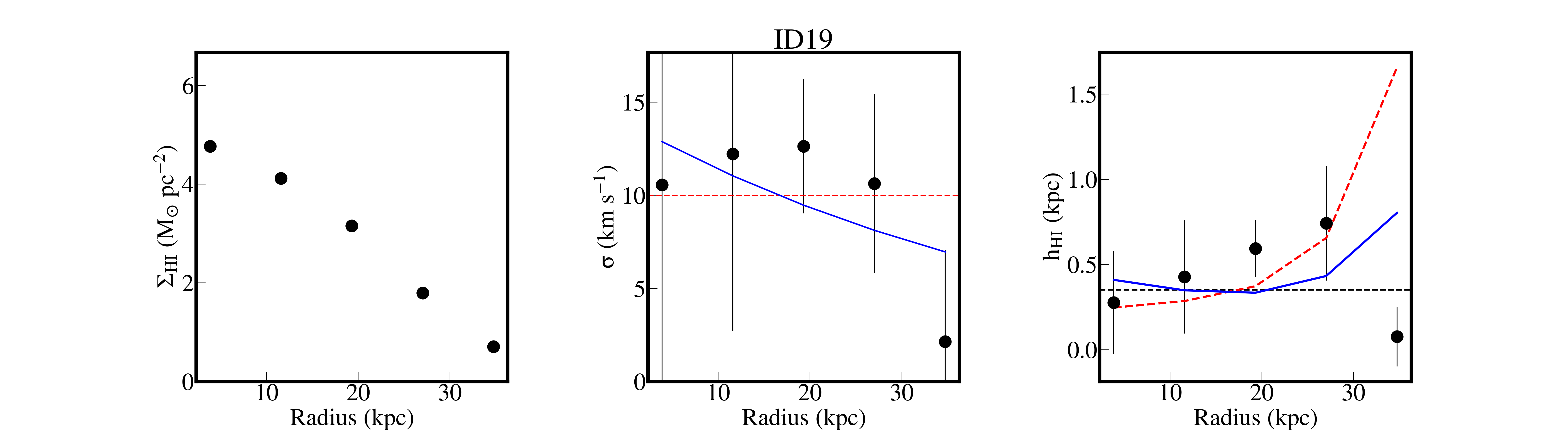}
\includegraphics[width=18.5cm]{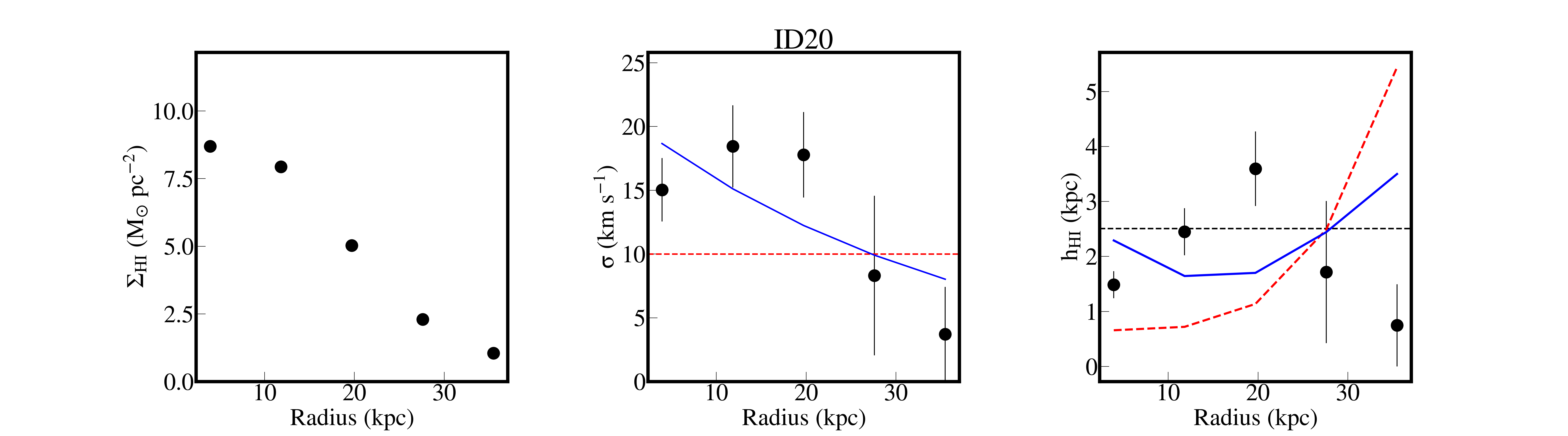}
\includegraphics[width=18.5cm]{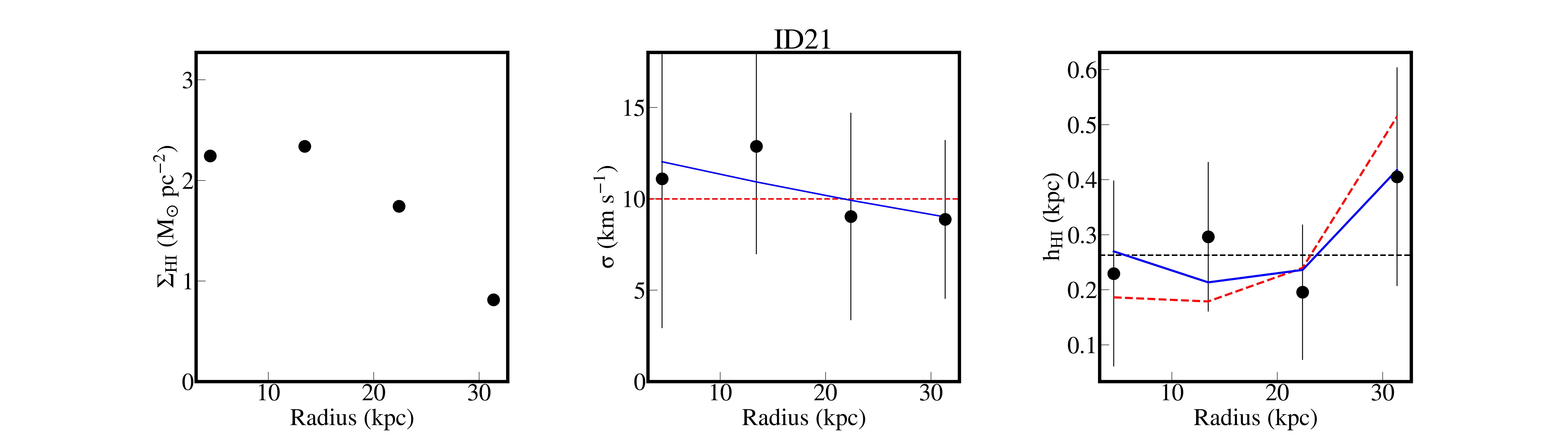}

\caption{HI scale height of the remaining HI-rich galaxies (cont.)}
\label{figb3}
\end{center}
\end{figure*}

\begin{figure*}
\begin{center}
\includegraphics[width=18.5cm]{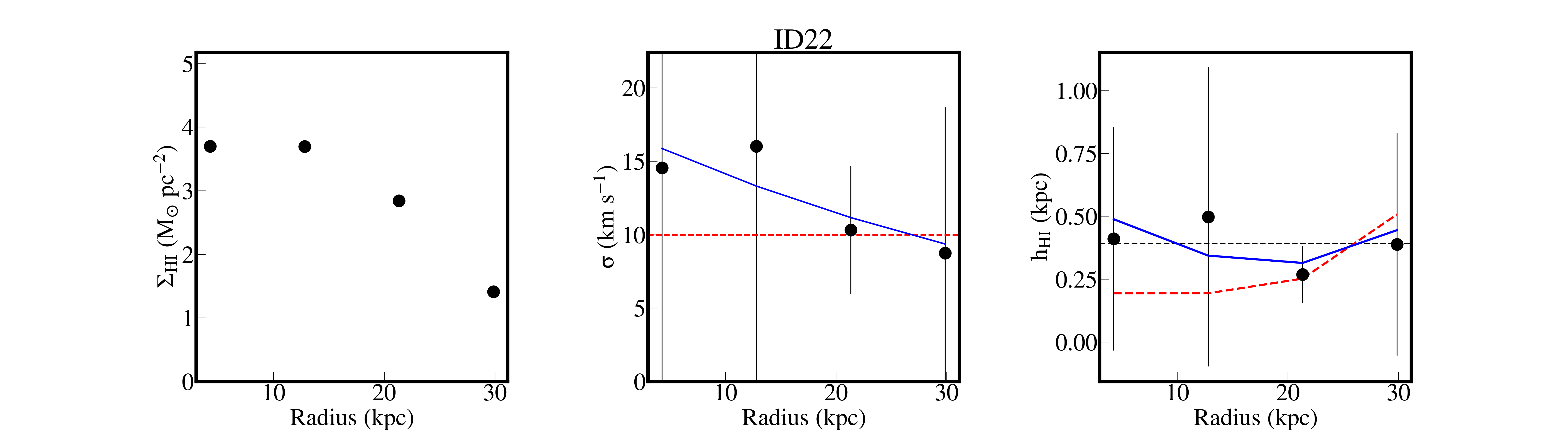}
\includegraphics[width=18.5cm]{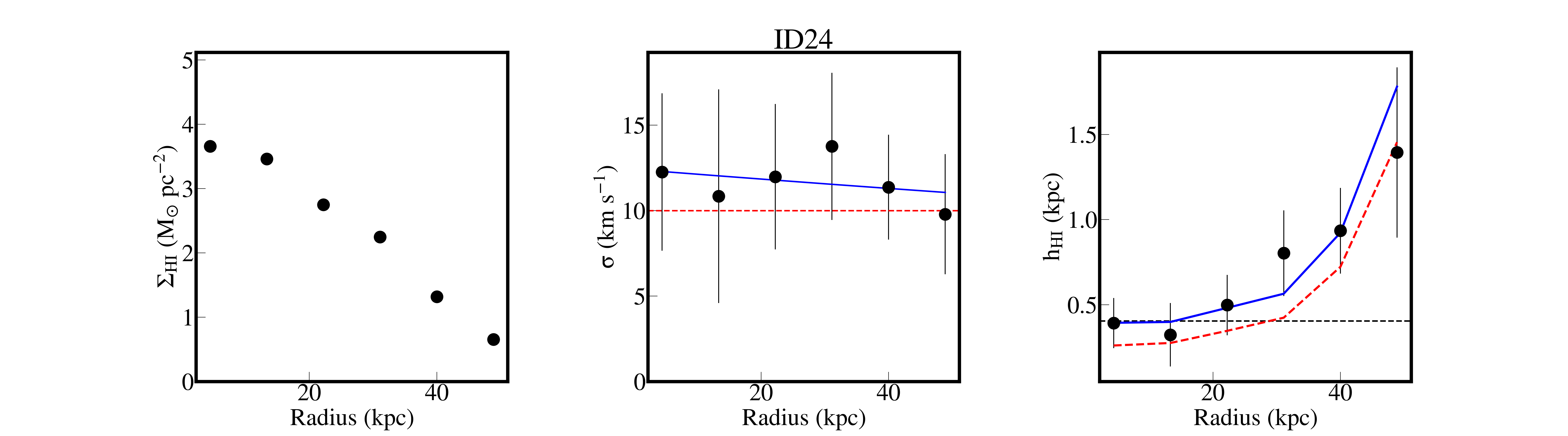}
\includegraphics[width=18.5cm]{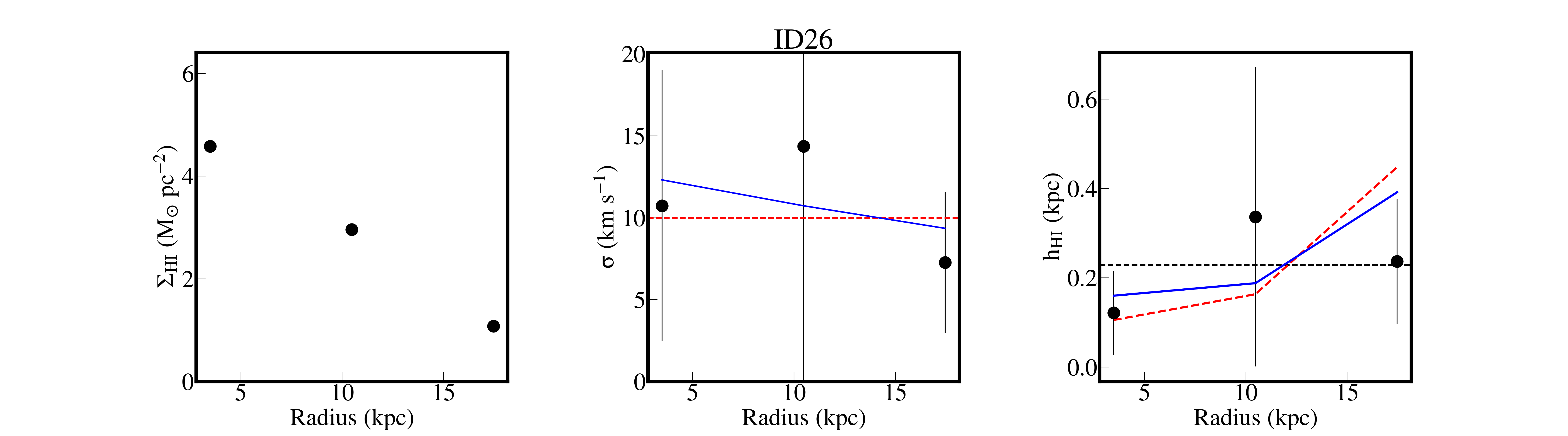}
\includegraphics[width=18.5cm]{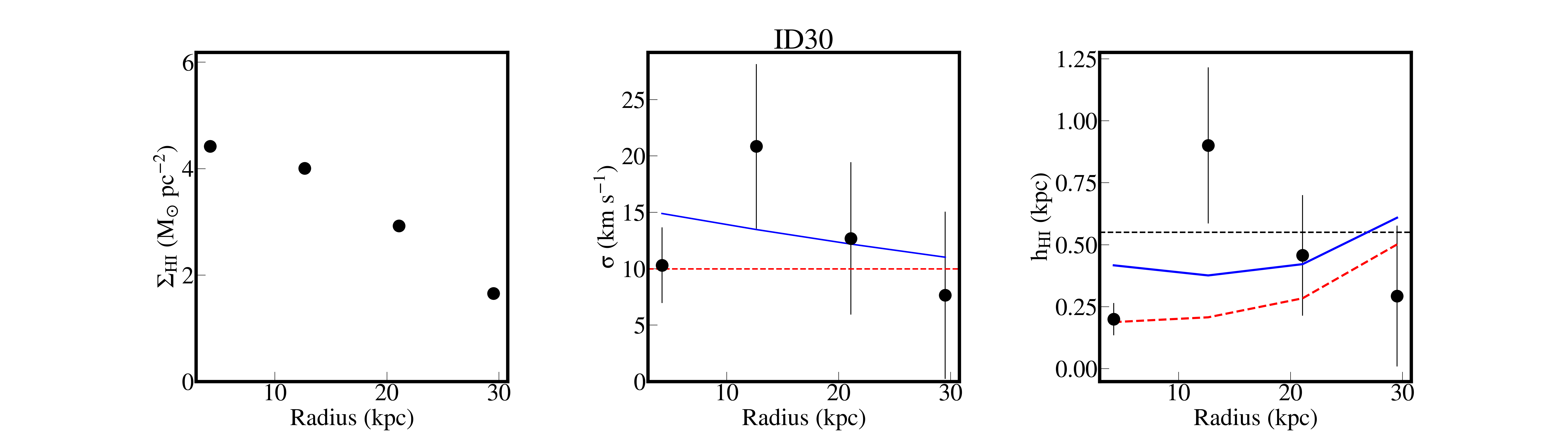}

\caption{HI scaleheight of the remaining HI-rich galaxies (cont.)}
\label{figb4}
\end{center}
\end{figure*}

\begin{figure*}
\begin{center}
\includegraphics[width=18.5cm]{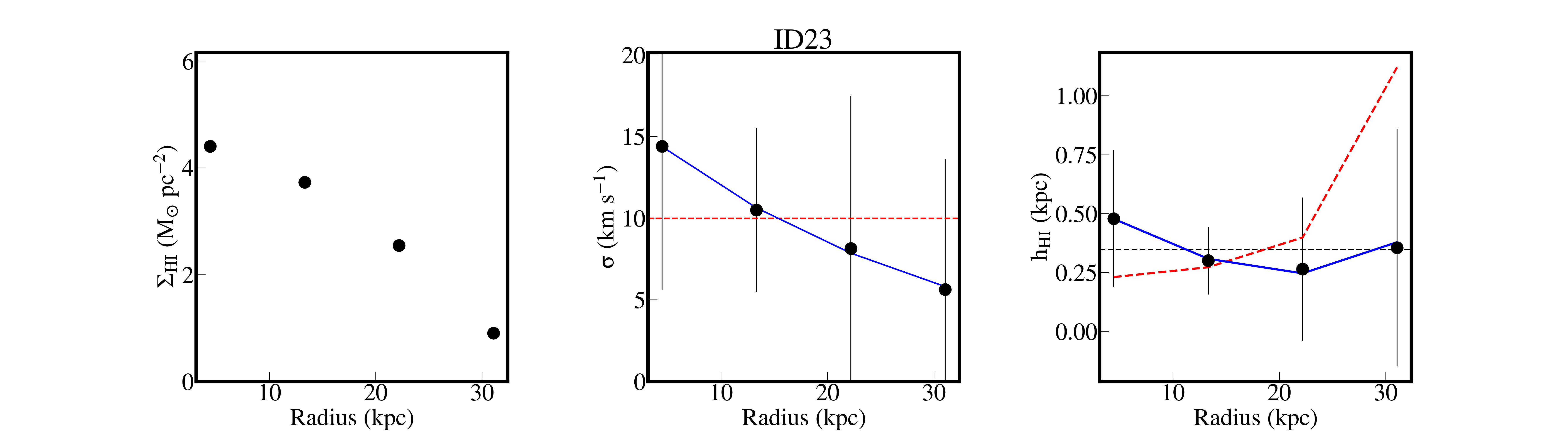}
\includegraphics[width=18.5cm]{ID25scale_height_n.png}
\includegraphics[width=18.5cm]{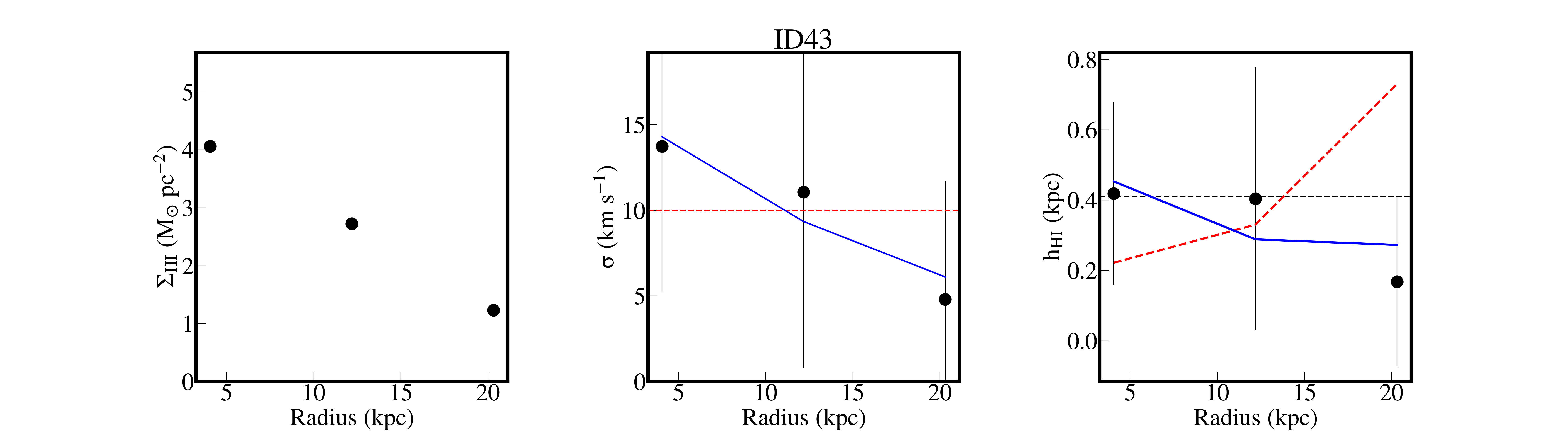}
\caption{HI scaleheight of the remaining control galaxies. See Figure 5 for details.}
\label{figb5}
\end{center}
\end{figure*}

\end{document}